\documentclass[12pt]{article}  % styl for ZAMP
\def\users{us}
\def\users{world}

\usepackage{a4}
\usepackage{epsf}
\usepackage{graphicx}
\usepackage{amsmath}
\usepackage{amssymb}
\usepackage{mathrsfs}
\numberwithin{equation}{section}
\usepackage{srcltx}
\usepackage{color}

 \numberwithin{equation}{section}

\newtheorem{theorem}{Theorem}[section]
\newtheorem{lemma}[theorem]{Lemma}
\newtheorem{proposition}[theorem]{Proposition}
\newtheorem{example}[theorem]{Example}
\newtheorem{remark}[theorem]{Remark}
\newtheorem{definition}[theorem]{Definition}

\usepackage{amsmath} 

\usepackage{bookmark,fancyvrb} 
\VerbatimFootnotes

\usepackage{color}
\definecolor{gt}{rgb}{0.1,0.2,0.1}
\definecolor{tr}{rgb}{0.1,0.1,0.3}
\newcounter{mycounter}
 0

\newcommand{\vep}{\varepsilon}

\numberwithin{equation}{section}

\newcommand\mathbbm{\mathbb}

\usepackage{amsfonts}
\usepackage{amssymb}
\usepackage{mathrsfs} %needed for script fonts
\usepackage[normalem]{ulem} %for insert-replace macros. the normalem option keeps italics

\newcommand\kt{^k_\tau}

\renewcommand{\r}{\varrho}

\renewcommand{\d}{{\rm d}}

\newcommand{\R}{\mathbb R}

\newcommand{\N}{\mathbb N}

\newcommand{\bbC}{\mathbb C}

\newcommand{\bbD}{\mathbb D}
\newcommand{\bbE}{\mathbb E}

\newcommand{\Ee}{{\boldsymbol E}}
\newcommand\bfS{{\boldsymbol S}} 
\newcommand{\EEtau}{{\Ee}_{\tau}}
\newcommand{\DTEE}{\DT{\Ee}} 
\newcommand{\DTEEtau}{\DT{\Ee}_{\tau}}
\newcommand{\barEEtau}{\barEe_{\tau}}

\renewcommand{\O}{{\Omega}}

\newcommand{\QED}{\ \hfill$\ \hfill\Box$}

 \newcommand{\DD}[2]{\begin{array}[t]{c}#1\vspace*{-.8em}\\_{#2}\end{array}}

 \newcommand{\dd}[2]{\DD{\begin{array}[t]{c}\underbrace{#1}\vspace*{1mm}\end{array}}{\mbox{\footnotesize\rm #2}}}

\newcommand{\barEe}{{\hspace*{.2em}\overline{\hspace*{-.2em}\Ee\hspace*{-.1em}}\hspace*{.1em}}}
\newcommand{\underlineEe}{{\hspace*{.1em}\underline{\hspace*{-.1em}\Ee\hspace*{-.15em}}\hspace*{.15em}}}
\newcommand{\barchi}{{\hspace*{.1em}\overline{\hspace*{-.1em}\chi\hspace*{-.1em}}\hspace*{.1em}}}
\newcommand{\barc}{{\hspace*{.0em}\overline{\hspace*{-.0em}\CHI\hspace*{-.0em}}\hspace*{.0em}}}
\newcommand{\barsigma}{{\hspace*{.07em}\overline{\hspace*{-.07em}\sigma\hspace*{-.07em}}\hspace*{.07em}}}
\newcommand{\barmu}{{\hspace*{.1em}\overline{\hspace*{-.1em}\mu\hspace*{-.1em}}\hspace*{.1em}}}
\newcommand{\bard}{{\hspace*{.1em}\overline{\hspace*{-.1em}d\hspace*{.1em}}\hspace*{-.1em}}}
\newcommand{\barf}{{\hspace*{.15em}\overline{\hspace*{-.15em}f\hspace*{.0em}}\hspace*{-.0em}}}
\newcommand{\bartheta}{{\hspace*{.1em}\overline{\hspace*{-.1em}\theta\hspace*{.0em}}\hspace*{-.0em}}}
\newcommand{\barw}{{\hspace*{.15em}\overline{\hspace*{-.15em}w\hspace*{-.1em}}\hspace*{.1em}}}
\newcommand{\barbff}{{\hspace*{.15em}\overline{\hspace*{-.15em}\bff\hspace*{.0em}}\hspace*{-.0em}}}
\newcommand{\barbfv}{{\hspace*{.1em}\overline{\hspace*{-.1em}\bfv\hspace*{-.05em}}\hspace*{.05em}}}

 \newcommand\dott[1]{\DT{#1}}

 \definecolor{gray}{rgb}{0.9,0.9,0.9}
 \newcommand\DT[1]{\mathchoice
                  {{\buildrel{\hspace*{.1em}\text{\LARGE.}}\over{#1}}}
                  {{\buildrel{\hspace*{.1em}\text{\Large.}}\over{#1}}}
                  {{\buildrel{\hspace*{.1em}\text{\large.}}\over{#1}}}
                  {{\buildrel{\hspace*{.1em}\text{\large.}}\over{#1}}}}
 \newcommand\DDT[1]{\mathchoice
    {{\buildrel{\hspace*{.1em}\text{\LARGE.\hspace*{-.1em}.}}\over{#1}}}
    {{\buildrel{\hspace*{.1em}\text{\Large.\hspace*{-.1em}.}}\over{#1}}}
    {{\buildrel{\hspace*{.1em}\text{\large.\hspace*{-.1em}.}}\over{#1}}}
    {{\buildrel{\hspace*{.1em}\text{\large.\hspace*{-.1em}.}}\over{#1}}}}

 \newcommand\dx{\,{\rm d}x}
 \newcommand\dS{\,{\rm d}S}
 \newcommand\dt{\,{\rm d}t}

 \newcommand\bfeps{{\boldsymbol\vep}}

 \newcommand\bfz{\boldsymbol z}
 \renewcommand\kt{^{k}_\tau}

 \newcommand\kkt{^{k-1}_\tau}

 \newcommand\bfh{\mathbf h}
\newcommand\bfq{{\boldsymbol q}}

 \newcommand\CHI{c}
 \newcommand\bbm{d}
 
 \newcommand\dtt{\mathfrak d}
 \usepackage{mathrsfs}
 \renewcommand\mathcal{\mathscr}

 \newcommand\bff{{\boldsymbol f}}
 \newcommand\bfu{{\boldsymbol u}}
 \newcommand\bfv{{\boldsymbol v}}
\newcommand\bfs{{\boldsymbol s}}

 \def\deltauu{\frac{\bfu\kt-\bfu\kkt}\tau}
 \def\deltau{\dtt\kt\bfu}
 
 \def\deltam{\dtt\kt\bbm}
 \def\deltamm{\frac{\bbm\kt-\bbm\kkt}\tau}

 \def\deltaxx{\frac{\CHI^k_\tau-\CHI\kkt}\tau}
 \def\deltax{\dtt\kt c}

\newcommand\phiterm{\varphi_{_{\mathrm{TH}}}^{}}
\newcommand\phimech{\varphi_{_{\mathrm{ME}}}^{}}
\newcommand\phichem{\varphi_{_{\mathrm{CH}}}^{}}
\newcommand\phichemstar{\varphi_{_{\mathrm{CH}}}^*}
\newcommand\phimechchem{\varphi_{_{\mathrm{ME/CH}}}^{}}
\newcommand\phitot{\varphi_{_{\mathrm{TOT}}}^{}}
\newcommand\eterm{e_{_{\mathrm{TH}}}^{}}

 \newcount\hour \newcount\minute
 \hour=\time 
 \divide \hour by 60
 \minute=\time
 \loop \ifnum \minute > 59 \advance \minute by -60 \repeat

 \usepackage{ulem}
 \usepackage{cancel}
 \ifthenelse{\equal{\users}{world}}{

         \newcommand{\DELETE}[1]{}

         \newcommand{\COMMENT}[1]{}
         \newcommand{\TCOMMENT}[1]{}
         \newcommand{\COL}[1]{{}{#1}}
         \newcommand{\REM}[1]{}
\def\TTT{\color{black}} 
\def\EEE{\color{black}} 

 \topmargin=-2cm\textheight=24cm
}	
 {
 
 \usepackage[notcite,notref]{showkeys}
 \definecolor{brown}{rgb}{0.6,0.2,0.2}

  \newcommand{\COMMENT}[1]{{\color{blue}{}\uuline{#1}{}}} 
  \newcommand{\DELETE}[1]{{\color{brown}\cancel{#1}{}}}

  \newcommand{\TCOMMENT}[1]{\COMMENT{\parbox{5em}{\tiny #1}}}
 
 \newcommand{\REM}[1]{\marginpar{\color{blue}\bfseries\tiny{{}#1}}}
 \newcommand{\COL}[1]{{\color{blue}{#1}}}
 \def\TTT{\color{magenta}} 
\def\EEE{\color{black}} 
}

\ifthenelse{\equal{\users}{gt}}{
         \newcommand{\GT}[1]{#1}
}
{\newcommand{\GT}[1]{}}

\title{Thermomechanics of damageable
 materials\\ under diffusion:modeling and analysis\thanks{This research has been  supported by GA\,\v CR through
the projects 201/10/0357 ``Modern mathematical and computational 
models for inelastic processes in solids'',
13-18652S
``Computational modeling of damage and transport processes 
in quasi-brittle materials'',
14-15264S 
``Experimentally justified multiscale modelling of shape memory alloys'',
by the CENTEM project no.\ CZ.1.05/21.00/03.0088 
 (cofounded from ERDF within the OP RDI programme, M\v SMT \v CR) at 
New Technologies Research Centre (Univ.\ of West Bohemia, Plze\v n), 
and by INdAM-GNFM through projects
``Mathematical Modeling of Morphing Processes'' and  ``Mathematical Modeling of Shape Changes in Soft Tissues''.
}
}

\author{Tom\'a\v s Roub\'\i\v cek\thanks{Mathematical Institute, Charles University,
Sokolovsk\'a 83, CZ-186~75~Praha~8, Czech Republic,
and
Institute of Thermomechanics of the ASCR,
Dolej\v skova 5, CZ-182 00 Praha 8, Czech Republic,
email: \texttt{tomas.roubicek@mff.cuni.cz}.}\,\, and Giuseppe Tomassetti\thanks{Universit\`a di Roma ``Tor Vergata'' 
- Dipartimento di Ingegneria Civile e Ingegneria Informatica,
Via Politecnico, I-00133~Roma, Italy,
email:\texttt{tomassetti@ing.uniroma2.it}.}}

\date{July 30, 2014}

\begin{document}
\begin{sloppypar}
\allowdisplaybreaks
\maketitle

\def\pippo{\thanks{This research has been  supported by GA\,\v CR through
the projects 201/10/0357 ``Modern mathematical and computational 
models for inelastic processes in solids'',
%201/12/0671 
%``Variational and numerical analysis in nonsmooth continuum mechanics'',
13-18652S
``Computational modeling of damage and transport processes 
in quasi-brittle materials'',
14-15264S 
``Experimentally justified multiscale modelling of shape memory alloys'',
by the CENTEM project no.\ CZ.1.05/21.00/03.0088 
 (cofounded from ERDF within the OP RDI programme, M\v SMT \v CR) at 
 New Technologies Research Centre (Univ.\ of West Bohemia, Plze\v n), 
and by INdAM-GNFM through projects
``Mathematical Modeling of Morphing Processes'' and  ``Mathematical Modeling of Shape Changes in Soft Tissues''.
}}

\begin{abstract}
We propose a thermodynamically consistent general-purpose 
model describing diffusion of a solute or a fluid in a solid undergoing
possible phase transformations and damage, beside possible visco-inelastic processes. 
Also heat generation/consumption/transfer is considered. Damage is modelled as rate-independent. 
The applications include metal-hydrogen systems with metal/hydride 
phase transformation, poroelastic rocks, 
structural and ferro/para-magnetic 
phase transformation, water and heat transport in concrete,
and, if diffusion is neglected, 
plasticity with damage and viscoelasticity, etc. For the ensuing system of partial differential equations 
and inclusions, we prove existence of solutions by a carefully devised semi-implicit 
approximation scheme of the fractional-step type. 
\end{abstract}

\noindent \textbf{AMS classification:} 35K55, %Nonlinear parabolic equations
 35Q74, %PDEs in connection with mechanics of deformable solids
 74A15, % solids, thermodynamics
 %74A30, %Nonsimple materials.
 74R20, % Anelastic fracture and damage
 74N10, % PT in solids, displacive transformation
 74F10, %Fluid-solid interactions(including aero- and hydro-elasticity, porosity
 76S99, %Flows in porous media; filtration; seepage: None of the above, but in this section
 80A17, %thermodynamic of continua.
 80A20. %Heat and mass transfer, heat flow.
\medskip

\noindent\textbf{Keywords:} {visco-elastic porous solids, incomplete damage, 
diffusion driven by chemical-potential gradient, phase transformations, 
poroelastic materials, plasticity, swelling, creep,
quasilinear parabolic systems, 
time-discretisation, a-priori estimates, convergence.}

%\COL{\texttt{NOTE: Added Remark 3.9!}}

\section{Introduction}
%        ~~~~~~~~~~~~
This paper addresses damage processes in visco-elastic materials undergoing
various diffusion or inelastic processes with the goal to develop
a general, multi-purpose model with wide application range and covering,
in particular, several particular models occuring in the literature 
in a unifying way.

We use the concept of internal variables, here specifically damage $d$ 
and a phase-field $\chi$. For the damage phenomenon, we consider the simplest 
scalar concept and therefore introduce an additional scalar 
variable $\bbm$, which we regard as an additional microscopic kinematical 
descriptor. This phase-field may, however, involve a lot of 
components, interpreted e.g.\ as content of a solute or a fluid, plastic
strain, porosity, viscous strain causing creep, etc. 
The primary fields in our model will be:
\begin{itemize}
\vspace*{-.3em}\item displacement $\bfu:[0,T]\times\Omega\to
\R^3$,
\vspace*{-.0em}\item phase field $\chi
:[0,T]\times\Omega\to K\subset\R^N$,
\vspace*{-.0em}\item concentration $c
:[0,T]\times\Omega\to\R^+$,
\vspace*{-.0em}\item damage variable $\bbm
:[0,T]\times\Omega\to [0,1]$,
\vspace*{-.0em}\item temperature $\theta
:[0,T]\times\Omega\to\R^+$,
\end{itemize}
where $\Omega\subset\R^3$ is a considered domain occupied by the solid body, $T>0$ is a fixed time horizon.
We work in the small-strain approximation and we assume
that the linear strain has the following form:
 \begin{equation}\label{eq:1}
   \bfeps(\bfu)=\Ee+\bbE\chi,
 \end{equation} 
where $\bfeps(\bfu)=\frac12(\nabla\bfu)^\top+\frac12\nabla\bfu$ is the total
linear strain, $\Ee$ is its elastic part, and $\bbE\chi$ is its anelastic part 
dependent linearly on
 the phase field $\chi$. Here $\mathbb E$ is a third-order tensor, which we think as the mapping of $\mathbb R^N$ into $\mathbb R^{3\times 3}$ defined by $(\mathbb E\chi)_{ij}=\mathbb E_{ijk}\chi_k$. Using the above decomposition we can cover the concept that
the free energy as well as the dissipation energy are let to depend on the 
elastic strain $\Ee$ rather than the total strain $\bfeps(\bfu)$. 
For a more general ansatz see Remark
%\TTT
{s}~\ref{rem-nonlin} and 
%\EEE 
\ref{rem:general-ansatz} below.

The main difficulty is the combination of thermo-visco-(in)elastic-diffusive
model with damage. As always, some simplifications are necessary mainly for 
analytical reasons. In particular, beside the mentioned small-strain concept, 
we admit only an incomplete damage. Furthermore, we perform analysis on a 
special ansatz, see \eqref{def-of-psi} below.  In particular, among other 
approximations, we neglect thermal expansion effects, direct coupling between 
concentration and strain (see Remark \ref{rem-dec1} below), direct coupling 
between concentration and damage, and coupling between damage and temperature. 
We also neglect cross-effects between dissipation in $\chi$ and $\bbm$, 
cf.\ \eqref{dissip-pot-m}, as well as the cross-effects
in the transport processes, i.e.\ the Soret/Peltier effects. On the other
hand, in case of a multi-component diffusant (when $c$ would be vector
valued) some cross effect could simply be modelled by a nondiagonal {$\mathbf M$}
in \eqref{zero-cross-effects}. 
Let us emphasize that the complete damage would bring serious mathematical 
difficulties; in the isothermal case it has been addressed in 
\cite{HaiKra14DCHS} using a formulation by a time-dependent domain.

Typical application of damage in engineering models assumes
that this destruction process is activated, irreversible, and much faster
than the time scale of the other internal processes or than external loading.
This is reflected by the dissipation potential being nonsmooth 
at zero rate, not everywhere finite, and homogeneous degree 1, 
cf.\ \eqref{def-of-xi} below and 
\cite{FraMar98RBFE,MiRoZe10CDEV,Thom10PhD,ThoMie10DNEM}, and also \cite{Fr2002a}. 
This also substantially distinguishes
between the character and the subsequent treatment of the damage $d$ and 
of the phase field $\chi$ which otherwise occur in a lot of places 
in a position of an overall internal variable $(\chi,d)$ if $\bbE$ in 
\eqref{eq:1} would just ignore the component $d$ and thus one could simply merge 
\eqref{qui} with \eqref{pluto}, or \eqref{chi-eqb} with \eqref{damage-eqb}, 
etc., similarly like e.g.\ in \cite{rocca2014degenerating}. 
%We point out that 
%\TTT
There are many variants of models which combined diffusion and
elasticity under damage in literature, some of them only isothermal 
and regularizing damage by rate-dependent terms like 
\cite{BHKS??MAPF} and/or involving a gradient term for the concentration of 
diffusant leading to a diffusion equation of Cahn-Hilliard type 
\cite{BHKS??MAPF,HaiKra14DCHS}, and of course very many 
other models not considering damage, like 
\cite{rocca2014degenerating,RocRos??ESTC}. \EEE
%A similar model has been recently studied in \cite{BHKS??MAPF} \TTT
%where, in contrast
%to the model here, \EEE inertia 
%and viscosity are neglected, damage evolution is modeled as a rate-independent 
%process, the free energy contains a gradient term for the concentration of 
%diffusant leading to a diffusion equation of Cahn-Hilliard type, and where 
%undamaged and damaged material are characterized by different coercivity 
%properties of the elastic energy, see \cite[Eq.\,(2)]{BHKS??MAPF}.

The main analytical result here is the proof of existence of weak solutions
of the initial-boundary-value problem for the thermodynamical system.
To this goal, a rather constructive method is used which 
%\TTT 
yields a conceptual 
algorithm \EEE that may serve for designing efficient
numerical strategy 
%\TTT 
after a suitable spatial discretisation \EEE 
(not performed here, however). 

It should be emphasized that there are several important differences from 
\cite{RoubT2014DCDS}: the elastic
strain (instead of the total strain) is systematically used also 
for dissipative processes and the internal variables are 
{split} into
two sets, both subjected to a 
substantially different treatment 
and allow for 
qualification of the free energy $\psi$ well fitted to damage. On top of it, 
the mathematical treatment of the heat-transfer equation is simplified like in 
by holding the formulation in terms of the couple temperature-enthalpy 
(instead of mere temperature). 

The paper is organized as follows: In Section~\ref{sec:model} we formulate 
the general model and briefly outline the underlying physics, referring mainly 
to the previous work \cite{RoubT2014DCDS}.
Then, in Section~\ref{sec:examples} we present a menagerie of various 
specific examples illustrating a wide range of applications. Eventually, in 
Section~\ref{sec:assumptions} summarizes the main existence result,
whose proof is performed in Section~\ref{sec:analysis}.

Thorough the whole article, we use the convention that 
small italics stands for scalar, small and capital bold stand, respectively, for elements of $\mathbb R^3$ and elements of $\mathbb R^{3\times 3}$. We shall occasionally use Greek letters to denote elements of $\mathbb R^N$, where $N$ is the number of internal variables incorporated in the phase field $\chi$. Consistent with this rule is our usage of the letter $\sigma$ to denote the conjugate variable of $\chi$, instead of stress.\color{black} For the reader's convenience we summarize in the following table the symbols mostly used in the rest of this paper
\begin{center}
\ \ \ \ \ \ \ \begin{minipage}[t]{0.40\linewidth}

$\bfu$ displacement,

$\chi$ phase field,

$c$ concentration of diffusant,

$d$ damage variable,

$\boldsymbol S$ stress,

$\boldsymbol f$ bulk force,

$\boldsymbol f_{\rm s}$ surface force,

$\sigma$ internal microforce associated to $\chi$,

$\boldsymbol\Sigma$ microstress,

$\gamma$, $g$ bulk microforces (eventually set to 0),

$e$ internal energy,

\end{minipage}
\hfill 
\begin{minipage}[t]{0.52\linewidth}

$\bfeps(\bfu)=\frac12\left(\nabla\bfu+\nabla\bfu^\top\right)$ small-strain tensor,

$\psi$ free energy,

$\theta$ temperature,

$\eta$ entropy,

$\mu$ chemical potential,

$\bfh$ flux of diffusant,

$h$ bulk supply of diffusant (eventually set to 0),

$h_{\rm s}$ surface supply of diffusant, 

$\bfq$ heat flux,

$q$ bulk heat supply (eventually set to 0).
\end{minipage}
\end{center}

\section{The general model: balance laws and dissipation inequality}\label{sec:model}
The model derivation follows quite closely that of \cite{RoubT2014DCDS}.
In order to derive a system of partial differential equations governing the 
evolution of the primary fields, we adopt the point of view of 
\cite{Gurtin1996a}. 
{The balance equations are considered as:}
\begin{subequations}\label{eq:98}
\begin{align}
 &&&\varrho\DDT\bfu-\textrm{div}\,{\bfS}=\bff,&&\text{{\sf(Force balance)}}
\label{paperino}
\\ &&&\sigma-\textrm{div}\,{\boldsymbol\Sigma}=\gamma,&&\text{{\sf(Microscopic balance for phase field)}}\label{qui}
\\
 &&&\DT\CHI+\textrm{div}\,\bfh=h,&&\text{{\sf(Solute balance)}}\label{quo}
\\ &&&s-\textrm{div}\,\bfs=g.&&\text{{\sf(Microscopic balance for damage)}}\label{pluto}
\intertext{The 
second and the fourth equations involve systems of microscopic forces $(\sigma,s)$ and microscopic stresses $(\Sigma,\bfs)$ that perform work, respectively, on $(\DT\chi,\DT\bbm)$ and $(\nabla\DT\chi,\nabla\DT\bbm)$. Consistent with this interpretation is the expression of the internal power expenditure that appears on the right-hand side of the 
energy equation:
}
&&&\DT e+\textrm{div}\,\mathbf q=
q+
{\bfS}\!:\!\bfeps(\DT\bfu)
 +\sigma\cdot\DT\chi
\nonumber \\&&&\qquad\qquad\ \ 
+\boldsymbol{\Sigma}\!:\!\nabla\DT\chi+s\DT\bbm+\bfs\cdot\nabla\DT\bbm
 +\mu\DT\CHI-\bfh{\cdot}\nabla\mu,&&\ \text{{\sf (Energy balance)}}
 \label{qua}   
 \end{align}
\end{subequations}
where also the energetic contribution from the diffusion of a chemical species is accounted for by the last two terms. The choice of constitutive prescriptions, which we shall make into is guided by the dissipation inequality:
 \[
 \DT\psi+\eta\DT\theta-\mu\DT\CHI\le \bfS\!:\!\bfeps(\DT\bfu)
 +\sigma\!\cdot\!\DT\chi+{\boldsymbol\Sigma}\!:\!\nabla\DT\chi+s\DT\bbm+\bfs\!\cdot\!\nabla\DT\bbm
 -\bfh\cdot\nabla\mu-\frac1\theta\mathbf q\!\cdot\!\nabla\theta,
 \] 
where $\psi=e-\eta\theta$ is the free energy. The dissipation inequality can 
be rewritten by exploiting the decomposition \eqref{eq:1}:
 \begin{equation}\label{eq:5b}
 \DT\psi+s\DT\theta-
 %(
 \mu
 %-\bfS\!:\!\EC)
\DT c
 \le \bfS\!:\!\DTEE
 +(\sigma+\bfS\!:\!\bbE)\cdot\DT\chi
 +\boldsymbol\Sigma\!:\!\nabla\DT\chi+s\DT\bbm+\bfs\cdot\nabla\DT\bbm-\bfh\cdot\nabla\mu
 -\frac 1 \theta\mathbf q\cdot\nabla\theta.
 \end{equation}
By using standard arguments, one can show that the free energy is ruled by 
a constitutive equation of the following type:
 \begin{equation}
   \label{eq:2}
   \psi=\varphi(\Ee,\chi,c,\bbm,\theta,\nabla\chi,\nabla\bbm).
 \end{equation}
Moreover, if entropy $\eta$ does depend on the same list of variables, then 
necessarily
 \begin{equation}
   \label{eq:3}
   \eta=-\partial_\theta^{}\varphi(\Ee,\chi,c,\bbm,\theta,\nabla\chi,\nabla\bbm).
 \end{equation}
 The dissipation inequality can be rendered in a more compact form by defining:
 \begin{subequations}\label{split-all}
 \begin{align}
 &\bfS_{\rm d}=\bfS-\partial_\Ee^{}\varphi,\label{split-S}
\\\label{split-sigma}
 &\sigma_{\rm d}=\sigma+\bfS\!:\!\bbE-\partial_\chi^{}\varphi,
\\
 &\mathbf\Sigma_{\rm d}={\boldsymbol\Sigma}-\partial_{\nabla\chi}^{}\varphi,\\
 &s_{\rm d}=s-\partial_{\bbm}\varphi,\\
 &\bfs_{\rm d}=\bfs-\partial_{\nabla\bbm}^{}\varphi,\\
\label{def-of-mu} &\mu_{\rm d}=\mu-\partial_c^{}\varphi.
\end{align}
\end{subequations}
In \eqref{split-sigma}, $\bfS\!:\!\bbE$ means the vector $\sum_{i,j=1}^3\bfS_{ij}\bbE_{ijk}$.
 With that splitting, one indeed obtains the 
\emph{reduced dissipation inequality}
 \begin{equation}\label{zg}
 0\le \bfS_{\rm d}:\DTEE+\sigma_{\rm d}\cdot\DT\chi+{\boldsymbol\Sigma}_{\rm d}\!:\!\nabla\DT\chi+s_{\rm d}\DT\bbm+\bfs_{\rm d}\cdot\nabla\DT\bbm+\DT c\mu_{\rm d}-\bfh\cdot\nabla\mu
 -\frac 1 \theta\bfq\cdot\nabla\theta.
 \end{equation}

\begin{remark}[{\sl Rheology of the model}]\label{rem-rheol}
From \eqref{split-S}, we can see that the total mechanical stress is 
$\bfS=\partial_\Ee^{}\varphi+\bfS_{\rm d}$, in what follows considered as
$\bfS=\partial_\Ee^{}\varphi+\bbD\DTEE$, cf.\ \eqref{dissip-pot-m}.
This is the classical Kelvin-Voigt rheological model. Through the 
internal parameters contained possibly in $\chi$, we can easily 
combine it to get some more complicated solid-type rheological models 
as Jeffreys' material or a so-called 4-parameter solid etc.
\end{remark}

\begin{remark}[{\sl An alternative understanding}]\label{rem-alternative}
%It is worth noticing that 
In the expression of the free energy we could 
%also
%\TTT 
alternatively 
%\EEE
 use as state variables the total strain $\bfeps(\bfu)$ in place of the elastic 
strain 
%\TTT by making a substitution $\Ee=\bfeps(\bfu){-}\bbE\chi$
%\EEE. 
In particular, the mechanical energy in \eqref{def-of-psi} could be written as $\phimech(\bfeps(\bfu){-}\bbE\chi,d)$, which would highlight the different roles played by $\chi$ and $d$; note that, in this case, the term $\mathbf S:\bbE$ in \eqref{split-sigma} would be incorporated in $\partial_\chi\varphi$, and thus \eqref{split-sigma} would be replaced by $\sigma_{\rm d}=\sigma-\partial_\chi\varphi$. This is ultimately the reason why we can afford viscous dissipation for $\chi$ and not for $d$. 
%\TTT 
Note that then
$\partial_\bfeps^{}\varphi$, which is standardly understood as the 
elastic part of the stress tensor, is indeed equal to $\partial_\Ee^{}\varphi$
we used in \eqref{split-S}. On the other hand, $-\partial_\chi^{}\varphi$,
which is standardly understood as the driving force for
evolution of $\chi$, is what is
$\bfS\!:\!\bbE-\partial_\chi^{}\varphi$ in \eqref{split-sigma}. 
\end{remark}

%\COMMENT{WE DEFINED $\sigma$ ONLY IMPLICITLY VIA 
%\eqref{split-sigma} with \eqref{dissip-pot-m} -- IS IT WHAT WE WANT?
%WHY NOT $\sigma=\partial_\chi\varphi$??? }

\begin{remark}[
%\TTT
{\sl Difficulties with nonlinear ansatz}\EEE]\label{rem-nonlin}
%Similarly, 
One can 
%also 
%\TTT 
attempt to \EEE generalize the linear term $\bbE\chi$ \TTT in \eqref{eq:1} 
\EEE but a strong nonlinear dependence may affect the semi-convexity 
assumption \eqref{assum-conve-e-chi} below. To better clarify this last point, 
assume that we replace the decomposition \eqref{eq:1} with 
$\bfeps(\bfu)=\Ee+\Ee_{\rm tr}(\chi)$, where $\Ee_{\rm tr}$ is now a non-linear 
function. Suppose also that the mechanical energy has the simple form 
$\phimech(\Ee)=\frac 12 \mathbb C\Ee:\Ee$. Then, the energy functional pertaining 
to $\bfu$ and $\chi$ would be
\begin{equation}
  \label{eq:102}
  (\bfu,\chi)\mapsto\int_\Omega \frac 12 \mathbb C(\bfeps(\bfu)-\Ee_{\rm tr}(\chi)):(\bfeps(\bfu)-\Ee_{\rm tr}(\chi))\,\dx
\end{equation}
\TTT{but} \EEE a simple calculation shows that if the function $\Ee_{\rm tr}$ is 
non-linear, then the integrand in \eqref{eq:102} cannot be convex (and not even 
semiconvex) with respect to the variables $\bfeps(\bfu)$ and $\chi$.
\end{remark}

\section{Specialization and examples}\label{sec:examples}
%        ~~~~~~~~~~~~~~~~~~~~~~~~~~~
The system of partial differential equations governing the evolution of the primary fields is obtained by combining the balance equations \eqref{eq:98} with appropriate constitutive prescriptions for the dissipative parts of the auxiliary fields, consistent with the reduced dissipation inequality \eqref{zg}. We make the following choice:
 \begin{subequations}\label{interm}
 \begin{align}
 &&&\bfS_{\rm d}=\bbD\DTEE,&&
 \sigma_{\rm d}\in \partial_{\DT\chi}\zeta(\Ee,\chi,c,\bbm,\theta,\DT\chi),
&& s_{\rm d}\in \partial_{\DT\bbm}{\xi}(\chi,\DT\bbm),&&
\label{dissip-pot-m}
\\\label{zero-cross-effects}
 &&&\bfq=-\mathbf K(\Ee,\chi,\bbm,\CHI,\theta)\nabla\theta,\hspace*{-1em}
 &&\bfh=-\mathbf M(\Ee,\chi,\bbm,\CHI,\theta)\nabla\mu,\hspace*{-1em}&&&&
 \\
 &&&\boldsymbol{\Sigma}_{\rm d}=0,&& \bfs_{\rm d}=0,&&\mu_{\rm d}=0.&&
\label{zero-dissip}\end{align}
 \end{subequations}
We also set to null bulk sources in \eqref{eq:98}, except for the force 
balance \eqref{paperino}, that is:
\begin{equation}
  \label{eq:89}
  \gamma=0,\qquad h=0,\qquad g=0,\qquad q=0.
\end{equation}
Here $\zeta$ and $\xi$ are (possibly non-smooth) dissipation potentials,
both $\zeta(\Ee,\chi,\bbm,\cdot)$ and $\xi(\chi,\cdot)$ being convex
and possibly nonsmooth at 0. 
More specifically, $\xi(\chi,\cdot)$ is assumed convex positively 
homogeneous degree-1, so it is always nonsmooth at 0. Assuming irreversible
(i.e.\ unidirectional) damage, $\xi$ will be in the form 
\begin{align}\label{def-of-xi}
\xi(\chi,\DT d)=\begin{cases}\alpha(\chi)|\DT d|&\text{if }\DT d\le0\\
\infty&\text{otherwise}\end{cases}
\end{align}
with $\alpha:\R^N\to\R^+$ denoting a phenomenological fracture-toughness 
coefficient, meaning the specific energy in J/m$^3$ need (and dissipated) by 
accomplishing the damage. As we shall see later, the degree-1 homogeneity 
of the dissipation potential $\xi$ in $\DT d$ has the consequence that $\DT d$ 
will be a measure in general. On the other hand, neither $\Ee$ nor $d$ are 
continuous. Thus, $\alpha$ in \eqref{def-of-xi} cannot be made to depend on 
$\Ee$ and $\bbm$ neither on $\theta$ but, on the other hand, its dependence on 
$\bbm$ is legitimate because we will have a regularity of $\bbm$ at disposal, 
cf.\ \eqref{est-of-Delta-m} below.

The involvement of non-trivial dissipative terms $\sigma_{\rm d}$ and 
$s_{\rm d}$ is essential. In fact, the  non-smooth and unbounded dissipation 
potential for $d$ allows us to incorporate the fenomenology of irreversibility 
of damage and its character to be an activated process. Moreover, we benefit 
from the dissipative term for $\chi$ to control the terms containing $\DT\chi$ 
and on the right-hand side of the heat equation, which would anyway be present
because of an adiabatic coupling. Moreover, the specification of a non-trivial 
viscous stress in the first of \eqref{dissip-pot-m} 
%\TTT 
which leads to 
Kelvin-Voigt-type visco-elastic rheological models as mentioned in 
Remark~\ref{rem-rheol} \EEE is needed to handle inertia, otherwise it can be 
neglected, cf.\ Remark~\ref{D=0} below. Clearly, we may also include non-trivial 
viscous-like contributions $\mathbf\Sigma_{\rm d}$ and $\mathbf s_{\rm d}$ in the 
stress-like terms under divergence in the balance equations (\ref{eq:98}c,d). 
Including these contributions would not significantly extend the range of 
applications of our treatment and would make all notation heavier (moreover 
they would make our analysis more trivial to some extent, as far as the equation 
governing $\chi$). Similar considerations hold for a dissipative contribution to 
chemical potential.

To 
%enable
%\TTT 
facilitate 
%\EEE
\COL{the} mathematical analysis of \eqref{eq:98} with \eqref{split-all} and 
\eqref{interm}, we must restrict the generality of the free energy\eqref{eq:2}. 
In particular:
\begin{enumerate}
\item If $\theta$ and $\Ee$ would generally be coupled in \eqref{eq:2},
in general a concept of nonsimple material
would be needed to have a control on $\nabla\Ee$,
cf.\ \cite{Roub13NCTV}.
An exception would be if $\theta$ would appear linearly, i.e.\ 
a term like $\theta\varphi(\Ee)$, which then would not contribute to the
heat capacity, cf.\ e.g.\ \cite{RoubT2014DCDS}. 

\item If $c$ and $\Ee$ would generally be coupled in \eqref{eq:2},
then the chemical potential $\mu=\partial_c^{}\varphi$ from 
\eqref{def-of-mu} with \eqref{zero-dissip} would depend on $\Ee$ and
again the estimate \eqref{est-of-nabla-concentration} of $\nabla c$ would 
need a control of $\nabla\Ee$, i.e.\ the concept of nonsimple materials.
Some alternative option would be introduce a gradient of $c$ 
into the free energy, i.e.\ capillarity concept, 
see also Rem~\ref{rem-dec1} below.
\item If $d$ and $\theta$  would  be coupled in \eqref{eq:2},
than the adiabatic heat term $\partial_{c\theta}^2\varphi(....,\theta)\DT d$
would occur but $\DT d$ is a measure and $\theta$ tends typically to jump
exactly at the points where $\DT d$ concentrates, and such term analytically
would not be well defined.
\item If $c$ and $\theta$ would generally be coupled in \eqref{eq:2},
the adiabatic heat term $\partial_{c\theta}^2\varphi\DT c$ would occur in
the right-hand side of the heat-transfer equation, cf.\ \eqref{heat-eq-cont}
below, but we will not have any estimate on $\DT c$ except 
the ``dual'' estimate \eqref{a-priori-IIe} below, 
so this term would not be controlled as a measure.\color{black}
\item If $c$ and $d$ would generally be coupled in \eqref{eq:2}, we would see a 
term $\partial_\bbm^{}\phichem(\chi,\bbm,\CHI)\DT\bbm$ in \eqref{chain-rule-for-c}. 
However, the term $\partial_\bbm^{}\phichem(\chi,\bbm,\CHI)$ would be not continuous, 
because of the aforementioned lack of continuity of $\CHI$, and hence its 
combination with the measure $\DT\bbm$ would not be well defined.
\end{enumerate}

These five requirements led us to 
make the following partly decoupled ansatz 
for the free energy \eqref{eq:2}, distinguishing mechanical, chemical, and 
thermal parts in the sense:
 \begin{align}
 \psi=\!\!\!\!\!\dd{\phimech(\Ee,\chi,\bbm)+\phichem(\chi,c)+
 \phiterm(\chi,
\theta)}
{$\ \ \ \ \ \ \ =:\phitot(\Ee,\chi,c,\bbm,\theta)$}\!\!\!\!\!
 +\frac12\kappa_1|\nabla\chi|^2
+\frac12\kappa_2|\nabla\bbm|^2+\delta^{}_K(\chi)+\delta_{[0,1]}(\bbm).
 \label{def-of-psi}
 \end{align}
 
 Here $K$ is a convex set where values of $\chi$ are assumed to lie,
while  $\bbm$ is valued at $[0,1]$, 
and $\delta$ stands for the indicator function, i.e.\ 
$\delta^{}_K(\cdot)$ is equal to 0 on
$K$ and $+\infty$ otherwise and similarly for $\delta_{[0,1]}(\cdot)$. 
The consequence 
of the decoupling \eqref{def-of-psi} is that the {\it internal energy}, 
defined by Gibbs' relation $e=\psi-\theta\eta$ and occurring already in \eqref{qua},
can be written as:
\begin{align}
e&=\eterm(\chi,\theta)
+\!\!\!\!\!\dd{\phimech(\Ee,\chi,\bbm)+\phichem(\chi,c)}
{$\ \ =:\phimechchem(\Ee,\chi,c,\bbm)$}\!\!\!\!\!
+\frac12\kappa_1|\nabla\chi|^2
+\frac12\kappa_2|\nabla\bbm|^2+\delta^{}_K(\chi)+\delta_{[0,1]}(\bbm),
\label{eq:4}
\end{align}
where the thermal part of the internal energy is
 \begin{equation}
   \label{eq:5}
   \eterm(\chi,\theta)=\phiterm(\chi,\theta)-\theta\partial_\theta\phiterm(\chi,\theta).
 \end{equation}
Altogether, from \eqref{eq:98} with \eqref{split-all},
\eqref{interm}, and \eqref{def-of-psi}, we obtain the following system:
\begin{subequations}\label{systemb}
\begin{align}\label{mombalb}
&\varrho\DDT{\bfu}-\mathrm{div}\,\big(\partial_{\Ee}^{}\phimech(\Ee,\chi,\bbm)
 %\mathbb C(\chi,\bbm){\Ee}
 {+}\bbD\DTEE
 %-\mathrm{div}\,\mathfrak{h}
 \big)=\bff&&\text{with } {\Ee}=\bfeps(\bfu)-\bbE\chi
,
 %-c\EC,
 %-\theta\ET,\label{bal-mech}
 %\ \ \mathfrak{h}=\mathbb H\nabla{\Ee}+\mathbb G\nabla\DT{\Ee}_{\rm e},
 %\end{equation}\begin{equation}
 \\
 %\nonumber
 &\partial_{\DT\chi}\zeta(\Ee,\chi,c,\bbm,\theta,\DT\chi)-\kappa_1\Delta \chi
 +\partial_{\chi}\phitot(\Ee,\chi,c,\bbm,\theta)
 \nonumber \\ & \qquad\qquad\qquad\qquad\qquad
-\bbE^\top{:}(\partial_{\Ee}^{}\phimech(\Ee,\chi,\bbm){+}\bbD\DT\Ee)
 +{\sigma_{\rm r}}\, \ni0&&\textrm{with }
{\sigma_{\rm r}}\in \partial \delta^{}_K(\chi),\label{chi-eqb}
 \\
 &\label{NS}
 \DT c-\mathrm{div}\big(\mathbf M(\Ee,\chi,c,\bbm,\theta)\nabla\mu\big)
=0&&\textrm{with }\mu=\partial_c^{}\phichem(\chi,c),
 \\
 &\partial_{\DT\bbm}{\xi}(\chi,\DT{\bbm})-\kappa_2\Delta \bbm
+\partial_{\bbm}\phimech(\Ee,\chi,\bbm)
+s_{\rm r}
\ni 0\!\!\!&&\textrm{with } s_{\rm r}\in \partial \delta_{[0,1]}(\bbm),
\label{damage-eqb}
 \\\nonumber
 &\DT w-\textrm{div}(\mathbf K(\Ee,\chi,c,\bbm,\theta)\nabla\theta)
=%\partial_{\bbm}\phiterm(\chi,\bbm,\theta)\DT\bbm+
\partial_{\DT\chi}\zeta(\Ee,\chi,c,\bbm,\theta,\DT\chi)\cdot\DT\chi{+}\partial_{\chi}\phiterm(\chi,\theta)\cdot\DT\chi
\hspace{-8em}
 \\
&\qquad\qquad\qquad\qquad\qquad
+
\bbD\DTEE\!:\!\DTEE
 %\bbD(\chi,\bbm)\DTEE
 %+\mathbb G\nabla\DTEE{\vdots}\nabla\DTEE
-\alpha(\chi)\DT d
+\mathbf M(\Ee,\chi,c,\bbm,\theta)\nabla\mu{\cdot}\nabla\mu
 %\end{multline}\begin{equation}
 %-\EC{:}(\partial_{\Ee}^{}\phimech(\Ee,\chi,\bbm)+\bfS_{\rm d})
&&\textrm{with}\quad w=\eterm(\chi,\theta).
\label{heat-eq-cont}
\end{align}
\end{subequations}
Note that $\zeta(\Ee,\chi,c,\bbm,\theta,\cdot)$ is possibly nonmooth so 
that $\partial_{\DT\chi}\zeta(\Ee,\chi,c,\bbm,\theta,\DT\chi)$ may be 
multi-valued, but anyhow we assume in \eqref{single-valued} below that 
the term 
$\partial_{\DT\chi}\zeta(\Ee,\chi,c,\bbm,\theta,\DT\chi)\cdot\DT\chi$ 
appearing in the right-hand side of \eqref{heat-eq-cont} and also in 
\eqref{engr-equality} below is well-defined. 
This occurs e.g.\ if $\zeta(\Ee,\chi,c,\bbm,\theta,\DT\chi)$
has a form $a(\Ee,\chi,c,\bbm,\theta)|\DT\chi|
+b(\Ee,\chi,c,\bbm,\theta)|\DT\chi|^2$ with some $a,b\ge0$.
We complete \eqref{systemb} 
with the following boundary condition:
 \begin{subequations}\label{bccb}
  \begin{align}
 &\big(\partial_{\Ee}^{}\phimech(\Ee,\chi,\bbm)
 %\mathbb C(\chi,\bbm){\Ee}
 %+\bbD(\chi,\bbm)\DTEE
 {+}
\bbD\DTEE
 %\big(\mathbb C(\chi,\bbm){\Ee}
 %+\bbD(\chi,\bbm)\DTEE
 %-\mathrm{div}\,\mathfrak{h}
 \big)\mathbf n
 %-\divS(\mathfrak{h}\mathbf n)
 =\bff_{\rm s},
 \label{BC-t-1bb}
 %\\\label{BC-t-1+}&\,\mathfrak{h}{:}({\mathbf n}\otimes{\mathbf n})=0,
 \\\label{BC-chi}
 &
 \,\nabla\chi\cdot{\mathbf n}=0,
 \\
 &\label{bc34bb}
 \,\mathbf M(\Ee,\chi,c,\bbm,\theta)\nabla \mu
 \cdot\mathbf n=h_{\rm s},
\\&\,\nabla\bbm\cdot{\mathbf n}=0,
\\ &\label{BC-t-2bb}
 \mathbf K(\Ee,\chi,c,\bbm,\theta)
 \nabla\theta\cdot\mathbf n=q_{\rm s}
 \end{align}
\end{subequations}
to be valid on the boundary $\Gamma$ of $\Omega$.
Using the convention like $\bfu(\cdot,t)=:\bfu(t)$,
we complete the system by the initial conditions:\\[-1.3em]
 \begin{align}
 &\label{ICb}
 \bfu(0)=\bfu_0,\qquad\DT\bfu(0)=\bfv_0,\qquad\chi(0)=\chi_0,
\qquad\CHI(0)=\CHI_0,\qquad\bbm(0)=\bbm_0,\qquad 
w(0)=w_0:=\phiterm(\chi_0,\theta_0).
 \end{align} 
The above formulated model is very general and can be understood as 
truly multi-purpose. As some important examples from very diverse physically 
well justified models, let us mention the 
 following:

\begin{example}[{\sl Metal/hydride phase transformation}]\label{exa-hydro}
\upshape
Some metals or intermetallics 
allows for relatively easy diffusion of hydrogen and undergo a transformation
into hydrides accompanied by markable volume changes (even up to 30\%), cf.\ 
\cite{Latro2004JPCS}. Several continuum models have been proposed, accompanied by their analytical study \cite{Bonetti2012,Bonetti2007,Chiod2011MMAS,RoubT2014DCDS}. 
Such big volume variation then 
often cause damage. In this case, we have $N=1$ and the (scalar-valued) 
 phase field $\chi$ stands for the volume fraction related to 
the metal/hydride transformation, while the variable $c$ 
is the hydrogen concentration. The free energy can be considered as
\begin{align}
\phitot(\Ee,\chi,c,\bbm,\theta)
 &=\!\!\!\!\!\dd{\frac12\bbC(\bbm)\Ee{:}\Ee}
{$=:\phimech$}\!\!\!\!\!
+\!\!\!\!\!\dd{\frac k2\big|a(\chi){-}\CHI\big|^2+\phi_1(\CHI)}{$=:\phichem$}\!\!\!\!\!
\label{hydrid-example}
+\!\!\!\!\!\dd{%\frac12\theta^2\bbC(\bbm)\bfalpha{:}\bfalpha+
\,\phi_2(\chi,\theta)_{_{_{_{_{_{_{}}}}}}}\!}
{$=:\phiterm$}
\!\!\!\!\!
\end{align} 
where $k>0$ is a (typically large)  coefficient and $a(\chi)$ a function that maps values of phase field into values of concentration, which reflects the fact that different phases typically manifest themselves at different concentrations. \color{black}As marked, \eqref{hydrid-example} is consistent with our splitting ansatz \eqref{def-of-psi}.
Without damage but considering still thermal expansion, this model has been treated in 
\cite{RoubT2014DCDS}. Thus, the present paper generalizes \cite{RoubT2014DCDS} towards this important damage phenomenon. In particular we may take $a(\chi)=\chi$. In this case, $\chi$ is somehow related to the concept of \emph{nonlocal species concentration}, a concept introduced in \cite{ubachs2004nonlocal}  to handle the difficulties related to the numerical solution of the Cahn-Hilliard equation and further elaborated in \cite{Anand2014cahn}, where the extra field (here $\chi$) is called \emph{micromorphic concentration}. 
As $\partial_{cc}^2\phichem(\chi,c)=k+\phi_1''(c)$, the uniform convexity in $c$ requires just a semi-convexity 
of $\phi_1$ if one admits $k$ large. 
Moreover, the physically relevant 
non-negativity of $c$ can then be ensured by taking for $\phi_2$ a function having a controlled singularity at zero $\phi_1$. An example may be the coarse-grain chemical free energy obtained from regular solution theory:
\begin{equation}\label{eq:104}
 \phi_2(\chi,\theta)=\phi_3(\theta)+A\theta(\chi\log(\chi)
+(1{-}\chi)\log(1{-}\chi))+B\chi(1{-}\chi),
\end{equation}
with $A$ and $B$ positive constants, cf.\ \cite{jones2002soft}. 
Of course, 
%give 
%\TTT 
due to 
%\EEE
the large strains 
%\TTT 
sometimes 
%\EEE 
involved in the hydration 
phase transformation, 
%our choice of 
developing the model in the small-strain setting 
%\TTT 
might be 
%\EEE
%is 
debatable. For a model coupling diffusion and damage in the large setting, 
we refer to \cite{duda2007modeling}.
\end{example}

\begin{example}[{\sl Poroelastic rocks}] 
\upshape
\def\porosity{\varpi}
Geophysical models of lithospheres in short time scale (less that 1 mil. yrs)
counts small strains. Typical phenomena to capture are damage, 
 plasticity (considered here rate-dependent), and water propagation through 
porous rocks,
%\TTT, 
cf.\ e.g.\ 
%\EEE
\cite{wang2004poroelasticity}. The phase field 
is then composed from plastic strain $\pi_{\rm pl}$ and porosity $\porosity$, 
while $c$ is the water concentration, cf.\ e.g.\ 
\cite{HaLyAg04EDPP,HaLyAg05DRDC,LyaHam07DEFF}. The free energy is considered as
$\psi=\frac12\lambda(\porosity,\bbm)
 ({\rm tr}\,{\Ee})^2+
 G(\porosity,\bbm)|{\Ee}|^2+\frac12M(\porosity,\bbm)|\beta{\rm tr}\,{\Ee}{-}c{+}\porosity|^2
+c_{\rm v}\theta({\rm ln}\theta{-}1)
$ with $\Ee=\bfeps(\bfu)-\pi_{\rm pl}$ and
with 
$\lambda$ and $\mu$ the first Lam\'e coefficient, 
$G$ the shear modulus (also called the second Lam\'e coefficient
and denoted by $\mu$ which here denotes, however, the chemical potential)
$M$ the Biot modulus, and $\beta$ the Biot coefficient.
The chemical potential $\mu$ and the Nernst-Plack equation \eqref{NS} play the role of a 
pressure and of the Darcy equation. 
This however does not directly fit with the splitting 
\eqref{def-of-psi} because $\CHI$ and $\Ee$ are directly coupled 
unless the Biot modulus or the Biot coefficient would be zero.
To make it consistent with \eqref{def-of-psi}, similarly 
as in Example~\ref{exa-hydro} we again distinguish 
between the water concentration $c$ and the water content $\gamma$;
in fact, \cite{HaLyAg04EDPP,HaLyAg05DRDC,LyaHam07DEFF}
speak about water content, although consider the water flow 
governed by the Darcy equation like if it would be a water concentration.
Therefore, we consider the augmented phase field as $\chi=(\pi_{\rm pl},\porosity,\gamma)$
and then the free energy as 
 \begin{align}\nonumber
 %\psi
\phitot
({\Ee},\porosity,\gamma,c,\bbm,\theta
)
 %\\\nonumber&
&=%\frac12\lambda(\porosity,\bbm)
 \!\!\!\!\!\dd{
\frac12\lambda(\porosity,\bbm)
 ({\rm tr}\,{\Ee})^2+
 G(\porosity,\bbm)|{\Ee}|^2+\frac12M(\porosity,\bbm)|\beta{\rm tr}\,{\Ee}{-}\gamma{+}\porosity|^2}{$=:\phimech$}
\!\!\!\!\!
 \\[-.4em]\label{poro-example}&\ \ \ \ \ 
 +\!\!\!\!\!\dd{\frac12k|\gamma-c|^2}{$=:\phichem$}\!\!\!\!\!
+\!\!\!\!\!\dd{c_{\rm v}\theta({\rm ln}\theta{-}1)_{_{_{_{_{_{_{}}}}}}}\!}{$=:\phiterm$}
\!\!\!\!\!
 %+\frac12\kappa_1|\nabla{\Ee}|^2
\qquad\ \text{ with }\ \Ee=\bfeps(\bfu)-\pi_{\rm pl}.
 \end{align}
Clearly, \eqref{eq:1} now uses $\bbE\chi=\bbE(\pi_{\rm pl},\porosity,\gamma):=\pi_{\rm pl}$.
 In fact, typically considered $\porosity$-dependence of the Biot modulus $M$ 
 is affine (cf.\ \cite{HaLyAg04EDPP,LyaHam07DEFF})
 which would make the functional $\porosity\mapsto\psi$ non-convex (and even
 equi-semiconvexity \eqref{assum-conve-e-chi}  below cannot be expected, 
which would make technical 
 troubles in using time discretisation below). Yet, \eqref{assum-conve-e-chi}
does not necessarilly mean $M$ or $G$ or $\lambda$ to be constant.
On the other hand, a small modification 
complying with the growth/coercivity restriction \eqref{ass-fundamental} made in 
what follows is needed for facilitating the analysis, cf.\ 
Remark~\ref{rem-poro-elast} below.
However, \cite{HaLyAg04EDPP,HaLyAg05DRDC,LyaHam07DEFF}
in fact (although not explicitly) consider cross-effects in dissipation of 
damage and porosity, which here has to be neglected here otherwise 
it would require coupling of \eqref{disc-u-chi-c}
and \eqref{c-mu-disc} and stronger separate semi-convexity qualification 
likely not compatible with \eqref{poro-example},
and moreover
%, like in Remark~\ref{rem-healing} below, 
damage was considered reversible, which is 
important in particular for longer-time scale processes in rocks. Also, 
\cite{HaLyAg04EDPP,HaLyAg05DRDC,LyaHam07DEFF} consider a phenomenological flow 
rule for $\pi_{\rm pl}$ not governed directly from the free energy.
\end{example}

\begin{example}[{\sl 
Magnetic and hydride transformations in intermetallics}]
\upshape
Some intermetallic compouds exposed to hydrogen (or deuterium) exhibit not 
only metal/hydride phase transformations as mentioned above 
but also a dramatic structural trasformation analogous to the martensitic
ferro-to-para magnetic transformation, 
both mutually coupled. This is the case of Uranium- 
or rare-earth-based alloys, cf.\ \cite{havela2009f,havela1999}. Experiments show 
that hydrogenation implies a substantial increase of the magnetic ordering 
temperature and noticeable increase of specific heat, cf.\ e.g.\  
\cite{fujita2003itinerant,kolwicz2005specific}. 
There does not seem any models for it in mathematical literature to exist,
however. The structural cubic-to-cubic phase transformation in
particular single-crystal grains is similar e.g.\ to cubic-to-tetragonal 
so-called martensitic transformation which may also exhibit magnetic phase 
transformation in intermetalics like NiMnGa. Here, densely packed cubic 
configuration is typically paramagnetic (because electron orbits of particular
atoms over-cover each other) while sparsely packed cubic configuration 
is typically ferromagnetic) in analog with cubic austenite and tetragonal
martensite in NiMnGa. Counting this analogy,
one can assembly a model from already existing particular models for 
the martensitic  transformation of the Souza-Aurichio type as e.g.\ in
\cite{AuReSt09MMSM,KreSte16ENFT}, for the ferro-to-para magnetic 
transformation as 
in \cite{PoRoTo10TCTF} in combination (and ignoring gyromagnetic effects) as in
\cite{RouSte??MSMA}, and for the metal/hydride transformation under diffusion 
as above in \eqref{hydrid-example}. When considering the magnetic variation 
slow and thus neglecting all induced electrical effects and when neglecting 
also the self-induced demagnetising field, one can consider the (vector-valued) 
phase field $\chi=(
\lambda,m)$ composed from 
the volume fraction $\lambda$ related to the metal/hydride transformation
and $m$ the magnetization vector. Magnetization hysteresis effects due to
magnetic-domain pinning effects can be accounted for through the nonsmooth 
potential $\zeta$, cf.\ 
\cite{tomassetti2004evolution,PodioT2006ITM,RoubT2011MMMAS,RoubTZ2009JMAA}. Again, damage is an 
important phenomenon which can even pulverize such materials due to markable 
volume changes during the cubic-to-cubic structural transformation. Combining 
\cite{RouSte??MSMA,roubivcek2013phase,RoubT2014DCDS}, the free energy can be considered as
\begin{align}\nonumber 
&\phitot(\Ee,\chi,c,\bbm,\theta)=\frac12\bbC(\bbm)
\Ee{:}\Ee+\frac k2\big|a(\lambda
){-}c\big|^2\!+\phi_1(\CHI)+\phi_2(\theta)
+\frac{a_0}2(\theta{-}\theta_{\rm c})|m|^2\!+\frac{b_0}4|m|^4\!
+\alpha_1(\theta)\lambda
\\%\nonumber
 &\quad=\!\!\!\!\!\dd{\frac12\bbC(\bbm)\Ee{:}\Ee+\frac{b_0}4|m|^4}
{$=:\phimech$}\!\!\!\!\!
+\!\!\!\!\!\dd{\frac k2\big|a(\lambda
){-}\CHI\big|^2+\phi_1(\CHI)}{$=:\phichem$}\!\!\!\!\!
+\!\!\!\!\!\dd{
\phi_2(\theta)
+\frac{a_0}2(\theta{-}\theta_{\rm c})|m|^2+\alpha_1(\theta)\lambda}
{$=:\phiterm$}\!\!\!\!\!
\label{ferro-parra-example}
\end{align} 
and with 
$\theta_{\rm c}>0$ the Curie temperature
above which the ferromagnetic phase does not exists, $a_0,\,b_0>0$,
and $\alpha_1$ is the temperature-dependent latent-heat density of
the medium. Now $\frac{\kappa_1}2|\nabla\chi|^2$ occurring in 
\eqref{def-of-psi} involves also $\frac{\kappa_1}2|\nabla m|^2$ which is 
called an exchange energy in magnetism. Further considerations may go beyond 
the ansatz in Section~\ref{sec:model} by involving global effects through 
a demagnetising field as in \cite[Remark~11]{PoRoTo10TCTF} 
or dynamical electro-magnetic effects including Joule
heating through the Maxwell system possibly in an eddy-current approximation 
as in \cite{RouSte??MSMA}.
\end{example}

\begin{example}[{\sl Water and heat propagation in concrete}]
\upshape
An important application in civil engineering is water/vapor and heat 
transport in concrete undergoing damage and creep. The creep strain is most 
influenced by the moisture and temperature distribution. 
The model would be quite similar to
\eqref{poro-example} although some microscopical mechanisms
behind the free energy and some dissipation mechanism are different. 
In particular, $c$ would be again intrepreted as water concentration and $\mu$ as pressure. Moreover, the decomposition \eqref{eq:1} would be adapted to the form considered in models of concrete that take into account creep, shrinkage and thermal strains \cite{bazant2004temperature,gawin2007modelling}, namely, $\bfeps(\bfu)=\Ee+\Ee_{\rm cr}+\Ee_{\rm sh}+\Ee_{\rm th}$, where $\Ee_{\rm cr}$ would be the creep strain, $\Ee_{\rm sh}$ is the shrinkage strain caused by change of 
{moisture}, and $\Ee_{\rm th}$ would be the thermal strain. Of course, since direct coupling with $c$ and $\theta$ is not included in our analysis, we would resort to a penalization like in \eqref{hydrid-example} for $c$ and a similar one for $\theta$, cf.\ Remark~\ref{rem:general-ansatz} below. Furthermore, in this case the constraint $\chi\in K$ would be dropped, that is, we would set $K=\R^N$. Moreover, the dissipation potential $\zeta$ would be smooth at $\DT\pi_{\rm pl}=0$
because, in contrast to plasticity, the creep is typically not an
activated processes. Also the ageing mechanisms are play a role,
influencing the activation threshold $\alpha$ of damage.
There is a lot of phenomenological models 
in literature, although typically not based on rational thermomechanics
to be directly fitted in the framework \eqref{systemb}; cf.\ e.g.\ 
\cite{KrKoKr12MPAS}.
\end{example}

\begin{remark}[{\sl A general treatment of swelling or thermal expansion}]
\label{rem:general-ansatz}
\upshape
Some models would rather need, instead of \eqref{eq:1}, \EEE a more general 
form 
\begin{align}\label{swelling-thermo-expand}
\bfeps(\bfu)=\Ee+\bbE\chi+s(c)\bbE^{}_\mathrm{sw}+r(\theta)\bbE^{}_\mathrm{ex}
\end{align}
with some  $\bbE^{}_\mathrm{sw}$ and $\bbE^{}_\mathrm{ex}$
swelling and thermal-expansion tensors and some \EEE
(possibly even nonlinear) mappings 
$s,r:\R\to\R$ modelling  swelling or thermal expansion effects, respectively.
 Thermal expansion would influence also the entropy: in fact, \eqref{eq:3} 
is no longer valid and, in the spirit of Remark~\ref{rem-alternative},
we should rather consider $\phimech(\bfeps(\bfu){-}\bbE\chi{-}
%\theta\mathbf A_{\rm th}
r(\theta)\bbE^{}_\mathrm{ex},\chi,d)$ which gives an additional contribution 
$r(\theta)'\bbE_\mathrm{ex}^\top
\partial_\Ee\phimech(\bfeps(\bfu){-}\bbE\chi{-}r(\theta)\bbE^{}_\mathrm{ex},\chi,d)$
to the enthalpy and also a corresponding contribution to the heat capacity
$-\theta\partial_{\theta\theta}^2\psi$. Additional troubles have already 
been mentioned in Remark~\ref{rem-nonlin}.
% Note also that, in the alternative setting outlined in 
%Remark~\ref{rem-alternative}, %in the case of thermal expansion, 
%the expression of the mechanical energy would read: $\phimech(\bfeps(\bfu){-}\bbE\chi{-}
%%\theta\mathbf A_{\rm th}
%\bbE_2\theta,d)$, where 
%%$\mathbf A_{\rm th}$
%$\bbE_2$ is the termal-expansion tensor. ......................................
\EEE
To fit such more general situations to our ansatz, we can little modify
the particular problem by introducing additional phase-field variables,
say $\chi_1$ and $\chi_2$, and augment appropriately the free-energy parts 
$\phichem$ or $\phiterm$. More specifically, $\phichem$ can be augmented by
$\frac12k|s^{-1}(\chi_1){-}c|^2$ or $\phiterm$ by 
$\frac12k|r^{-1}(\chi_2){-}\theta|^2-\frac12k\theta^2$. In fact, we already used 
the former term in \eqref{hydrid-example}, 
\eqref{poro-example}, and \eqref{ferro-parra-example}, too.
Considering a presumably large constant $k$ makes $\chi_1$ mostly nearly 
equal to $s(c)$ and $\chi_2$ nearly equal to $r(\theta)$  and, instead 
of \eqref{swelling-thermo-expand}, we then take
\begin{align}
\bfeps(\bfu)=\Ee+\bbE\chi+\chi^{}_1\bbE^{}_\mathrm{sw}+\chi^{}_2\bbE^{}_\mathrm{ex},
\end{align}
which complies with \eqref{eq:1}. \EEE
Note that the positive definiteness $\partial_{cc}^2\phichem$
is kept and the heat capacity $-\theta\partial_{\theta\theta}^2\phiterm(\chi,\theta)$
is not affected by this modification.  The energy balance \eqref{qua} 
is then affected through non-thermal terms, which exhibits similar modelling 
effects but makes the analysis easier\EEE. 
\end{remark}

 %\begin{remark}[Simple variant]\upshape
 %Omitting the hyperstress by putting $\bbH=0$ and $\bbG=0$,
 %the driving force $\frac12\partial_\chi^{}\mathbb C(\chi,\bbm){\Ee}{:}{\Ee}$ and $\frac12\partial_\bbm^{}\mathbb C(\chi,\bbm){\Ee}{:}{\Ee}$ in \eqref{chi-eq} and \eqref{damage-eq}
 %would become bounded only in $L^\infty(0,T;L^1(\Omega))$-type space
 %which which would destroy the regularity argument ......................... 
 %below. Alternatively to the non-simple material variant, we could 
 %consider a nonquadratic contribution $\mathcal C({\Ee})$
 %to the free energy \eqref{def-of-psi}, 
 %assuming $\mathcal C({\Ee})\ge\epsilon|{\Ee}|^4$
 %for some $\epsilon>0$. Then the mentioned driving terms
 %would again got an $L^\infty(0,T;L^2(\Omega))$-structure but 
 %the terms ..... ..... ..... ..... ..... 
 %would additionally occur in \eqref{chi-eq} and \eqref{damge-eq}.
 %\COMMENT{the remark to omit in simple-material variant!!!}\end{remark}

%\TTT
\begin{example}[{\sl Damage by freezing water}]
\upshape
%After having a certain model 
Application of thermal expansion discussed in Remark~\ref{rem:general-ansatz}
%\eqref{swelling-thermo-expand}, we can model 
can be modelling a very common phenomenon that water propagating in porous medium
expands during water-ice phase transformation and may cause damage. This 
happens e.g.\ in concrete or in poroelastic rocks discussed above.
Other occurrence is in polymer membranes in fuel cells \cite{YTLL06ESTP}, etc.
%\COMMENT{Tomas, I don't know this reference. PLEASE COMPLETE!}, etc.
%The vector of internal variables is now $\chi=(\gamma,\lambda)$ with $\gamma$ 
%the total water content and
The thermal expansion is now 
%depends rather
(approximately) proportional to 
the overall amount of ice, 
i.e.\ $c\lambda$ where $c$ is the water concentration 
and  $\lambda$ the volume fraction of ice versus 
liquid water. Naturally, $\lambda=\lambda(\theta)$.
%More specifically, in the so-called enthalpy
%formulation of the heat equation, cf.\ also Definition~\ref{def} below,
%we can consider $\lambda=\lambda(w)$ with $w$ denoting the enthalpy.
Like in Remark~\ref{rem:general-ansatz}, we consider rather the linear
splitting $\bfeps(\bfu)=\Ee+\beta\bbE^{}_\mathrm{ex}$ with $\beta$ and $\gamma$ 
new phase-field variables which are expectedly close to $\lambda(\theta)c$ and
$c$, respectively. This can be achieved e.g.\ by adding terms like 
$\frac12k|\theta-\lambda^{-1}(\beta/\gamma)|^2-\frac12k\theta^2
+\frac12k|\gamma-c|^2$ to $\phiterm+\phichem$. 
%$\lambda$ than temperature
%itself and is naturally proportional to the water+ice content $\gamma$,
%i.e.\ $\bfeps(\bfu)=\Ee+r(\lambda)\gamma\bbE^{}_\mathrm{ex}$.
%\COMMENT{ BUT IT IS JUST THE SITUATION \eqref{eq:102}!!!}
%\COMMENT{Tomas, I don't know what to do here. Please fix it.}
As ice practically cannot move, the mobility of water 
$\mathbf M=\mathbf M(\theta)$ falls to very small values if $\theta$ is below
freezing point $\theta_{_{\rm F}}$. The heat capacity $c_{\rm v}(\gamma,\theta)=
-\theta\partial_{\theta\theta}^2\phiterm(\gamma,\theta)$ also depends also on the 
water content $\gamma$ and, moreover, may contain a Dirac distribution supported 
at the freezing point, i.e.\ $c_{\rm v}(\chi,\theta)=c_{\rm v,0}
+\gamma L\delta_{\theta_{_{\rm F}}}\!(\theta)$ with 
$L$ the latent heat of the water/ice phase transition. This is called the Stefan 
problem, cf.\ e.g.\ \cite{roubivcek1989stefan,visintin1996models},
%\COMMENT{STILL SOME MORE CITATIONS??} 
and then the upper bound $C$ in the first
condition in \eqref{eq:67} and also in \eqref{eq:21} is $\infty$ and the graph 
of $\vartheta(\chi,\cdot)$ 
from \eqref{eq:20} below has a horizontal segment of the length $\gamma L$
at the height $\theta_{_{\rm F}}$. Some arguments we use below must be
then slightly generalized as such $\vartheta(\chi,\cdot)$ with 
now $\chi=(\beta,\gamma)$ is not invertible 
and smooth; important, \eqref{eq:71} and the corresponding 
$\nabla\theta$-estimate holds for such generalization, too.
%More specifically, in the so-called enthalpy
%formulation of the heat equation, cf.\ also Definition~\ref{def} below,
%we can consider $\lambda=\lambda(w)$ with $w$ denoting the enthalpy.
In fact, $\lambda$ should better depend on the enthalpy $w$ rather than
on temperature $\theta$, cf.\ also Definition~\ref{def} below, but anyhow
a fine reqularization of the Stefan problem seems needed to overcome
singular character of this problem.
\end{example}

\begin{remark}[{\sl Other applications}]
\upshape
Beside, there are a lot of applications without considering any diffusion
or without any damage, i.e.\ $c$ or $d$ is void (not used). For example
magnetostriction in magnetic shape-memory alloys as in \cite{RouSte??MSMA}
with no damage and no diffusion,
or plasticity in metals with damage but no diffusion, and or a combination 
of inelastic processes with 
some more complex rhelogies involving additional internal variables in $\chi$
as eg.\ Jeffreys' model involving a creep strain, %\COL
i.e.\ it combines the 
Maxwell rheology (responsible for creep) with the Kelvin-Voigt one, 
etc.,
%\TTT 
cf.\ also Remark~\ref{rem-rheol}.\EEE 
 %\begin{itemize}\item
 %{\bf plasticity with damage:} $\chi=\pi_{\rm pl}=\,$plastic strain.
 %\item {\bf Jeffreys visco-elastic solid with plasticity:} 
 %$\chi=(\pi_{\rm pl},\pi_{\rm Mx})$ with $\pi_{\rm Mx}$ Maxwell strain. 
 %%while $c$ is again void. 
 % \COMMENT{REFERENCES?}\end{itemize}
 \end{remark}

 \begin{remark}[{\sl Some restrictions}]
 \upshape
Within our approach, we cannot unfortunately handle the dependence of 
 the coefficients in the gradient terms, i.e.\ of 
 %$\mathbb H$, 
 $\kappa_2$
 and $\kappa_1$, on damage $\bbm$ or on $\chi$, which would be natural
 in some application. This dependence would give rise the higher-order
 $L^1$-type terms in \eqref{chi-eq} and \eqref{damage-eq} which would
 destroy the regularity \eqref{est-of-Delta-m} below without
which also \eqref{by-part-for-chi} could not be proved.\color{black}
 \end{remark}

 %\COMMENT{collision of notation: $\mu$!!}

\section{Data qualification, weak formulation, and the main result}
%        ~~~~~~~~~~~~~~~~~~~~~~~~~~~~~~~~~~~~~~~~~~~~~~~~~~~~~~~~~
\label{sec:assumptions}
Beside the standard notation for the Lebesgue $L^p$-space, we will use 
$W^{k,p}$ for Sobolev spaces whose $k$-th derivatives 
are in $L^p$-spaces, the abbreviation $H^k=W^{k,2}$. We consider a fixed time interval $I=[0,T]$ and we denote
by $L^p(I;X)$ the standard Bocher space of Bochner-measurable mappings $I\to X$ with $X$ a Banach space. 
Also, $W^{k,p}(I;X)$ denotes the Banach space of mappings from $L^p(I;X)$ 
whose $k$-th distributional derivative in time is also in $L^p(I;X)$.
Also, $C(I;X)$ and $C_\text{weak}(I;X)$ will denote the Banach space of 
continuous and weakly continuous mappings $I\to X$, respectively.
Moreover, we denote by ${\rm BV}(I;X)$ the Banach space
of the mappings $I\to X$ that have 
bounded variation on $I$, and by ${\rm B}(I;X)$ the space of 
Bochner measurable, everywhere defined, and bounded mappings $I\to X$.
By ${\rm Meas}(I;X)$ we denote the space of $X$-valued measures on
$I$. 

Let us collect our main assumptions on the data: 
\begin{subequations}\label{assum-smoot}
\begin{align}
&\Omega\subset\R^3\ \text{ is a bounded domain with }\ 
\Gamma:=\partial\Omega\in C^2,
\\
  \label{eq:11}
&\phimech\in C^2(\R^N{\times}\R{\times}\R^{3{\times} 3}),\qquad\phichem\in C^2(\R^N{\times}
\R),\qquad
\phiterm\in C^2(\R^N{\times}
\R^+), 
\\
&\mathbf M\in C(\R^{3{\times} 3}{\times}\R^N{\times}\R{\times}\R{\times}\R),\qquad \mathbf K\in C(\R^{3{\times} 3}{\times}\R^N{\times}\R{\times}\R{\times}\R).\label{regul-M}\\
\label{ass-fundamental}
&\phimech(\Ee,\chi,\bbm)\ge\epsilon|\Ee|
^2-C,\ \ \ \ |\partial_{(\Ee,\chi)}\phimech(\Ee,\chi,\bbm)|\le C(1+|\Ee|),
\\
\label{growth-of-phi-c-d}
&|\partial_\chi^{}\phichem(\chi,c)|\le C(1+|c|^3
),%\\
\\
\label{eq:67}
&0<\epsilon\le -\theta\partial^2_{\theta\theta}\phiterm(\chi,\theta)\le C
\ \ \ \text{ and }\ \ \ 
\big|\partial_\chi^{}\phiterm(\chi,\theta)-\theta\partial^2_{\chi\theta}\phiterm(\chi,\theta)\big|\le C.
\\
\label{eq:27}
&\phichem(\chi,c)\ge \epsilon c^2-C,\\
\label{eq:88}
&
\Ee\mapsto\phimech(\Ee,\chi,\bbm)\ 
\textrm{ is strongly convex uniformly with respect to $\chi$ and $\bbm$}, 
\\\label{assum-conve-e-chi}
\color{blue}
&(\Ee,\chi)\mapsto
{\phimech(\Ee,\chi,\bbm)}
+
M|\chi|^2\ \textrm{ is convex for 
$M$ sufficiently large,}
\\
&c\mapsto \phichem(\chi,c), \textrm{ 
 is strongly convex uniformly with respect to }\chi,
\label{eq:43}
\\
\label{eq:14}
&\mathbf M(\Ee,\chi,c,\bbm,\theta)\ \text{ and }\ 
\mathbf K(\Ee,\chi,c,\bbm,\theta)\ \text{ 
are uniformly positive definite and bounded},
\\
\label{eq:76}
  &\exists C>0,\ \epsilon>0:\ \ \ 
|\partial_c^{}\phichem(\chi,c)|\le C+\epsilon\phichem(\chi,c),
\\[-.5em]\label{assum-dissi-poten-a}&
\zeta(\Ee,\chi,c,\bbm,\theta,\cdot):\R^N\to\R^+\ 
\textrm{ is convex 
and }\ \exists 
\epsilon>0:\ \ 
\epsilon|\DT\chi|^2\le\zeta(\Ee,\chi,c,\bbm,\theta,\DT\chi)\le\frac{1+|\DT\chi|^2}\epsilon,
\\[-.2em]\label{single-valued}&
\DT\chi\mapsto
\partial_{\DT\chi}\zeta(\Ee,\chi,c,\bbm,\theta,\DT\chi){\cdot}\DT\chi:
\R^N\to\R^+\ \text{ is single-valued and strictly convex, while 
%moreover
\EEE}
\\&\nonumber
\xi\ \text{ satisfies \eqref{def-of-xi} with }\ \alpha:\R^N\to\R^+\ 
\text{{ smooth and}} 
\\[-.2em]\label{assum-dissi-poten-b}&\hspace{10em}
\text{{ bounded together with its second gradient}
and }\
\inf\alpha(\R^N)>0,
\\
\label{assum-mphi}
  &|\partial^2_{\chi c}\phichem(\chi,c)|\le C,
\\
\label{eq:16}
&|\partial_\chi^{}\phiterm(\chi,\theta)|\le C\sqrt{1+\phimechchem(\Ee,\chi,c,\bbm)+{\eterm(\chi,\theta)}},\ \text{ and}
\\
&\big|\partial_\chi^{} \eterm(\chi,\theta)\big|\le C\,\sqrt{1+{\eterm(\chi,\theta)}}.
\label{assum-holder-1}
\end{align}
\end{subequations}
Then, we add the qualification of 
the initial data:
\begin{subequations}\label{assum-initi}
\begin{align}
&
\bfu_0\!\in\!H^1(\Omega;\R^3),\ \ \bfv_0\!\in\!L^2(\Omega;\R^3),\ \ 
\CHI_0\!\in\!H^1(\Omega),\ \ \chi_0\!\in\!H^1(\Omega;\R^N),
\ \ \bbm_0\!\in\!H^1(\Omega
),\ \ w_0\!\in\!L^1(\Omega),
\\
&
\bbm_0\!\in\![0,1],\ \ \chi_0\!\in\!K,\ \ w_0\ge0\ \ \text{ a.e.\ in }\Omega,
\ \text{ and}
\label{IC-ass-1}
\\\nonumber
&\int_\Omega
\phimech(\Ee_0,\chi_0,\bbm_0)
+\frac{\kappa_2}2|\nabla\bbm_0|^2\,\d x
\\[-.5em]&\hspace{2em}
\le\int_\Omega\Big(
\phimech(\Ee_0,\chi_0,\widetilde\bbm)
+\frac{\kappa_2}2|\nabla\widetilde\bbm|^2
+\alpha(\chi(t))(\widetilde\bbm{-}\bbm_0)\Big)\d x
\quad
\forall \widetilde\bbm\!\in\!H^1(\Omega),\ 0\le\widetilde\bbm\le\bbm_0
\text{ a.e.\ on }\Omega,
\label{damage-IC}
\end{align}
\end{subequations}
with $\Ee_0=\bfeps(\bfu_0)-\bbE\chi_0$,
and the qualification of the outer mechanical, chemical, and thermal loading:
\begin{align}\label{eq:19}
&\bff\!\in\!L^2(Q;\R^3) ,\quad \bff_{\rm s}\!\in\!L^2(\Sigma\EEE;\R^3),
\quad q_{\rm s}\!\in\!L^1(\Sigma\EEE),
\quad h_{\rm s}\!\in\!L^2(\Sigma\EEE)\ \textrm{ with }q_{\rm s}\ge0\text{ and }h_{\rm s}\ge 0 \textrm{ a.e. on }\Sigma,\
\end{align}
where we used the abbreviation $Q=(0,T)\EEE\times\Omega$ and 
$\Sigma=(0,T)\EEE\times\Gamma$ for a fixed time horizon $T>0$.
Later, ${\,\overline{\!Q\!}\,}$ will denote the closure of $Q$.

The uniform strong convexity means e.g.\ in  \eqref{eq:88} that
\begin{align}
\exists\epsilon>0\ \forall (\Ee,\tilde\Ee,\chi,\bbm)\!\in\!
\R^{3\times 3}\!\!\times\!\R^{3\times 3}\!\!\times\!K\!\times\![0,1]:\quad
\big(\partial_\Ee^{}\phimech(\tilde\Ee\tilde\Ee,\chi,\bbm)
-\partial_\Ee^{}\phimech(\Ee,\chi,\bbm)\big):(\tilde\Ee{-}\Ee)
\ge\epsilon|\tilde\Ee{-}\Ee|^2_{}.
\label{eq:88+}
\end{align}
Analogous meaning has also \eqref{eq:76}. The statament \eqref{eq:14} is as
usual understood that there exists $\epsilon>0$ such that, for all 
$\bfv\in\R^3$ and 
$(\Ee,\chi,c,\bbm,\theta)\in\R^{3\times 3}\times K\times[0,1]\times\R\times\R^+$,
it holds $\epsilon|\bfv|^2\le
\mathbf M(\Ee,\chi,c,\bbm,\theta)\bfv\cdot\bfv\le|\bfv|^2/\epsilon$ and 
$\epsilon |\bfv|^2\le\mathbf K(\Ee,\chi,c,\bbm,\theta)\bfv\cdot\bfv
\le|\bfv|^2/\epsilon$. In \eqref{eq:16}--\eqref{assum-holder-1}, we understand 
that $\phimechchem\ge0$ and $\eterm\ge0$, as we always can without loss of 
generality. The qualification \eqref{damage-IC} represents a semi-stability 
of the initial damage profile; the adjective ``semi'' refers to 
that $\Ee_0$ and $\chi_0$ are fixed on both the left- and the
right-hand sides of \eqref{damage-IC}.

Let us briefly comment the main aspects of the above assumptions:
\begin{enumerate}
\item {The smoothness of $\partial\Omega$ is here needed to ensure $H^2$ regularity of the solutions of the elliptic problems considered in Lemma \ref{lem-1+} below (see in particular the regularized equation \eqref{eq:92}), where we shall prove $L^2$ regularity of $\Delta\chi$ (see \eqref{est-of-Delta-m}). To this aim, we note that the second growth assumption in \eqref{ass-fundamental} and assumption \eqref{growth-of-phi-c-d}
on $\partial_{\chi}\phimechchem$, together with assumption
\eqref{assum-dissi-poten-a} on $\zeta$ facilitates this estimate.}
\item
In principle we could admit a general $p$-growth/coercivity with $p\le2$ in 
\eqref{ass-fundamental}. However, we cannot handle $p>2$. The main reason is 
the integrability of terms in \eqref{balance-of-d} and, if $\varrho>0$, also 
the duality in \eqref{quality-of-u}. For simplicity, we confine ourselves to 
$p=2$ only, without restricting substantially possible applications. 
\item Assumptions \eqref{eq:14} are needed to have coercivity in the diffusion 
and the heat-conduction equation.
\item 
It follows from 
the first condition in \eqref{eq:67}
 that the thermal part of the internal 
energy $\eterm$ defined in \eqref{eq:5} is continuously differentiable, and satisfies: 
\begin{align}\label{eq:29}
  0<\epsilon\le\partial_\theta \eterm(\chi,\theta)\le C.
\end{align}
Hence, the function $\theta\mapsto \eterm(\chi,\theta)$ is invertible for all $\chi$ 
and its inverse is continuously differentiable. Thus, there exists a function 
$\vartheta\in C^1(\R^N{\times}\R^+;\R^+)$ such that
\begin{align}\label{eq:20}
  \eterm(\chi,\vartheta(\chi,w))=w
\end{align}
\begin{equation}
  \label{eq:21}
  0<{1/C}\le\partial_w \vartheta(\chi,w)\le {1/\epsilon}
\end{equation}
with $\epsilon$ and $C$ referring to \eqref{eq:29}.
Moreover, we observe that 
the second condition in \eqref{eq:67} entails that
\begin{equation}
  \label{eq:64}
  |\partial_\chi^{}\vartheta(\chi,w)|\le C.
\end{equation}
\item {The semi-stability of the initial damage \eqref{damage-IC} is 
needed for the energy conservation which, in turn, is vitally needed for
the convergence in the heat equation, cf.\ Step~9 in the proof of 
Proposition~\ref{prop-conv}.}
\end{enumerate}

We can now state the weak formulation of the initial-boundary-value
problem \eqref{systemb}--\eqref{ICb}. As for the damage part, we use the 
concept of the so-called energetic solution devised by Mielke and Theil 
\cite{MieThe04RIHM}, based on the energy (in)equality and the so-called 
stability, cf.\ \eqref{damage-eq} below, and further employed in the 
thermodynamical concept in \cite{Roub10TRIP}. This formulation is 
essentially equivalent to the conventional weak formulation but excludes 
time derivatives of rate-independent variables, i.e.\ here the damage $d$.

\begin{definition}\label{def}
Given initial conditions \eqref{assum-initi} and the bulk and boundary data 
\eqref{eq:19}, the seven-tuple $(\bfu,\chi,\CHI,\bbm,\theta,\mu,w)$ 
with 
\begin{subequations}\label{weak-sln}
\begin{align} 
&\bfu\in L^\infty(I;W^{1,p}(\Omega;\R^3))\cap H^1(I;H^1(\Omega;\R^3)),
\\&\chi\in 
C_\mathrm{weak}(I;H^1(\Omega;\R^N))\cap H^1(I;L^2(\Omega;\R^N))\cap C({\,\overline{\!Q\!}\,};\R^N),
\\&c\in L^\infty(I;L^2(\Omega)),
\\&
\bbm\in {\rm B}(I;H^1(\Omega))
\cap
{\rm BV}(I;L^1(\Omega)),\ \ \ d\ge0\ \text{a.e.\ on}\ Q,
\hspace*{-5em}
\\&\label{def-of-r}
\theta\in L^r(I;W^{1,r}(\Omega))\qquad 
\text{ for any }r\in[1,5/4),\hspace*{-5em}
\\&\mu\in L^\infty(I;H^1(\Omega)),
\\&w\in L^\infty(I;L^1(\Omega))
\end{align}
\end{subequations}
{is called a weak solution to the initial-boundary-value problem 
\eqref{systemb}--\eqref{ICb} if}
\begin{subequations}\label{def-of-weak-sln}
\begin{align}\nonumber
&
\int_Q(\partial_\Ee^{}\phimech(\Ee,\chi,\bbm)+\bbD\DT\Ee)\!:\!\bfeps(\bfz)
-\varrho\DT\bfu{\cdot}\DT\bfz\dx\dt=\int_Q\bff{\cdot}\bfz\dx\dt
+\int_\Sigma \bff_{\rm s}{\cdot}\bfz\,\dS\dt
+\int_\Omega\bfv_0{\cdot}\bfz(0)\dx
\\&\hspace{9em}
\forall \bfz\!\in\!L^2(I;H^1
(\Omega;\R^3))\cap H^1(I;L^2(\Omega;\R^3)),\ \bfz(T)=0,
\\
\nonumber
&\int_Q\!\zeta(\Ee,\chi,c,\bbm,\theta,z)
+\big(
\partial_\chi^{}\phitot(\Ee,\chi,c,\bbm,\theta)
-\bbE^\top\!:\!
(\partial_\Ee^{}\phimech(\Ee,\chi,\bbm){+}\bbD\DT\Ee)+{\sigma_{\rm r}}\big)
{\cdot}(z{-}\DT\chi)+\kappa_1\nabla\chi\!:\!\nabla z\dx\dt
\\[-.3em]\label{chi-eq}
&\hspace{1em}
+\int_\Omega\!\frac{\kappa_1}2\big|\nabla\chi_0\big|^2\dx\ge
\int_Q\!\zeta(\Ee,\chi,c,\bbm,\theta,\DT\chi)\dx\dt
+\int_\Omega\!\frac{\kappa_1}2\big|\nabla\chi(T)\big|^2\dx
\quad\ \forall z\!\in\!L^2(I;H^1(\Omega;\R^M)),
\\
&
\int_Q\!\mathbf M(\Ee,\chi,c,\bbm,\theta)\nabla\mu\cdot\nabla z-c\DT z\dx\dt
=
\int_{\Sigma}h_{\rm s} z\,\dS\dt+\int_\Omega c_0z(0)\d x
\qquad\ \ \forall z\!\in\!H^1(Q),\ z(T)=0,
\\\nonumber
&\int_\Omega
\phimech(\Ee(t),\chi(t),\bbm(t))
+\frac{\kappa_2}2|\nabla\bbm(t)|^2\,\d x
\le\int_\Omega\Big(
\phimech(\Ee(t),\chi(t),\widetilde\bbm)
+\frac{\kappa_2}2|\nabla\widetilde\bbm|^2
\\[-.3em]&\hspace{9em}
+\alpha(\chi(t))(\widetilde\bbm{-}\bbm(t))\Big)\d x
\qquad\qquad\ \ \ \ \ \forall\,t\!\in\!I\ 
\forall \widetilde\bbm\!\in\!H^1(\Omega),\ 0\le\widetilde\bbm\le\bbm(t)
\text{ on }\Omega,
\label{damage-eq}\\
&%-\int_Q w \DT z\dx\dt+
\int_Q \mathbf K(\Ee,\chi,c,\bbm,\theta)\nabla\theta\cdot\nabla z
-w\DT z\, \d x\dt=\int_\Omega\big(w_0{+}\alpha(\chi_0)d_0\big)z(0)\dx
+\int_\Sigma q_{\rm s} z\, \d S\dt
\nonumber\\[-.3em]
&\qquad\qquad
+\int_Q\Big(\big(\partial_{\DT\chi}\zeta(\Ee,\chi,c,\bbm,\theta,\DT\chi)
{+}\partial_\chi^{}\phiterm(\chi,\theta)
{+}\alpha'(\chi)d\big)
{\cdot}\DT\chi
+\mathbf M(\Ee,\chi,c,\bbm,\theta)\nabla\mu{\cdot}\nabla\mu
\nonumber\\[-.3em]\label{heat-eq-weak}
&\qquad\qquad\qquad\qquad\qquad\quad
+\bbD\DTEE\!:\!\DTEE\Big)z+\alpha(\chi)d\DT z\,\d x\dt
\qquad\qquad\forall z\!\in\!W^{1,\infty}(Q),\ \ z(T)=0,
\\\nonumber
&\mathcal E_{_{\rm MC}}(T)
+\int_\Omega\alpha(\chi(T))d(T)\,\d x
+\int_Q\!\Big(\bbD\DT\Ee{:}\DT\Ee+
\big(
\partial_{\DT\chi}\zeta(\Ee,\chi,c,\bbm,\theta,\DT\chi)
{+}\partial_\chi^{}\phiterm(\chi,\theta)
{+}\alpha'(\chi)d\big){\cdot}\DT\chi
\\[-.3em]
&\hspace{1em}
+{\mathbf M}(\Ee,\chi,c,\bbm,\theta)\nabla\mu{\cdot}\nabla\mu\Big)\,\d x\d t
=\mathcal E_{_{\rm MC}}(0)+\int_\Omega\alpha(\chi_0)d_0\,\d x
+\int_Q\bff{\cdot}\DT\bfu\,\d x\d t
+\int_\Sigma\!\bff_{\rm s}{\cdot}\DT\bfu{+}q_{\rm s}{+}\mu h_{\rm s}\,\d S\d t
\label{engr-equality}
\intertext{where $\Ee\in 
H^1(I;L^2(\Omega;\R^{3\times 3}_{\rm sym}))$, $\mu$ and $w$ from 
(\ref{weak-sln}f,g), and ${\sigma_{\rm r}}\!\in\!L^2(Q;\R^N)$, 
satisfy}
&\Ee=\bfeps(\bfu){-}\bbE\chi,
\ \ \ \ \ \ \ \ \mu=\partial_c^{}\phichem(\chi,c),
\ \ \ \ \ \ \ \ w=\eterm(\chi,\theta)
,
\ \ \ \ \ \ \ \ \sigma_{\rm r}\!\in\!\partial \delta^{}_K(\chi)
\end{align}
\end{subequations}
a.e.\ in $Q$ and
where the energy occurred in \eqref{engr-equality} at time $t=0$ and ${t=}T$ is:
\begin{align}\nonumber
\mathcal E_{_{\rm MC}}(t):=\int_\O
\phimechchem(\Ee(t),\chi(t),\bbm(t),c(t))
+\frac\varrho2\big|\DT\bfu(t)\big|^2
+\frac{\kappa_1} 2\big|\nabla\chi(t)\big|^2
+\frac{\kappa_2} 2\big|\nabla\bbm(t)\big|^2\,\d x,
\end{align}
where $\phimechchem$ is defined in \eqref{eq:4}.
Eventually, the remaining three initial conditions holds:
\begin{align}\label{eq:39}
 \bfu(0)=\bfu_0,\qquad 
\chi(0)=\chi_0,\qquad
\bbm(0)=\bbm_0\qquad\text{ a.e. in $\Omega$}.
\end{align}
\end{definition}

The above definition uses several tricky points, which seems in general
quite inevitable for this sort of complicated problems, cf.\ also
\cite{RocRos??ESTC} for even more tricky definition. 
Here, \EEE let us note that we made the by-part integration in time in 
(\ref{def-of-weak-sln}a,c,e) to avoid abstract duality pairing due to only 
weak estimates on $\varrho\DDT\bfu$, $\DT c$, and $\DT w$ we will have at 
disposal, cf.\ (\ref{apriori-II+}a-c) below.
The other three initial conditions $\DT{\bfu}(0)=\bfv_0$, $c(0)=c_0$, and 
$w(0)=w_0$ are thus involved already in (\ref{def-of-weak-sln}a,c,e).
The  by-part integration used in \eqref{chi-eq} is to avoid 
the term $\int_Q\nabla\chi{\cdot}\nabla\DT\chi\,\dx\dt$ where
$\nabla\DT\chi$ would not have a good sense.
Also note that we used the by-part integration in time for 
the heat term induced by damage, i.e.\ 
\begin{align}\label{by-part-d}
\int_{\,\overline{\!Q\!}\,}\alpha(\chi)z\DT d(\d x\d t)=\int_\Omega \alpha(\chi(T))z(T)d(T)\,\d x
-\int_{Q}(\alpha'(\chi)\DT \chi z
+\alpha(\chi)\DT z)d\,\d x\d t
-\int_\Omega \alpha(\chi(0))z(0)d(0)\,\d x
\end{align}
for $z$ smooth with $z(T)=0$ in \eqref{heat-eq-weak}
or with $z=1$ in \eqref{engr-equality}
to avoid usage of
the measure $\DT d$ in Definition~\ref{def}
although here it would be well possible because 
 $\alpha(\chi)$ is a continuous function on ${\,\overline{\!Q\!}\,}$. 
More importantly, we balanced only the mechano-chemical energy 
$\mathcal E_{_{\rm MC}}$ to avoid usage of $\theta(T)$ or of $w(T)$ which 
would not have a specified meaning as the temperature as well as the thermal
part of the internal energy may
be discontinuous in time instances where $\DT d$ concentrates.

\begin{theorem}\label{th:main}
Let 
the assumptions \eqref{assum-smoot}--\eqref{eq:19} 
be valid.
Then there exists a weak solution to the initial-boundary-value 
problem \eqref{systemb}--\eqref{ICb} in the sense of Definition \ref{def}
such that also $\varrho\DDT\bfu\in L^2(I;H^1(\Omega;\R^3)^*)$, 
$\DT c\in L^2(I;H^1(\Omega)^*)$, $\DT w\in \mathrm{Meas}(I;W^{1,r/(r-1)}(\Omega)^*)$, 
$\theta\ge0$ a.e.\ in $Q$.
Moreover, the total-energy balance is satisfied ``generically'' in the
sense that, for a.a.\ $t\in I$, it holds that 
\begin{align}\label{eq:99}
&\mathcal  E_{_{\rm TOT}}(t)-\mathcal E_{_{\rm TOT}}(0)=\int_0^t\Big(\int_\Omega
\bff{\cdot}\DT\bfu 
\,\dx
+\int_\Gamma\!\bff_{\rm s}{\cdot}\DT\bfu+q_{\rm s}+\mu h_{\rm s}\,\d S\Big)\d t,
\end{align}
where, referring to $\phimechchem$ and $\eterm$ from \eqref{eq:4}--\eqref{eq:5},
the total energy is:
\begin{align}\nonumber
\mathcal E_{_{\rm TOT}}(t)&=\int_\Omega\Big(
\phimechchem\big(\Ee(t),\chi(t),\bbm(t),c(t)\big)+\eterm\big(\chi(t),\theta(t)\big)
+\frac\varrho2\big|\DT u(t)\big|^2+\frac 12 \kappa_1\big|\nabla\chi(t)\big|^2
+\frac 12\kappa_2\big|\nabla\bbm(t)\big|^2\Big)\dx.
\end{align}
\end{theorem}

We will prove this theorem in the next section by a constructive 
way which also suggest a conceptual numerical strategy which, after another
discretisation in space, would allow for an efficient computer implementation.

\begin{remark}[Poro-elastic model revisited]\label{rem-poro-elast}
\upshape
\def\porosity{p}
The poro-elastic example \eqref{poro-example} does not comply 
with 
the second condition in \eqref{ass-fundamental} 
unless $\lambda$, $G$, and $M$ is independent of $\bbm$.
Also, \eqref{poro-example} is incompatible with \eqref{assum-conve-e-chi}
in general, although one could weaken this condition and 
then make finer splitting in the time-discretisation in 
Section~\ref{sec:analysis} 
by considering $(\porosity,\gamma)$ separately from $(\Ee,\pi_\mathrm{pl})$
in \eqref{disc-u-chi-c}; here it however depends on the dissipation which 
should allow this separation.
Anyhow, due to \eqref{ass-fundamental}, the example
\eqref{poro-example} has to be slightly modified. 
One option is to pose:
 \begin{align}\nonumber
\phitot({\Ee},\porosity,\gamma,c,\bbm,\theta
)
=
\frac12
\frac{
 \lambda(\porosity,\bbm)
 ({\rm tr}\,{\Ee})^2}
 {\sqrt{1{+}\epsilon({\rm tr}\,{\Ee})^2}} 
 +G(\porosity,\bbm)\frac{|{\Ee}|^2}
 {\sqrt{1{+}\epsilon|{\Ee}|^2}}
 +\frac12M(\porosity,\bbm)\frac{|\beta{\rm tr}\,{\Ee}{-}\gamma{+}\porosity|^2}
 {\sqrt{1{+}\epsilon({\rm tr}\,{\Ee})^2}}
 \\[-.2em]\ \ \ +
 \frac12\lambda_0
 %\Big(\frac12K_0{+}\frac13G_0\Big)
({\rm tr}\,{\Ee})^2+
G_0|{\Ee}|^2
+\frac12k|\gamma-c|^2
 +c_{\rm v}\theta({\rm ln}\theta{-}1)
 %+\frac12\kappa_1|\nabla{\Ee}|^2
 %+\frac12\kappa_1|\nabla\porosity|^2
 %+\frac12\kappa_2|\nabla\bbm|^2,
\label{poro-elastic-modif} 
\end{align}
 with $\epsilon>0$ presumably small regularing parameter and with 
$\lambda_0\ge0$ and $G_0>0$. Then \eqref{ass-fundamental} is satisfied.
\end{remark}

\section{%The analysis: 
{Time discretisation and the}
proof of existence of weak solutions}\label{sec:analysis}

Our proof of Theorem \ref{th:main} is based on the following strategy:
\begin{itemize}
\item
A semi-implicit discretisation of the system \eqref{systemb} of a 
fractional-step type which decouples this system at each time level and is 
numerically stable under quite weak convexity requirements on $\phitot$,
cf.\ also \cite[Remark\,8.25]{Roub13NPDE} for a general discussion.
Namely, we use four steps in such a way that 
one solves separately (\ref{systemb}a-b), then (\ref{systemb}c), then 
(\ref{systemb}d),
and eventually \eqref{heat-eq-cont}. 
\item
The above specified splitting 
allows for only
the {relatively weak (partial)} semi-convexity assumption 
\eqref{assum-conve-e-chi}. 
In particular, we thus do not require 
any (semi)-convexity of $\phimechchem$ or even of 
$\phimech$ itself which would exclude interesting 
applications like the examples in Section~\ref{sec:examples}.

\item Further key ingredient is an $L^2$-estimate of the driving 
force 
for \eqref{chi-eq} 
together with the qualification of the dissipation potential $\zeta$
for 
$\chi$ 
to have a bounded subdifferential
so that one can obtain the additional estimate 
$\Delta\chi\in L^2(Q;\R^N)$.
This facilitates the by-part integration formula \eqref{by-part-for-chi} and, if $\Omega$ is smooth,
also the $H^2$-regularity of $\chi$ is needed to give a good sense to the 
dissipative-heat term $\alpha(\chi)\DT d$.
\end{itemize}

We use an equidistant partition of the time interval $I=I$ with a time 
step $\tau>0$, assuming $T/\tau\in\N$, and denote by 
$\{\bfu\kt\}_{k=0}^{T/\tau}$ an approximation of the desired values 
$\bfu(k\tau)$, and similarly $\bbm\kt$ is to approximate $\bbm(k\tau)$, etc.
Further, let us abbreviate by $\dtt\kt$ the backward difference
operator, i.e.\ e.g.\  $\deltau:=\deltauu$, and similarly 
also $[\dtt\kt]^2\bfu=\dtt\kt[\deltau]=\frac{{\bfu}\kt{-}
2{\bfu}\kkt{+}{\bfu}_\tau^{k-2}}{\tau^2}$, or 
$\deltam:=\deltamm$, $\deltax:=\deltaxx$, etc. When evaluating 
$[\dtt\kt]^2\bfu$ for $k=1$, we let $\bfu_\tau^{-1}=\bfu_\tau^0-\tau\bfv_0$.

%\COMMENT{THE FOLLOWING PARTLY REPEATS THE ABOVE POINTS:}
%Then, we devise a semi-implicit discretisation using a philosophy of fractional
%steps, based on the splitting of the state variables $(\bfu,\bbm,\chi,c,\theta)$.
From {a} conceptual ``algorithmic'' viewpoint that may 
serve for a possible numerical implementation 
%on computers) 
and for making the free-energy qualification as weak as possible,
it is advantageous to make as fine splitting as possible. The finest 
splitting is essentially dictated by the 
%coupling 
%\TTT 
considered de-coupled form 
%\EEE 
of the dissipation energy, cf.\ also 
\EEE 
Rem.~\ref{D=0} below. 
%\COMMENT{I am not sure: what do we mean by ``dissipation energy''? Does this make sense?}
%Here, it means that we need to keep $(\bfu,\bbm)$ coupled
%because of the $\bbD$-term. Therefore, we come to the following 
%4-step recursive scheme to be solved for $k=1,...,T/\tau$:

\noindent Step 1:
We seek weak solution  $\bfu\kt$, 
and $\chi\kt$ to
the following boundary-value problem (written in the classical formulation)
\begin{subequations}\label{disc-u-chi-c}
\begin{align}
&\varrho[\dtt\kt]^2\bfu-\mathrm{div}\big(
\partial_{\Ee}^{}\phimech(\EEtau^k,\chi\kt,\bbm\kkt)
{+}
\bbD\dtt\kt\EEtau
\big)=\bff\kt
\label{TGMd-1}
\\\nonumber
&%\sigma_{d,\tau}^k
\partial_{\DT\chi}\zeta(\Ee\kkt,\chi\kkt,\bbm\kkt,c\kkt,\theta\kkt,\dtt\kt\chi)
-\kappa_1\Delta\chi\kt
+\partial_\chi^{}\phimechchem(\EEtau^k,\chi\kt,\bbm\kkt,c\kkt)
\\
\label{TGMd-2}&\hspace{5em}
+{\partial_\chi^{}\phiterm}(\chi\kkt,\theta\kkt)
-\bbE^\top{:}\big(\partial_{\Ee}^{}\phimech(\EEtau^k,\chi\kt,\bbm\kkt)
+
\bbD\dtt\kt\EEtau\big)
+\sigma_{\rm r,\tau}^k\ni0,
\\&\text{with }\ \EEtau^k=\bfeps(\bfu\kt){-}\bbE\chi\kt
\ \text{ and some }\ 
\sigma_{\rm r,\tau}^k\in \partial \delta^{}_K(\chi\kt)\ 
\ \text{ on }\Omega,
\intertext{and with boundary conditions}
  &\big(\partial_{\Ee}^{}\phimech(\Ee_\tau^k,\chi\kt,\bbm\kkt)
 {+}
\bbD\dtt\kt\EEtau\big)\mathbf n
 =\bff_{\rm s,\tau}^k
 \label{BC-t-1b}
 %\\&
\ \ \ \text{ and }\ \ \
 \,\nabla\chi\kt\cdot{\mathbf n}=0\ \ \text{ on }\Gamma.
\end{align}
\end{subequations}
\noindent Step 2. We seek a weak solution $c\kt$ and $\mu\kt$ to the 
boundary-value problem:
\begin{subequations}\label{c-mu-disc}
\begin{align}
\label{eq:23}
& \dtt\kt c
-\mathrm{div}\big(\mathbf M(\EEtau^k,\chi\kt,\bbm\kkt,c\kkt,\theta\kkt)
\nabla\mu\kt\big)=0
\ \ \text{ with }\ \mu\kt=\partial_c^{}\phichem(\chi\kt,c\kt)\ \ \text{ on }\Omega,\\
\intertext{with boundary condition}
\label{eq:24}&\nabla\mu\kt\cdot\mathbf n=0\ \ \text{ on }\Gamma.
\end{align}
\end{subequations}
  
\noindent Step 3.  We seek $\bbm\kt\in H^1(\Omega)$ as a (global) minimizer of the functional
\begin{align}\label{dam-disc}
\bbm \mapsto\int_\Omega
{\phimech(\EEtau^k,\chi\kt,d)}
-\alpha(\chi\kt) d+\frac{\kappa_2}2|\nabla d|^2\d x
\end{align}
on the set $\{d\!\in\!H^1(\Omega);\ 0\le d\le d\kkt\}$.

\noindent Step 4. 
Eventually, we seek a weak solution $\theta\kt$ and $w_\tau^k$ to the 
boundary-value problem:
\begin{subequations}\label{heat-disc}
\begin{align}
\label{eq:25}
&\frac{w^k_\tau{-}w^{k-1}_\tau}\tau-\mathrm{div}\big(
 \mathbf{K}(\EEtau^k,\chi\kt,\bbm\kt,\CHI\kt,\theta\kt)\nabla \theta\kt\big)\nonumber
=%q\kt
\big(\partial_{\DT\chi}\zeta(\Ee\kkt,\chi\kkt,\bbm\kkt,c\kkt,\theta\kkt,\dtt\kt\chi){+}\partial_\chi^{}\phiterm(\chi\kt,\theta\kt)\big){\cdot}\dtt\kt\chi
  \nonumber
\\[-.4em]
&\quad
-\alpha(\chi\kt)\dtt\kt d
+\frac{\mathbbm D\dtt\kt\Ee\!:\!\dtt\kt\Ee}
{1+\tau|\dtt\kt\Ee|^2}
 +\frac{\mathbf M(\EEtau^k,\chi\kt,\bbm\kkt,\CHI\kkt,\theta\kkt)
 \nabla\mu\kt\cdot\nabla\mu\kt}{1+\tau|\nabla\mu\kt|^2}
\ \ \ \ \ \text{with }\ \ w\kt=\eterm(\chi\kt,\theta\kt)
\intertext{with the boundary condition}
&\mathbf K(\EEtau^k,\chi\kt,\bbm\kt,c\kt,\theta\kt)\nabla\theta\kt=q_{s,\tau}^k\ \ \text{ on }\Gamma.
\label{heat-disc-bc}
\end{align}
\end{subequations}

{
Of course, this recursive scheme is to be started for $k=1$ by 
putting 
\begin{align}
\bfu_\tau^0=\bfu_0,\ \ \ \ \ \ \ \ 
\bfu_\tau^{-1}=\bfu_0{-}\tau\bfv_0,\ \ \ \ \ \ \ \ 
\chi_\tau^0=\chi_0,\ \ \ \ \ \ \ \ 
c_\tau^0=c_0,\ \ \ \ \ \ \ \ 
\bbm_\tau^0=\bbm_0,\ \ \ \ \ \ \ \ 
w_\tau^0=w_0
\end{align}
with $w_0$ from \eqref{ICb}.} An important feature of this scheme is that 
it decouples to four boundary-value problems, which (after a further spatial 
discretisation) can facilitate a numerical treatment and 
which is advantageously used even to show existence of
approximate solutions. 
We also remark that the term $\partial_\chi^{}\phiterm$ 
is treated as lower-order terms. 
Note also that $d\kt$ obtained in Step 3 is a weak solution to the 
boundary-value problem:
\begin{align}  
&\partial_{\DT \bbm}\xi(\chi\kt,\dtt\kt{\bbm})
+\partial_\bbm^{}
{\phimech(\EEtau^k,\chi\kt,\bbm\kt)}
+s_{\rm r,\tau}^k\ni\kappa_2\Delta\bbm\kt\ \ \text{ on }\ \Omega\ \ \text{ with }\ \
\nabla\bbm\kt\cdot\mathbf n=0\ \ \text{ on }\ \Gamma,
\end{align}
with some $s_{\rm d,\tau}^k\!\in\!$
and $s_{\rm r,\tau}^k\!\in\!\partial\delta_{[0,\infty]}(\bbm\kt)$ on $\Omega$.
We however rely on the fact that $d\kt$ is a special weak solution
which is also the minimizer of the underlying potential \eqref{dam-disc}.
This need not be the same if 
$\phimech(\Ee,\cdot)$ is not convex,
i.e.\ if damage may undergo weakening effects.

\addcontentsline{toc}{subsection}{existence of the discrete solutions} 
\phantomsection
\begin{lemma}[{\sc Existence of the discrete solution}]\label{lem-1}
Let \eqref{assum-smoot}--\eqref{eq:19} hold. 
Then, for any $k=1,...,T/\tau$, \eqref{disc-u-chi-c}--\eqref{heat-disc} 
has a solution $\bfu\kt\in 
H^1(\Omega;\R^3)$, $\chi\kt\in H^1(\Omega;\R^N)$, $\bbm\kt\in H^1(\Omega)$,  
$\sigma_{\rm r,\tau}^k\in L^2(\Omega;\R^N)$,  
$c\kt\in H^1(\Omega)$, $\mu\kt\in H^1(\Omega)$, $\theta\kt\in H^1(\Omega)$
such that 
$\theta\kt\ge0$. 
\end{lemma}

\noindent{\it Proof}.
Step 1: Let us consider the space $V=
H^1(\Omega;\R^3)\times H^1(\Omega)$. The 
boundary-value problem \eqref{disc-u-chi-c} is of the form 
$A(\bfu,\chi)\ni 0$, with $A$ 
a set-valued mapping from $V$ to its dual $V^*$ such that $A=\partial\Phi$, 
where
\begin{align}\nonumber
\Phi(\bfu,\chi)&=\int_\Omega\bigg(
\phimechchem(\bfeps(\bfu)-\bbE\chi,\chi,\bbm\kkt,c\kkt)+
\frac12\kappa_1|\nabla\chi|^2+\delta^{}_K(\chi)
\\
&\nonumber
\qquad+\frac{\tau^2}2\varrho
\Big|\frac{\bfu-2\bfu\kkt+\bfu_\tau^{k-2}}{\tau^2}\Big|^2
+\frac\tau2\bbD\Big(\bfeps\big(\frac{\bfu{-}\bfu\kkt}\tau\big)-\bbE\chi\Big):
\Big(\bfeps\big(\frac{\bfu{-}\bfu\kkt}\tau\big)-\bbE\chi\Big)
\\&\qquad
\nonumber
+\tau\zeta\Big(\Ee\kkt,\chi\kkt,\bbm\kkt,c\kkt,\theta\kkt,\frac{\chi-\chi\kkt}\tau\Big)
\\&\qquad
+\partial_{\chi}\phiterm(\chi\kkt,\theta\kkt)\cdot \chi
\label{minimization-u-m}
+\bff\kt{\cdot}\bfu\bigg)\,\d x+\int_\Gamma\bff_{\rm s,\tau}^k{\cdot}\bfu\,\d S.
\end{align}
Here it is understood that $\varrho=0$ in the quasistatic case. \color{black}By 
\eqref{assum-conve-e-chi} and \eqref{assum-dissi-poten-a}, the potential 
$\Phi$ is weakly lower-semicontinuous 
although not necessarily convex here;
in fact, later it will be convex 
{if $\tau>0$ is small enough, cf.\ Lemma~\ref{lem-2} below}.
Due to the terms $\delta^{}_K$ and $\zeta$, $\Phi$ is nonsmooth.
Moreover, by \eqref{eq:27} it is also coercive. Hence, by using the direct method, 
cf.\ e.g.\ \cite[Theorem 5.3]{Roub13NPDE}, we can see that \eqref{disc-u-chi-c} has at least a solution. Next we observe that the solution satisfies an inclusion:
\begin{equation}
\partial_{\chi}(\Phi_{1}+\Phi_{2})(\chi)+D\Phi_{3}(\chi)\ni 0,
\end{equation}
where the left-hand side is a subset of $H^1(\Omega;\R^N)^*$, with $\Phi_{1}(\chi)=\tau\int_\Omega\zeta(\Ee\kkt,\chi\kkt,\bbm\kkt,\theta\kkt,\frac{\chi-\chi\kkt}{\tau})\,\dx$, $\Phi_{2}(\chi)=\int_\Omega \delta^{}_K(\chi)\,\dx$ and with $\Phi_{3}$ the remaining part of the potential $\Phi$, which is Gateaux differentiable (we denote its Gateaux differential by $D\Phi_{3}$). Next, we observe that $\Phi_1$ is convex, and by \eqref{assum-dissi-poten-a} its domain is the whole space $H^1(\Omega;\R^N)$, and it is bounded from above in a bounded set of $H^1(\Omega;\R^N)$. Thus, by \cite[Thm. 4.8]{visintin1996models}, $\Phi_1$ is locally Lipschiz continuous. Next, since ${\rm dom}(\Phi_2)\neq \emptyset$, there exists $\chi_0\in {\rm dom}\Phi_1\cap{\rm dom}\Phi_2$ such that $\Phi_1$ is in particular continuous at $\chi_0$. As both $\Phi_1$ and $\Phi_2$ are convex and lower semicontinuous, we conclude \cite[Thm. 4.7]{visintin1996models} that:
\begin{equation}
  \label{eq:68}
  \partial\Phi_{1}(\chi)+\partial\Phi_{2}(\chi)=\partial_{\chi}(\Phi_{1}+\Phi_{2})(\chi).
\end{equation}
Now, we can take a measurable selection $\sigma_{\rm d,\tau}^k\in\tau\partial_{\chi}\zeta(\Ee\kkt,\chi\kkt,\bbm\kkt,\theta\kkt,\frac{\chi-\chi\kkt}{\tau})$ and by \eqref{assum-dissi-poten-a}, 
$\sigma_{\rm d,\tau}^k\in L^2(\Omega;\R^N)$. Then by comparison, we obtain
$\sigma_{\rm r,\tau}^k\in L^2(\Omega;\R^N)$, cf.\ also the arguments leading to \eqref{est-of-xi} below.

Step 2: We can treat \eqref{c-mu-disc} by a variational approach. We consider the function 
$\phichemstar(\chi\kt,
\cdot)$ defined as the Legendre  transform of $\phichem(\chi\kt,
\cdot)$, and we notice that (\ref{c-mu-disc}b) can be written as 
\begin{equation}\label{legen-trans}
c\kt=\partial_\mu\phichemstar(\chi\kt,
\mu\kt).
\end{equation} 
Then \eqref{c-mu-disc} is equivalent the following variational problem on 
$H^1(\Omega)$:
\begin{equation}
  \label{eq:26}
  \textrm{minimize}\ \mu\mapsto\int_\Omega\frac{\phichemstar(\chi\kt,
\mu)
-c\kkt\mu}{\tau}+\mathbf M(\EEtau^k,\chi\kt,\bbm\kkt,c\kkt
,\theta\kkt)\nabla\mu\cdot\nabla\mu\dx,
\end{equation}
which can be solved through the direct method. To this aim, we notice that, since the domain of $\varphi(\chi\kt,\bbm_\tau^{k-1},\cdot)$ is $\R$, its Legendre transform is coercive. Indeed, on writing for short $f(c)=\phichem(\chi\kt,
c)$, and on denoting by $f^*$ the Legendre transform of $f$, we have $\lim_{|\mu|\to+\infty}f^*(\mu)/{|\mu|}=\lim_{|\mu|\to+\infty}\sup_{c}(({\rm sign}\mu)c-f(c)/{|\mu|})\ge \lim_{|\mu|\to+\infty}({\rm sign}{\mu})\barc-f(\barc)/{|\mu|})=({\rm sign}{\mu})\barc$  for every $\barc\in\R$. Thus $\lim_{\mu\to+\infty}f^*(\mu)/\mu=+\infty$.

Step 3: The solution of \eqref{dam-disc} can be obtained 
simply by weak lower semicontinuity and coercivity arguments.
Note that we do not require convexity of $\phimech(\Ee,\bbm,\cdot)$ 
so that the solution of \eqref{dam-disc} does not need to be unique.

Step 4: In this final step we solve the time-discrete heat equation \eqref{eq:25} 
with boundary conditions \eqref{heat-disc-bc}. To this aim, we note that
$\nabla\mu\kt\in L^2(\Omega;\R^3)$. In particular, we have simply both 
$\mathbbm D\dtt\kt\Ee\!:\!\dtt\kt\Ee/(1{+}\tau|\dtt\kt\Ee|^2)
\in L^\infty(\Omega)$ and
$\mathbf M({\mathsf E}_{\rm e,\tau}^k,\chi\kt,\bbm\kkt,\CHI\kkt,\theta\kkt)
\nabla\mu\kt{\cdot}\nabla\mu\kt/(1{+}\tau|\nabla\mu\kt|^2)\in L^\infty(\Omega)$,
and thus the right-hand side of \eqref{eq:25} is in $L^2(\Omega)$.
Therefore, eventually, we are to solve \eqref{heat-disc}, 
which represents a semilinear heat-transfer equation with the right-hand side 
in $H^1(\Omega)^*$. The only nonlinearity with respect to $\theta\kt$ is in the terms 
$\mathbf{L}(\bfeps(\bfu\kt),\chi\kt,\bbm\kt,\CHI\kt,\theta\kt)$ and 
$\partial_\chi^{}\phiterm(\chi\kt,\theta\kt)$. 
The later 
is needed to guarantee 
$\theta\kt\ge0$. Anyhow, since this nonlinearity is of lower order, we can pass 
through it by compactness and strong convergence. Thus, it suffices for us to 
check coercivity of the underlying operator. To this aim, we test 
\eqref{heat-disc} by $\theta\kt$. The terms on the right-hand side of \eqref{eq:25} containing 
$\theta\kt$ are estimated standardly by using H\"older's and Young's inequalities, 
and using the qualification 
\eqref{eq:16}.

Having coercivity, we see that there exists at least one solution. 
Moreover, this solution satisfies $\theta\kt\ge0$, which can be seen by testing 
$\eqref{heat-disc}$ by the negative part of  $\theta\kt$ and using that 
${\partial_\chi^{}\phiterm}(\chi,\theta)=0$ 
for $\theta\le0$.\marginpar{$<=$} 
\QED

\begin{remark}\label{remar-delay}\upshape
The delayed term
{s $\Ee\kkt$ and $\bbm\kkt$ in \eqref{TGMd-2}
and}
 $c\kkt$ in \eqref{eq:23} allows us to treat 
{\eqref{disc-u-chi-c}
and }
\eqref{c-mu-disc} as variational problems. 
This does not create troubles in the limit passage, see also 
\hyperlink{limit-passage-diffusion-equation}{Step 5} in the 
proof of Proposition \ref{prop-conv} below.
\end{remark}

\medskip

Let us define the piecewise affine interpolant $\bfu_\tau$  by 
\begin{subequations}
\begin{align}
&\bfu_\tau(t):=
\frac{t-(k{-}1)\tau}\tau\bfu\kt
+\frac{k\tau-t}\tau \bfu\kkt
\quad\text{ for $t\in[(k{-}1)\tau,k\tau]\ $}
\intertext{with $\ k=0,...,T/\tau$. Besides, we define also the backward 
piecewise constant interpolant $\bar\bfu_\tau$ and $\underline\bfu_\tau$ by}
&\bar\bfu_\tau(t):=\bfu\kt,
\qquad\qquad\qquad\qquad\text{ for $t\in((k{-}1)\tau,k\tau]\ $,\ \ 
$k=1,...,T/\tau$},
\\
&\underline\bfu_\tau(t):=\bfu\kkt,
\qquad\qquad\qquad\quad\ \text{for $t\in[(k{-}1)\tau,k\tau)\ $,\ \ 
$k=1,...,T/\tau$}.
\intertext{Similarly, we define also $\bbm_\tau$, $\bard_\tau$, 
$\underline\bbm_\tau$, $\barw_\tau$, $w_\tau$,
$\bar g_\tau$, $\barbff_{\rm b,\tau}$, etc. 
We will also need the piecewise affine interpolant of the 
(piecewise constant) velocity 
$\DT\bfu_\tau$,
which we denote by 
$\big[\DT\bfu_\tau\big]^{\rm i}$, 
i.e.}
&\big[\DT\bfu_\tau\big]^{\rm i}(t):=
\frac{t{-}(k{-}1)\tau}\tau\,\frac{\bfu\kt{-}\bfu\kkt}\tau
+\frac{k\tau{-}t}\tau
\,\frac{\bfu\kkt{-}\bfu_\tau^{k-2}}\tau
\ \text{ for $t\in((k{-}1)\tau,k\tau]$}.
\end{align}
\end{subequations}
Note that $\DDT\bfu_\tau^{\rm i}:=\frac{\partial}{\partial t}
\big[\DT\bfu_\tau\big]^{\rm i}$ is piecewise constant with the values 
$(\bfu_\tau^k{-}2\bfu_\tau^{k-1}{+}\bfu_\tau^{k-2})/\tau^2$ on the 
particular subintervals $((k{-}1)\tau,k\tau)$. \hypertarget{2013-11-20-3}{}

In terms of interpolants, we can write the approximate system 
\eqref{disc-u-chi-c}--\eqref{heat-disc} 
and the semi-stability information we can get from \eqref{dam-disc}
in a more ``condensed'' form closer 
to the desired continuous system \eqref{systemb}, namely:

\begin{subequations}\label{system-disc}
\begin{align}\label{mombal-disc}
&\varrho\DDT\bfu_\tau^{\rm i}-\mathrm{div}\big(
\partial_\Ee^{}\phimech(\barEEtau,\barchi_\tau,\underline\bbm_\tau)
+\bbD\DTEEtau
\big)=\barbff_\tau\qquad\ 
\text{with }\ {\barEEtau}=
\bfeps(\bar\bfu_\tau){-}\bbE\barchi_\tau,
\\\nonumber
&\partial_{\DT\chi}\zeta(\underlineEe_\tau,\underline\chi_\tau,\underline\bbm_\tau,\underline c_\tau,\underline\theta_\tau,\DT\chi_\tau)-\kappa_1\Delta\barchi_\tau
+\partial_{\chi}\phimechchem(\barEe_\tau,\barchi_\tau,\underline\bbm_\tau,\underline c_\tau)
\nonumber
\\
&\qquad+\partial_\chi^{}\phiterm(\underline\chi_\tau,
\underline\theta_\tau)
-\bbE^\top{:}
\big(\partial_{\Ee}^{}\phimech(\barEEtau,\barchi_\tau,\underline\bbm_\tau)
{+}\bbD\DTEEtau\big)
{+}\barsigma_{\rm r,\tau}\ni0\quad\textrm{ with }\barsigma_{\rm r,\tau}\in \partial \delta^{}_K(\barchi_\tau),
\label{chi-eq-disc}\\
&\label{NS-disc}
\DT c_\tau-\mathrm{div}\big(\mathbf M(\barEEtau,
\barchi_\tau,\underline\bbm_\tau,\underline c_\tau,\underline\theta_\tau)\nabla\barmu_\tau\big)=0\qquad\textrm{with}\quad \barmu_\tau
=\partial_c^{}\phichem(\barchi_\tau,
\barc_\tau),
\\\nonumber
&\!\!
\int_\Omega\phimechchem(\barEe_\tau(t),\barchi_\tau(t),\bard_\tau(t),
\barc_\tau(t))+\frac{\kappa_2}2|\nabla\bard_\tau(t)|^2\,\d x
\le\int_\Omega\Big(\phimechchem(\barEe_\tau(t),\barchi_\tau(t),\widetilde\bbm,\barc_\tau(t))
\\[-.2em]&\hspace{1em}
+\frac{\kappa_2}2|\nabla\widetilde\bbm|^2+\alpha(\barchi_\tau(t))(\widetilde\bbm{-}\bard_\tau(t))\Big)\,\d x
\qquad
\forall\widetilde\bbm\!\in\!H^1(\Omega),\ 0\le\widetilde\bbm\le\bard_\tau(t)
\text{ on }\Omega,
\text{ 
for a.a.\ }t\!\in\!I,
\label{damage-eq-disc}
\\\nonumber
&\DT w_\tau-\textrm{div}\big(
\mathbf K(\barEEtau,\barchi_\tau,\bard_\tau,\barc_\tau,\bartheta_\tau)\nabla \bartheta_\tau\big)=\bar r_\tau
\qquad\textrm{with}\quad
\barw_\tau=\eterm(\barchi_\tau,\bartheta_\tau)
\quad\text{ and}
\\[-.3em]
&\quad\textrm{with }\ 
\bar r_\tau=
\big(\bar s_{\rm d,\tau}{+}\partial_\chi^{}\phiterm(\barchi_\tau,
\bartheta_\tau)\big){\cdot}\DT\chi_\tau
-\alpha(\barchi_\tau)\DT\bbm_\tau
+\frac{\bbD\DTEEtau{:}\DTEEtau}{1{+}\tau|\DTEEtau|^2}
+
\frac{\mathbf M(\barEEtau,\barchi_\tau,\underline\bbm_\tau,\underline c_\tau,\underline\theta_\tau)
\nabla\barmu_\tau{\cdot}\nabla\barmu_\tau}{1+\tau|\nabla\barmu_\tau|^2},
\label{heat-eq}
\\\nonumber
&
\int_\O
\phimechchem(\Ee\kt,\chi\kt,\bbm\kt,c\kt)
+\frac\varrho2\big|\DT\bfu\kt\big|^2
+\frac{\kappa_1} 2\big|\nabla\chi\kt\big|^2
+\frac{\kappa_2} 2\big|\nabla\bbm\kt\big|^2\,\d x
\\[-.3em]\nonumber&\hspace{2em}
+\int_0^{k\tau}\!\!\int_\Omega\!\Big(
\bbD\DT\Ee_\tau{:}\DT\Ee_\tau+
\big(
(\partial_{\DT\chi}\zeta(\underlineEe_\tau,\underline\chi_\tau,\underline\bbm_\tau,\underline c_\tau,\underline\theta_\tau,\DT\chi_\tau)-\sqrt\tau\DT\chi_\tau)
{+}\partial_\chi^{}\phiterm(\underline\chi_\tau,\underline\theta_\tau)
\big){\cdot}\DT\chi_\tau
-\alpha(\barchi_\tau)\DT d_\tau
\\[-.3em]
&\hspace{2em}
+{\mathbf M}(\barEEtau,
\barchi_\tau,\underline\bbm_\tau,\underline c_\tau,\underline\theta_\tau)
\nabla\barmu_\tau{\cdot}\nabla\barmu_\tau
\Big)\,\d x\d t
\label{engr-equality-disc}
\le
\int_0^{k\tau}\!\!
\int_\Gamma\!\barbff_{\rm s,\tau}{\cdot}\DT\bfu_\tau{+}q_{\rm s,\tau}{+}\barmu_\tau \bar h_{\rm s,\tau}\,\d S
+
\mathcal E_{_{\rm MC}}(0)
\end{align}\end{subequations}
holding for any $k=0,...,T/\tau$,
together with the corresponding boundary conditions
\begin{subequations}\label{BC-disc}
\begin{align}
&\big(\partial_\Ee^{}\phimech(\barEEtau,\barchi_\tau,\underline\bbm_\tau)
+\bbD\DTEEtau\big)\mathbf n=\barbff_{\rm s,\tau},
\label{BC-t-1}
\\
\label{neuma-condi-chi}
&\nabla\chi_\tau{\cdot}\mathbf n=0,
\\\label{neuma-condi-c}
&\mathbf M(\barEEtau,\barchi_\tau,\underline\bbm_\tau,\underline c_\tau,\underline\theta_\tau)
\nabla\barmu_\tau\cdot\mathbf n={\bar h_{\rm s,\tau}},
\\&\label{BC-t-2-3}
\big(
\mathbf{K}(\barEEtau,\barchi_\tau,\bard_\tau,\barc_\tau,\bartheta_\tau)\nabla\bartheta_\tau
\big){\cdot}\mathbf n={\bar q_{\rm s,\tau}}.
\end{align}\end{subequations}

The discrete semi-stability 
\eqref{damage-eq-disc} has been 
obtained from minimizing \eqref{dam-disc} by comparing 
a solution $d\kt$ against $\widetilde d$ and using the triangle inequality for 
$\xi(\chi\kt,\cdot)$; here the positive degree-1 homogeneity of  $\xi(\chi,\cdot)$ is used. Note that we just choose one global minimizer $d\kt$
of \eqref{dam-disc} from possibly many, if $\phimech(\Ee,\chi,\cdot)$ is
not convex. The energy inequality \eqref{engr-equality-disc} 
which we will prove in Lemma~\ref{lem-2} below is an analog 
of the mechanical/chemical energy \eqref{engr-equality} over the time interval 
$[0,k\tau]$ and without making the by-part integration like \eqref{by-part-d}.
Note also the term $\sqrt\tau|\DT\chi_\tau|^2$
in \eqref{engr-equality-disc} which facilitates handling of non-convex energies 
$\phimech$ 
{as admitted by the semi-convexity assumption 
\eqref{assum-conve-e-chi}  but which disappers in the limit for 
$\tau\to0$, cf.\ \cite[Rem.\,8.24]{Roub13NPDE} for this trick}.

\addcontentsline{toc}{subsection}{First estimates} 
\phantomsection
\begin{lemma}[{\sc First estimates}]\label{lem-2}
\slshape
Let again the assumptions of Lemma~\ref{lem-1} hold. Then the mecano-chemical
energy inequality \eqref{engr-equality-disc} holds and 
the following estimates 
hold uniformly with respect to the time-step provided 
$\tau\le\min(1/M^2,T,4\epsilon^2)$ 
with $M$ from \eqref{assum-conve-e-chi} (or simply $\tau\le T$ if $M=0$):
\begin{subequations}\label{apriori-I}
\begin{align}\label{apriori-Ia}
&\big\|\bfu_\tau\big\|_{W^{1,\infty}(I;L^2(\O;\R^3))\,\cap\,
H^1(I;H^1(\O;\R^3))}\le C,
\\
\label{aprio-elast-strai}&\big\|\Ee_\tau\big\|_{L^\infty(I;L^2(\Omega;\R^3))}\\
\label{eq:28}
&\big\|\chi_\tau\big\|_{L^\infty(I;H^1(\O;\R^N))\,\cap\,
H^1(I;L^2(\O;\R^N))\,\cap\,L^\infty(Q;\R^N)}\le C,
\\\label{apriori-Ia++}
\hypertarget{2013-11-20-2}{}&
\big\|\CHI_\tau\big\|_{L^2(I;H^1(\O))}\le C,
\\\label{apriori-Ia+++}
&\big\|\mu_\tau\big\|_{L^2(I;H^1(\Omega))}\le C,
\\\label{apriori-Ia+}
&\big\|\bbm_\tau\big\|_{L^\infty(I;H^1(\O))\,\cap\,
{\rm BV}(I;L^1(\Omega))
\,\cap\,L^\infty(Q)}\le C,
\\\label{eq:29f}
&\big\|w_\tau\big\|_{L^\infty(I;L^1(\O))}\le C.
\end{align}
Moreover, for every $1\le r<
\frac54$ there exists $C_r>0$, 
independent of $\tau$, such that
\begin{align}
\label{eq:73}
&\color{black}\big\|\nabla w_\tau\big\|_{{L^r(Q;\R^3)}}\le C_r,
\\
\label{eq:30}
&\big\|\nabla \theta_\tau\big\|_{{L^r(Q;\R^3)}}\le C_r,
\\
\label{eq:b58}
&\|w_\tau\|_{L^{r/(2-r)}(Q)}\le C_r,
\\
\label{eq:c58}
&\|\theta_\tau\|_{L^{r/(2-r)}(Q)}\le C_r.
\end{align}
\end{subequations}
\end{lemma}

\noindent{\it Proof}.
The strategy is to test the particular equations in \eqref{disc-u-chi-c}--\eqref{heat-disc} 
respectively by $\deltau$,  $\dtt\kt\chi$, $\mu\kt$, $\dtt\kt\bbm$, and $\frac12$.
For (\ref{disc-u-chi-c}a,b), we note that a standard argument using convexity of 
$(\Ee,\chi)\mapsto\phimechchem(\Ee,\chi,c,\bbm)+\delta^{}_K(\chi)$
composed with the linear mapping $(\bfu,\chi)\mapsto\Ee$,
and of $\DT\chi\mapsto\zeta(\Ee,\chi,c,\bbm,\theta,\DT\chi)$ yields:
\begin{align}\nonumber
  \label{eq:31}
&\int_\Omega\varrho[\dtt\kt]^2\bfu{\cdot}\dtt\kt\bfu+
\big(\partial_{\Ee}^{}\phimech(\EEtau^k,\chi\kt,\bbm\kkt)
+{\bbD}\dtt\kt\EEtau^k\big){:}\bfeps(\dtt\kt\bfu)
\\&\nonumber\qquad
+\big(\partial_{\DT d}\zeta(\Ee\kt,\chi\kt,\bbm\kt,\CHI\kt,\theta\kt,\dtt\kt\chi)+
\partial_\chi^{}\phimech(\EEtau^k,\chi\kt,\bbm\kkt,c\kkt)
\\&\nonumber\qquad
-\bbE^\top{:}\big(\partial_{\Ee}^{}\phimech(\EEtau^k,\chi\kt,\bbm\kkt)
+{\bbD}\dtt\kt\EEtau^k\big)
+\sigma_{\rm r,\tau}^k\big)
{\cdot}\dtt\kt\chi
+\kappa_1\nabla\chi\kt{:}\nabla\dtt\kt\chi\,\d x
\nonumber\\&
\ge
\int_\Omega\bbD\dtt\kt\EEtau
{:}\big(\bfeps(\dtt\kt\bfu)-\bbE\dtt\kt\chi\big)
+
s_{{\rm d},\tau}^k+\partial_\chi^{}\phiterm(\chi\kkt,
\theta\kkt))\dtt\kt\chi
\nonumber\\&\qquad
+\frac\varrho2|\dtt\kt\bfu|^2+
\phimech(\Ee\kt,\chi\kt,\bbm\kkt,c\kkt)
+\frac{\kappa_1}2|\nabla\chi\kt|^2
+\delta^{}_K(\chi\kt)
\nonumber\\&\qquad
-\frac\varrho2|\dtt\kkt\bfu|^2\!-
\phimech(\EEtau^{k-1},\chi\kkt,\bbm\kkt,c\kkt)
-\frac{\kappa_1}2|\nabla\chi\kkt|^2\!-\delta^{}_K(\chi\kkt)\,\d x.
\end{align}
Further, we execute the test of \eqref{c-mu-disc}
relying on the convexity of $c\mapsto\phichem(\chi,
c)$
\begin{align}
\int_\Gamma h_{{\rm s},\tau}^k\mu\kt\,{\rm d}S&=\int_\Omega\mu\kt\dtt\kt c
+\mathbf M(\EEtau^k,\chi\kt,\bbm\kkt,c\kt,\theta\kkt)
\nabla\mu\kt{\cdot}\nabla\mu\kt\,\d x\nonumber
\\&
\ge
\int_\Omega\phichem(\chi\kt,
c\kt)
-\phichem(\chi\kt,
c\kkt)
+\mathbf M(\EEtau^k,\chi\kt,\bbm\kkt,c\kt,\theta\kkt)
\nabla\mu\kt{\cdot}\nabla\mu\kt\,\d x.\label{eq:42}
\end{align}
Moreover, we test \eqref{dam-disc} with $\bbm\kkt$ to obtain
\begin{align}
&\int_\Omega
\phimech(\EEtau^k,\chi\kt,\bbm\kt)
+\frac {\kappa_2}2|\nabla\bbm\kt|^2
\,\d x 
\le\int_\Omega
\phimech(\EEtau^k,\chi\kt,\bbm\kkt)+\frac{\kappa_2}2|\nabla\bbm\kkt|^2-\alpha(\chi\kt)
(\bbm\kt-\bbm\kkt)
\,\d x.  \label{eq:2bis}
\end{align}
Relying on the semiconvexity \eqref{assum-conve-e-chi} of the mechano-chemical 
part $\phimech$ of the free energy (see the argument in 
\cite[Rem.\,8.24]{Roub13NPDE}) and adding \eqref{eq:31}, \eqref{eq:42}, and 
\eqref{eq:2bis}, and recalling that $\phimechchem=\phimech+\phichem$, 
we benefit with the telescopic cancellation 
of the terms $\pm\phimechchem(\EEtau^k,\chi\kt,\bbm\kkt,c\kt)$
and $\pm\phimechchem(\EEtau^k,\chi\kt,\bbm\kkt,c\kkt)$, and 
we obtain 
the following mechano-chemical energy balance:
\begin{align}\nonumber
&\int_\Omega\frac\r2\big|\dtt\kt\bfu\big|^2
+\phimechchem(\EEtau^k,\chi\kt,\bbm\kt,c\kt)+\frac{\kappa_2}2|\nabla\bbm\kt|^2+\frac{\kappa_1}2|\nabla\chi\kt|^2
+\delta^{}_K(\chi\kt)+\delta_{[0,1]}(\bbm\kt)
\,\d x
\\&\nonumber
\ \ 
+\tau\sum_{j=1}^k\int_\Omega\bbD\dtt_\tau^j\Ee:\dtt_\tau^j\Ee
+
(\partial_{\DT\chi}\zeta(\Ee\kkt,\chi\kkt,d\kkt,c\kkt,\theta\kkt,\dtt_\tau^j\chi)
-\sqrt\tau\dtt_\tau^j\chi)\cdot\dtt_\tau^j\chi
-\alpha(\chi_\tau^j)\cdot\dtt_\tau^j\bbm\nonumber\\
&\qquad\qquad\qquad
+{\mathbf M}(\Ee_\tau^{j-1},
\chi_\tau^{j-1},\bbm_\tau^{j-1},\CHI_\tau^{j-1},\theta_\tau^{j-1})
\nabla\mu_\tau^j{\cdot}\nabla\mu_\tau^j
\d x\d t\nonumber
 \\\nonumber&\ \le\int_\Omega\bigg(\frac\r2|\bfv_0|^2
+\phimechchem(\Ee_0,\chi_0,\bbm_0,c_0)+\frac{\kappa_2}2|\nabla\bbm_0|^2+\frac{\kappa_1}2|\nabla\chi_0|^2
\bigg)
\,\d x
\\\nonumber&\qquad\qquad\qquad
+\tau\sum_{j=1}^k\bigg(\int_\Omega
 \bff^j_\tau{\cdot}\dtt_\tau^j\bfu
\dtt_\tau^j\chi+\int_\Gamma\bff^j_{\rm s,\tau}{\cdot}\dtt_\tau^j\bfu+
  h^j_{\rm s,\tau}{\cdot}\mu_\tau^j
\d S\bigg).
\end{align} We also used that 
$\delta^{}_K(\chi_0)+\delta_{[0,1]}(\bbm_0)=0$.
This proves \eqref{engr-equality-disc}.

Finally, we test the heat equation \eqref{heat-disc}  by 1/2, and we add the 
resulting equation\ to \eqref{eq:31}--\eqref{eq:2bis}. This is not 
a physical test (which would be by 1 instead of 1/2) and thus, in this summation, 
{the adiabatic terms do not cancel out}.
This scenario simplifies the implicit discretisation to let is decoupled 
by using $\theta\kkt$ in $\phiterm$ in \eqref{TGMd-2} instead of $\theta\kt$
and also it allows to estimate $w$ with the other variables simultaneously
but it forces more restrictive assumption on the growth 
of heat capacity than physically necessary, cf.\ \cite[Exercise~12.9]{Roub13NPDE}. 
Upon summing over $k$, we arrive at the following estimate:
\begin{align}\nonumber
&\int_\Omega\frac12 w\kt+\frac\r2\big|\dtt\kt\bfu\big|^2
+\phimechchem(\EEtau^k,\chi\kt,\bbm\kt,c\kt)+\frac{\kappa_2}2|\nabla\bbm\kt|^2+\frac{\kappa_1}2|\nabla\chi\kt|^2
+\delta^{}_K(\chi\kt)+\delta_{[0,1]}(\bbm\kt)
\,\d x
\\&\nonumber
\qquad\ \ +\tau\sum_{j=1}^k\int_\Omega\bbD\dtt_\tau^j\Ee:\dtt_\tau^j\Ee
+
\partial_{\DT\chi}\zeta(\Ee\kkt,\chi\kkt,d\kkt,c\kkt,\theta\kkt,\dtt_\tau^j\chi)
\cdot\dtt_\tau^j\chi
-\alpha(\chi_\tau^j)\cdot\dtt_\tau^j\bbm\nonumber\\[-.5em]
&\qquad\qquad\qquad\qquad\qquad
+\frac12{\mathbf M}(\Ee_\tau^j,\chi_\tau^j,
\bbm_\tau^{j-1},\CHI_\tau^{j-1},\theta_\tau^{j-1})
\nabla\mu_\tau^j{\cdot}\nabla\mu_\tau^j
\d x\d t\nonumber
 \\\nonumber&\ \le\int_\Omega\bigg(
\frac12 \eterm(\chi_0,\theta_0)+\frac\r2|\bfv_0|^2
+\phichem(\chi_0,
c_0)+\phimech(\Ee_0,\bbm_0,c_0)
+\frac{\kappa_2}2|\nabla\bbm_0|^2+\frac{\kappa_1}2|\nabla\chi_0|^2
\bigg)\,\d x
\\\nonumber&
\ \ +\tau\sum_{j=1}^k\bigg(\int_\Omega
 \bff^j_\tau{\cdot}\dtt_\tau^j\bfu\,
+\Big(\frac12 \partial_\chi^{}\phiterm(\chi_\tau^j,
\theta_\tau^j)-\partial_\chi^{}\phiterm(\chi_\tau^{j-1},
\theta_\tau^{j-1})\Big)\dtt_\tau^j\chi
+\int_\Gamma\bff^j_{\rm s,\tau}{\cdot}\dtt_\tau^j\bfu+
  h^j_{\rm s,\tau}{\cdot}\mu_\tau^j+\frac12 q^j_{\rm s,\tau}\d S\bigg).
\end{align}
{Using \eqref{eq:16} and Holder's and Young's inequalities,} followed by the discrete Gronwall lemma, 
we see that the left-hand side {in the above inequality} is bounded uniformly with respect to $t$ 
for all $k=1,\dots,T/\tau$. We therefore obtain the following bounds:
%\COMMENT{some of them augments \eqref{apriori-I} -- why they are not listed in the statement of this Lemma?. BECAUSE THEY ARE NOT NEEDED IN THE PROPOSITIONS THAT FOLLOW, BUT ARE NEEDED TO OBTAIN THESE ESTIMATES}
\begin{subequations}\label{eq:33}
\begin{align}
  \label{eq:34}
&\|\DT\bfu_\tau\|_{L^\infty(L^2(\Omega;\R^3))}\le C,
\\
&\|\phimech(\barEe_\tau,\barchi_\tau,\bard_\tau)\|_{L^\infty(I;L^1(\Omega))}\le C,
\\
\label{eq:34b}
&\|\phichem(\barchi_\tau,
\barc_\tau)\|_{L^\infty(I;L^1(\Omega))}\le C,
\\
\label{eq:33c}
  &\|\nabla\chi_\tau\|_{L^\infty(L^2(\Omega;\R^{N\times 3}))}\le C,\\
\label{eq:34d}
  &\|\nabla\bbm_\tau\|_{L^\infty(L^2(\Omega;\R^{3}))}\le C,\\
&\chi_\tau\in K\textrm{ a.e. in }Q,\\
  &\|\DT\Ee_\tau\|_{L^2(Q;\R^{3\times 3})}\le C,\\
\label{eq:61i}
  &\|\DT\chi_\tau\|_{L^2(Q;\R^N)}\le C,
\\
\label{eq:61j}
  &\|\DT\bbm_\tau\|_{L^1(Q)}\le C,
\\
&\|\nabla\mu_\tau\|_{L^2(Q;\R^3)}\le C,\label{eq:75x}
  \end{align}
\end{subequations}
along with \eqref{eq:29f}. {From (\ref{eq:33}b) and \eqref{ass-fundamental} we immediately obtain the a-priori estimate \eqref{aprio-elast-strai}. Furthermore, from 
\COL{(\ref{eq:33}d,h)} we get \eqref{eq:28}. From (\ref{eq:33}e,i), combined with the initial condition for $d$ contained in \eqref{IC-ass-1}, and from the monotonicity of $d$ (see Step 3 of the discretization scheme at the beginning of this section) we obtain the bound \eqref{apriori-Ia+} on the damage variable. Moreover, since $\bfeps(\DT\bfu_\tau)=
\DTEEtau+\bbE\DT\chi_\tau\in L^2(Q;\R^{3\times3})$, the estimates (\ref{eq:33}g,h) and Korn's inequality entail
\begin{equation}
  \label{eq:35}
  \|\nabla\DT{\bfu}_\tau\|_{L^2(Q;\R^{3\times 3})}\le C.
\end{equation}
The estimate \eqref{apriori-Ia} is now recovered from \eqref{eq:31} and \eqref{eq:35}.} Next, from (\ref{c-mu-disc}b)
taking a gradient of {$\barmu_\tau
=\partial_c^{}\phichem(\barchi_\tau,\barc_\tau)$}, cf.\ \eqref{NS-disc}, {and using the strong convexity assumption \eqref{eq:43}, we obtain the following estimate on concentration gradient:}
\begin{align}\label{est-of-nabla-concentration}
\nabla\barc_\tau=[\partial_{\CHI\CHI}^2\phichem(\barchi_\tau,\barc_\tau)]^{-1}
\big(\nabla\barmu_\tau-\partial_{\CHI\chi}^2\phichem(\barchi_\tau,\barc_\tau)
\nabla\barchi_\tau\big),
\end{align}
and hence, by 
(\ref{assum-mphi}b), (\ref{eq:33}d,
{j}), and 
\eqref{apriori-Ia+++}, we obtain 
\begin{equation}
  \label{eq:36}
  \|\nabla c_\tau\|_{L^2(Q;\R^3)}\le C.
\end{equation}
Because of the coercivity assumption in \eqref{eq:27}, the bounds (\ref{eq:33}c) and \eqref{eq:36} imply 
\eqref{apriori-Ia++}. Using \eqref{eq:75x} and assumption \eqref{eq:76} we obtain \eqref{apriori-Ia+++}. 

It remains for us to prove 
\eqref{eq:73}--\eqref{eq:c58}. Let us fix $r\in [1,5/4)$  and let us set 
\begin{equation}\label{eq:58}
\phi(\omega)=\frac{1+\omega-(1+\omega)^{1-\eta}}{\eta(1-\eta)}\qquad\text{with}\quad\eta=\frac{5-4r}{3}.
\end{equation}
We are going to exploit the following properties of $\phi$:
\begin{subequations}  \label{eq:37}
\begin{align}
 &\forall\omega\ge 0:\quad \phi(\omega)\ge 0,\quad \phi'(w)\le C,\quad 0<\epsilon\le\phi''(\omega)<\frac 1 {1+\omega},\\ &\limsup_{\omega\to+\infty}\frac  1 {\phi''(\omega)\omega^{1+\eta}}\le C.
\end{align}
\end{subequations}
By the second inequality in (\ref{eq:37}a), the function $\phi'(\barw_\tau)$ is 
in $L^\infty(Q)$ and hence it is a legal test for \eqref{heat-eq}, whose 
right-hand side $\bar r_\tau$  is in $L^1(Q)$. By performing this test, and by 
exploiting the convexity of $\phi$ we arrive at:
\begin{equation}
  \label{eq:38}
  \int_\Omega \phi(\barw_\tau(T))\dx+\int_Q\phi''(\barw_\tau)\mathbf{K}(\barEEtau,\barchi_\tau,\bard_\tau,\barc_\tau,\bartheta_\tau)\nabla\bartheta_\tau\cdot\nabla \barw_\tau\dx\dt\le C.
\end{equation}
{F}rom the equation in the third line of \eqref{heat-eq}, recalling from \eqref{eq:20} that 
$w\mapsto\vartheta(\chi,w)$ is the inverse of $\theta\mapsto \eterm(\chi,\theta)$, 
we obtain
\begin{equation}
  \label{eq:66}
  \bartheta_\tau=\vartheta(\barchi_\tau,
\barw_\tau)
\end{equation}
a.e.\ in $Q$. Furthermore, since $\bartheta_\tau(t)\in H^1(\Omega)$, 
$\barchi_\tau(t)\in H^1(\Omega;\R^N)$, and $\bard_\tau(t)\in H^1(\Omega;\R)$ 
for all $t\in(0,T)$. By the chain rule for Sobolev functions, the equation 
\eqref{eq:66} together with assumptions \eqref{eq:21} and \eqref{eq:64} 
implies that $\bar \theta_\tau(\cdot,t)\in H^1(\Omega)$ for all 
$t\in (0,T)$ with
\begin{equation}
  \label{eq:65}
  \nabla \bartheta_\tau=\partial_\chi^{} \vartheta(\barchi_\tau,
\barw_\tau)\nabla\barchi_\tau
+\partial_w \vartheta(\barchi_\tau,
\barw_\tau)\nabla \barw_\tau
\end{equation}
holding a.e. in $Q$. 

Now, by \eqref{eq:14} and 
{\eqref{eq:21}}, we have
\begin{equation}
   \label{eq:71}
    \phi''(\barw_\tau)\mathbf{K}(\barEEtau,\barchi_\tau,\bard_\tau,\barchi_\tau,\bartheta_\tau)\partial_w\vartheta(\barchi_\tau,
\barw_\tau)\nabla \barw_\tau\cdot\nabla\barw_\tau\ge{\epsilon_1}\phi''(\barw_\tau)|\nabla \barw_\tau|^2, 
\end{equation}
{with some $\epsilon_1>0$}. Moreover, using, in the order, Holder's and Young's inequalities, the last inequality in (\ref{eq:37}a), the last line of \eqref{heat-eq}, and \eqref{eq:20}, we obtain, for \COL{$\delta$ sufficiently small},
%\COMMENT{WHY $\ll$ IS NEEDED? WASN'T $<<$ JUST A TYPO?}
\begin{align}
\nonumber
&\phi''(\barw_\tau)\mathbf{K}(\barEEtau,\barchi_\tau,\bard_\tau,\barchi_\tau,\bartheta_\tau)
\partial_\chi^{}\vartheta(\barchi_\tau,
\barw_\tau)\nabla\barchi_\tau\cdot\nabla\barw_\tau
\\
\nonumber
&\qquad\qquad \ge -\frac 1 {2\delta}{\phi''(\barw_\tau)|\mathbf{K}(\barEEtau,\barchi_\tau,\bard_\tau,\barc_\tau,\bartheta_\tau)\partial_\chi^{}\vartheta(\barchi_\tau,
\barw_\tau)\nabla\barchi_\tau|^2}-\frac \delta 2\phi''(\barw_\tau)|\nabla \barw_\tau|^2
   \\
  \nonumber
 &\qquad\qquad \ge -\frac 1 {2\delta}
 \Big|\frac{
 \mathbf{K}(\barEEtau,\barchi_\tau,\bard_\tau,\barc_\tau,\bartheta_\tau)
\partial_\chi^{}\vartheta(\barchi_\tau,
\barw_\tau)\nabla\barchi_\tau}{\sqrt{1+\barw_\tau}}\Big|^2-\frac \delta 2\phi''(\barw_\tau)|\nabla \barw_\tau|^2
 \\
 \label{eq:69}
 &\qquad\qquad
 =-\frac 1{2\delta}\Big|\frac{\mathbf K(\barEEtau,\barchi_\tau,\bard_\tau,\barc_\tau,\bartheta_\tau)\partial_\chi^{} e_\vartheta(\barchi_\tau,
\bar \theta_\tau)\nabla\barchi_\tau}{\partial_\theta \eterm(\barchi_\tau,\bar \theta_\tau)\sqrt{1+\eterm(\barchi_\tau,\bartheta_\tau)}}\Big|^2- \frac \delta 2 \phi''(\barw_\tau)|\nabla \barw_\tau|^2.
\end{align}
The above chain of inequalities, combined with Assumption \eqref{assum-holder-1} and with the lower bound in \eqref{eq:29} yields: 
\begin{align}
 \phi''(\barw_\tau)\mathbf{K}(\barEEtau,\barchi_\tau,\bard_\tau,\barc_\tau,\bartheta_\tau)
\partial_\chi^{}\vartheta(\barchi_\tau,
\barw_\tau)\nabla\barchi_\tau\cdot\nabla\barw_\tau
\ge -\frac C {\delta}|\nabla\barchi_\tau|^2-\frac \delta 2 \phi''(\barw_\tau)|\nabla \barw_\tau|^2.
\end{align}
By combining \eqref{eq:65} with \eqref{eq:69}, and \eqref{eq:71}, we arrive at
\begin{align}
  \label{eq:70}
\phi''(\barw_\tau)\mathbf{K}(\barEEtau,\barchi_\tau,\bard_\tau,\barc_\tau,\bartheta_\tau)\nabla\bartheta_\tau\cdot\nabla \barw_\tau\ge  (1-\delta)\phi''(\barw_\tau)|\nabla\barw_\tau|^2-\frac C{\delta}|\nabla\barchi_\tau|^2.
\end{align}
Since $\phi$ is bounded from below, see (\ref{eq:37}a), and since $\nabla\bard_\tau$ and $\nabla\barchi_\tau$ are bounded in $L^2(Q)$, see (\ref{eq:33}d,e), it follows from \eqref{eq:38} and \eqref{eq:70} that
\begin{equation}
  \label{eq:40}
  \int_Q \phi''(\barw_\tau)|\nabla\barw_\tau|^2\dx\dt\le C.
\end{equation}
Now, by Holder's inequality and by $(\ref{eq:37}b)$ we have the bound
\begin{align}\label{eq:41}
\nonumber  
\int_Q \big|\nabla\barw_\tau\big|^r\dx\dt&\le
\Big(\int_Q \phi''(\barw_\tau)\big|\nabla\barw_\tau\big|^2\dx\dt\Big)^{r/2}
\Big(\int_Q\Big(\frac 1{\phi''(\barw_\tau)^{r/2}}\Big)^{2/(2-r)}\dx\dt\Big)^{
%\frac {2-r}2
1-r/2}
\\
&\le C\Big(1+\int_Q\big|\barw_\tau\big|^{(2-r)r/(1+\eta)}\dt\Big)^{
%\frac {2-r}2
1-r/2}.
\end{align}
Now, let us set $\lambda=\frac {2-r}{1+\eta}$. The choice \eqref{eq:58} for $\eta$ entails that 
\color{black}$\frac \lambda {r}=1-\lambda+\frac \lambda {r^*}$, \color{black} with $r^*=\frac{3r}{3-r}$ 
the exponent of the Sobolev embedding $L^{r^*}(\Omega)\subset W^{1,r}(\Omega)$. Thus, a standard 
interpolation argument based on \color{black}Holder's inequality  entails
$\color{black}\|\barw_\tau\|_{L^{r(1+\eta)/(2-r)}(\Omega)}\color{black}
=\|\barw_\tau\|_{L^{r/\lambda}(\Omega)}\le 
\|\barw_\tau\|^{1-\lambda}_{L^1(\Omega)}\|\barw_\tau\|^{\lambda}_{L^{r^*}(\Omega)}\color{black}
=\|\barw_\tau\|^{r(1+\eta)/(2-r)}_{L^1(\Omega)}\|\barw_\tau\|^{(2-r)/(1+\eta)}_{L^{r^*}(\Omega)}
$. 
Hence, the Sobolev embedding and Poincar\'e's inequality \color{black}yield \color{black}
\[
\big\|\barw_\tau\big\|_{L^{r(1+\eta)/(2-r)}(\Omega)}\le C\big\|\barw_\tau\big\|^{(r-1+\eta)/(1+\eta)}_{L^1(\Omega)}
\Big(\big\|\barw_\tau\big\|_{L^1(\Omega)}+\big\|\nabla \barw_\tau\big\|_{L^r(\Omega)}\Big)^{(2-r)/(1+\eta)}.
\] 
Thus, on taking into account the bound 
$\|\barw_\tau\|_{L^\infty(I;L^1(\Omega))}\le C$, which has already been established, we obtain $
  \int_{Q}|\nabla \barw_\tau|^r\dx\dt\le 
C(1+\int_Q |\nabla\barw_\tau|^r\dx\dt)^{\color{black}1-r/2}$, whence 
\eqref{eq:73} and thence \eqref{eq:b58}. Finally, thanks to the boundedness of $\partial_\chi^{}\vartheta$, 
\emph{cf.} \eqref{eq:64}, by combining \eqref{eq:65} with (\ref{eq:33}d-e), we obtain \eqref{eq:30}. In view of \eqref{eq:21}, from \eqref{eq:b58} we obtain \eqref{eq:c58}.
\color{black}
\QED

\addcontentsline{toc}{subsection}{Further estimates} 
\phantomsection
\begin{lemma}[{\sc Further estimates}]\label{lem-1+}
\slshape
Under the assumption of Lemma~\ref{lem-1}, for some constant $C$ 
%\TTT 
and $C_r$ independent of $\tau$, it also holds:
\begin{subequations}\label{apriori-II+}
\begin{align}\label{apriori-IIc}
&\big\|\r\DDT\bfu_\tau^{\rm i}
\big\|_{L^2(I;H^1(\O;\R^3)^*)}\le C,
\\\label{a-priori-IId}
&\big\|\DT{w}_\tau\big\|_{L^1(I;W^{1,r/(r-1)}(\O)^*)}\le C_r
\ \ \ \text{ with $r$ from \eqref{def-of-r}},
\\\label{a-priori-IIe}
&\big\|\DT{c}_\tau\big\|_{L^2(I;H^1(\O)^*)}\le C,
\\\label{est-of-Delta-m}
&\big\|\Delta\chi_\tau\big\|_{L^2(Q;\R^N)}\le C,
\\\label{est-of-xi}
&{\big\|\sigma_{{\rm r},\tau}\big\|_{L^2(Q;\R^N)}\le C.}
\end{align}
\end{subequations}
%with $r$ from \eqref{def-of-r}.
\end{lemma}

\noindent{\it Proof}.
The ``dual'' estimates (\ref{apriori-II+}a-c) follow by comparison from the 
time-discrete equations (\ref{system-disc}a,c,e) with the corresponding 
boundary conditions (\ref{BC-disc}a,c,e).

In order to prove the remaining estimate (\ref{apriori-II+}d,e), we consider a measurable selection 
$\barsigma_{\rm d,\tau}\in\partial_{\DT\bbm}\zeta(\underlineEe_\tau,\underline\chi_\tau,\underline d_\tau,\underline c_\tau,\underline\theta_\tau,\DT d_\tau)$. {We notice that, thanks to the growth assumption in \eqref{assum-dissi-poten-a} and the estimate \eqref{eq:28}, we have $\barsigma_{\rm d,\tau}\in L^2(Q;\R^N)$.} Thus, the equation in \eqref{chi-eq-disc} can be written as 
\begin{equation}\label{eq:12}
\kappa_1\Delta\barchi_\tau+\barf_\tau\in\partial\delta^{}_K(\barchi_\tau),
\end{equation} 
where $\barf_\tau=-\barsigma_{{\rm d},\tau}
{-}\partial_\chi^{}{\phichem}(\barchi_\tau,
\underline c_\tau)
{-}\partial_\chi^{}{\phimech}(\barEe_\tau,\barchi_\tau,\underline\bbm_\tau)
{-}{\partial_\chi^{}\varphi_{\theta}}(\underline\chi_\tau,{\underline{\bbm}_\tau},\underline{\theta}_\tau)\in L^2(Q;\R^N)$. 
{At this stage, \eqref{chi-eq-disc} is understood in the weak sense, so that $\Delta\barchi_\tau$ (and hence also $\barsigma_{\rm r,\tau}$) are elements of $H^1(\Omega;\R^N)^*$}. 
%\COMMENT{maybe a reference to Pavia's papers and shortening.....}
Let us introduce the Yosida regularization of $\delta^{}_K$, defined by $
  \delta_{K,\varepsilon}(\chi)=\inf_{\widetilde\chi\in\R^N}\left(\delta^{}_K(\widetilde\chi)+\frac 12 |\chi-\widetilde\chi|^2\right)$. Consider the \emph{strictly convex} functional $\widetilde\chi\mapsto\int_\Omega
\big(\frac12|\nabla\widetilde\chi|^2+\frac 12 |\widetilde\chi{-\barchi_\tau}|^2+\delta_{\varepsilon}(\widetilde\chi)-{\barf_\tau\cdot\widetilde\chi}\big)\dx$ 
%\COMMENT{HERE I CHANGED $\barchi$ FOR $\widetilde\chi$, OK???} 
over $H^1(\Omega;\R^N)$. The functional has a unique minimizer $\barchi_{\tau,\varepsilon}$, and this minimizer is the \emph{unique solution} of the regularized elliptic equation
\begin{align}\label{eq:92}
-\kappa_1\Delta\barchi_{\tau,\varepsilon}+\delta'_{K,\varepsilon}(\barchi_{\tau,\varepsilon})+\barchi_{\tau,\varepsilon}=\barf_\tau+\barchi_{\tau},
\end{align}
with boundary condition $\partial_n\barchi_{\tau,\varepsilon}=0$. 

Testing \eqref{eq:92} by $\barchi_{\tau,\varepsilon}$ and using $\delta'_{K,\varepsilon}(\barchi_{\tau,\varepsilon})\cdot\barchi_{\tau,\varepsilon}\ge C(|\barchi_{\tau,\varepsilon}|^2-1)$, together with Holder's and Young's inequalities we get $\int_\Omega \kappa_1|\nabla\barchi_{\tau,\varepsilon}|^2+C|\barchi_{\tau,\varepsilon}|^2\dx\le C|\Omega|+\frac 1 \delta \int_\Omega |\barf_\tau|^2\dx+\delta\int_\Omega |\barchi_{\tau,\varepsilon}|^2\dx$
for any $\delta>0$. By the arbitrariness of $\delta$, we obtain the bound 
$\|\barchi_{\tau,\varepsilon}\|_{H^1(\Omega;\R^N)}\le C$. Moreover, since $\delta_\varepsilon'$ 
is Lipschitz continuous although not uniformly in $\varepsilon$, 
we have $\delta_\varepsilon'(\barchi_{\tau,\varepsilon})\le C(1+|\barchi_{\tau,\varepsilon}|){/\varepsilon}$, 
and hence $\|\delta'_{K,\varepsilon}(\barchi_{\tau,\varepsilon})\|_{L^2(\Omega;\R^N)}\le C{/\varepsilon}$. By 
comparison in \eqref{eq:92}, we obtain $\Delta\barchi_{\tau,\varepsilon}\in L^2(\Omega;\R^N)$. 
Thus we can test \eqref{eq:92} by $-\Delta\barchi_{\tau,\varepsilon}$ and use 
$$
\int_\Omega -\delta_{K,\varepsilon}'(\barchi_{\tau,\varepsilon}){\cdot}\Delta\barchi_{\tau,\varepsilon}\dx
=\int_\Omega \nabla \delta_{K,\varepsilon}'(\barchi_{\tau,\varepsilon}){:}\nabla\barchi_{\tau,\varepsilon}\dx
=\int_\Omega \delta''_{K,\varepsilon}(\barchi_{\tau,\varepsilon})|\nabla\barchi_{\tau,\varepsilon}|^2\dx\ge0
$$
to obtain the inequality $\int_\Omega \kappa_1|\Delta\barchi_{\tau,\varepsilon}|^2-\Delta\barchi_{\tau,\varepsilon}\cdot\barchi_{\tau,\varepsilon}\,\dx\le\int_\Omega(\barf_\tau+\barchi_\tau)\barchi_{\tau,\varepsilon}\,\dx$. Again, the application of H\"older's and Young's inequalities, yields the estimate $\|\Delta\barchi_{\varepsilon,\tau}\|_{L^2(\Omega;\R^N)}\le C$. By comparison in \eqref{eq:92}, we still have the estimate
$\|\delta_{K,{\varepsilon}}'(\barchi_{\tau,\varepsilon})\|_{L^2(\Omega;\R^N)}
=\|\kappa_1\Delta\barchi_{\tau,\varepsilon}{-}\barchi_{\tau,\varepsilon}
{+}\barf_\tau{+}\barchi_{\tau}\|_{L^2(\Omega;\R^N)}\le C$
now independent of $\varepsilon$. By the a priori bound derived above, there exists $\widetilde\chi_\tau\in H^1(\Omega;\mathbb R^N)$ and a subsequence of $\{\chi_{\tau,\varepsilon}\}_{\tau>0}$ such that $\barchi_{\tau,\varepsilon}\to \widetilde\chi_\tau$ weakly in $ H^1(\Omega;\R^N)$ and strongly in $L^2(\Omega;\R^N)$, $\Delta\barchi_{\tau,\varepsilon}\to \Delta \widetilde\chi_\tau$ weakly in $L^2(\Omega;\R^N)$, and $\delta'_{K,\varepsilon}(\barchi_{\tau,\varepsilon})\to {\widetilde{\sigma}_{\rm r}}$ weakly in $L^2(\Omega;\R^N)$, with  $\widetilde\sigma_{\rm r}=\kappa_1\Delta \widetilde\chi_\tau-\widetilde\chi_\tau+{\barf_\tau+\barchi_\tau}$. In order to identify the limit $\widetilde\chi_\tau$, we consider an arbitrary $z\in L^2(\Omega;\R^N)$, and we write
\begin{align}
  \int_\Omega \delta_{K,\varepsilon}(z)-\delta_{K,\varepsilon}(\barchi_{\tau,\varepsilon})\dx&\ge \int_\Omega \delta_{K,\varepsilon}'(\barchi_{\tau,\varepsilon}){\cdot}(z{-}\barchi_{\tau,\varepsilon})\dx
=\int_\Omega (\barf_\tau+\barchi_\tau+\kappa_1\Delta\barchi_{\tau,\varepsilon}-\barchi_{\tau,\varepsilon}){\cdot}(z{-}\barchi_{\tau,\varepsilon})\dx.
\label{eq:95}
\end{align}
% \GT{We recall the following properties of the Yosida regularization 
% (see \cite[Lemma 5.17]{Roub13NPDE}):
% \begin{subequations}\label{eq:94}
% \begin{align}
%   &v_\varepsilon\to v\textrm{ weakly in }L^2(\Omega;\R^N)\Rightarrow \liminf_{\varepsilon\to 0}\int_\Omega \delta_{K,\varepsilon}(v_\varepsilon)\dx \ge \int_\Omega \delta^{}_K(v)\dx;
% \\
%   &\forall v\in L^2(\Omega;\R^N):\quad \limsup_{\varepsilon\to 0}\int_\Omega \delta_{K,\varepsilon}(v)\dx \le \int_\Omega \delta^{}_K(v)\dx.
% \end{align}
% \end{subequations}}
Using standard properties of the Yosida regularization 
(see e.g.\ \cite[Lemma 5.17]{Roub13NPDE}), we can pass to the limit in 
\eqref{eq:95} to obtain
$\int_\Omega \delta^{}_K(z)-\delta^{}_K(\widetilde\chi_\tau)\dx \ge\int_\Omega (\barf_\tau+\barchi_\tau+\kappa_1\Delta \widetilde\chi_\tau-\widetilde\chi_\tau)\cdot(z-\widetilde\chi_\tau)\dx$.
Thus,  $\widetilde\chi_\tau$ solves the differential inclusion
\begin{align}\label{eq:96}
  \kappa_1\Delta\widetilde\chi -\widetilde\chi_\tau+\barf_\tau+\barchi_\tau\in \partial \delta^{}_K(\widetilde\chi_\tau)
\end{align}
with homogeneous Neumann condition on $\partial\Omega$. By comparing \eqref{eq:12} with \eqref{eq:96} and the corresponding boundary conditions, we see that $\barchi_\tau$ is a solution of \eqref{eq:96} as well. On the other hand solving \eqref{eq:96} is equivalent to minimizing a strictly 
convex functional, thus the solution of \eqref{eq:96} is unique. We therefore conclude that 
$\widetilde\chi_\tau=\barchi_\tau$ and ${\widetilde{\sigma}_{\rm r}}=\kappa_1\Delta\barchi_\tau+\barf_\tau=\barsigma_{\rm r}$.
Consequently, on using the strong convergence of $\barchi_{\tau,\varepsilon}$ and the weak convergence of $\delta'_{K,\varepsilon}(\widetilde\chi_\tau)$ we arrive at:
\begin{align}\nonumber
  0&\le \limsup_{\varepsilon\to 0}\int_\Omega \kappa_1 \delta_{K,\varepsilon}''(\barchi_{\tau,\varepsilon})\nabla\barchi_{\tau,\varepsilon}{:}\nabla\barchi_{\tau,\varepsilon}\dx=\limsup_{\varepsilon\to 0}\int_\Omega
\big(-\kappa_1 \Delta\barchi_{\tau,\varepsilon}\big){\cdot}\delta'_{K,\varepsilon}(\barchi_{\tau,\varepsilon})\dx\nonumber 
\\\nonumber
&=\limsup_{\varepsilon\to 0}\int_\Omega \big(\barf_\tau+\barchi_{\tau}-\delta'_{K,\varepsilon}(\barchi_{\tau,\varepsilon})-\barchi_{\tau,\varepsilon}\big){\cdot}\delta'_{K,\varepsilon}(\barchi_{\tau,\varepsilon})\dx
\\\nonumber&\le 
\int_\Omega\big(\barf_\tau+\barchi_\tau-\barsigma_{\rm r,\tau}-\barchi_\tau\big){\cdot} \barsigma_{\rm r,\tau}\dx
= \kappa_1\int_\Omega \Delta\barchi_\tau\cdot \barsigma_\tau\dx.
\end{align}
Since $\big\|\barf_\tau\big\|_{L^2(Q;\R^N)}\!\le C$, the test of \eqref{eq:43} with 
$\Delta\barchi_\tau$, with Holder's and Young's inequalities yields 
\eqref{est-of-Delta-m}. The bound \eqref{est-of-xi} follows by comparison. 
\QED

\medskip

\begin{proposition}[{\sc Convergence for $\tau\to0$}]\label{prop-conv}
\slshape
Let again the assumption of Lemma~\ref{lem-1} hold and let $\Omega$ be 
smooth. Then there is a subsequence such that
\begin{subequations}\label{15.-strong}
\begin{align}
\label{15.-u-strong}
&\bfu_\tau\to\bfu&&\text{strongly in }\ H^1(I;H^1(\O;\R^3)),
\\\label{15.-m}
&\chi_\tau\to \chi &&\text{strongly in }\ 
H^1(I;L^2(\O;\R^N))\,\cap\,C({\,\overline{\!Q\!}\,};\R^N),
\\\label{15.-chi2}
&\barc_\tau\to c\quad \&\quad \underline c_\tau\to c &&\text{strongly in }\ L^2(Q),
\\\label{15.-chi}
&
{\bbm_\tau(t)\to\bbm(t)}&&{\text{weakly in }H^1(\O)\ \ \forall t\!\in\!I},
\\[-.3em]\label{15.-chi+}
&{\DT\bbm_\tau\to\DT\bbm}&&{\text{weakly* in }
\mathrm{Meas}({\,\overline{\!Q\!}\,})\cong C({\,\overline{\!Q\!}\,})^*},
\\
\label{eq:29ff}
&\bar{\theta}_\tau\to {\theta},\qquad \underline{\theta}_\tau\to {\theta}\!\!\!\!\!\!\!\!
&&\text{strongly in }\ L^s(Q)\ \text{ with any }\ 
{1\le s<5/3},
\\\label{15.-mu-strong}
&{\barmu_\tau\to\mu}&&\text{strongly in }\ L^2(I;H^1(\O)),
\\
\label{15.-theta-strong}
&\bar{w}_\tau\to {w}\ \ \&\ \ \underline{w}_\tau\to {w}\!\!\!\!\!\!\!\!
&&\text{strongly in }\ L^s(Q)\ \text{ with any }\ 
{1\le s<5/3},
\end{align}\end{subequations}
and any 
$(\bfu,\chi,\CHI,\bbm,\theta,\mu,w)$ 
obtained in this way is a weak solution to {the initial-boundary-value 
problem \eqref{systemb}--\eqref{ICb} according 
Definition~\ref{def}}
which also preserves the total energy in the sense \eqref{eq:99}.
\end{proposition}

\noindent{\it Proof}. {For clarity of exposition, we divide the proof to
eleven particular steps.}
\newcounter{step}
\setcounter{step}1

\medskip

\noindent \emph{Step \thestep: Selection of a converging subsequence.} 
By Banach's selection principle, we select a weakly* converging subsequence with respect 
to the norms from the estimates \eqref{apriori-I} and \eqref{apriori-II+}, namely,
\begin{subequations}\label{eq:46}
\begin{align}
\label{eq:47}  &&&\bfu_{\tau}\to \bfu&&\text{weakly}^*\text{ in }W^{1,\infty}(I;L^2(\Omega;\R^3))\cap H^1(I;H^1(\Omega;\R^3))\\
\label{eq:48}  
&&&{\Ee}_\tau\to \Ee\!\!\! \!\!
&&\text{weakly}\text{ in } H^1(I;L^2(\Omega;\R^{3\times 3}))\\
&&&\chi_\tau\to\chi 
&&\text{weakly}^*\text{ in }L^\infty(I;H^1(\O;\R^N))\,\cap\,
H^1(I;L^2(\O;\R^N))\,\cap\,L^\infty(Q;\R^N),\label{eq:49}\\
\label{eq:50}
&&&\CHI_\tau\to\CHI
&&\text{weakly}\text{ in }L^2(I;H^1(\O)),\\
\label{eq:46e}
&&&\DT\CHI_\tau\to\DT\CHI&&\text{weakly}\text{ in }L^2(I;H^1(\O)^*),\\
\label{eq:51}
&&&\mu_\tau\to\mu&&\text{weakly}\text{ in }{L^2(I;H^1(\O))},\\
\label{eq:52}
&&&\bbm_\tau\to\bbm&&\text{weakly}^*\text{ in }L^2(I;H^1(\O))\,\cap\,
L^\infty(Q),
\\\label{15.-xi-weak}
&&&\barsigma_{\rm r,\tau}\to \sigma_{\rm r}&&\text{weakly in }\ L^2(Q;\R^N),
\\\label{15.2}
\end{align}
\end{subequations}
and also \eqref{15.-chi+}. Moreover, by the BV-estimate \eqref{apriori-Ia+} 
and by Helly's selection principle, we can rely also on \eqref{15.-chi} for a 
subsequence.
We introduce the shorthand notation
\begin{equation}
  \label{eq:54}
 \bfv_\tau=[\DT{\bfu}_\tau]^{\rm i}\qquad\text{and}\qquad\barbfv_\tau=\DT{\bfu}_\tau.
\end{equation}
We also define $t_\tau:=t-\tau[t/\tau]$. Then, we have 
$\barbfv_\tau-\bfv_\tau=(\tau-t_\tau)\DT{\bfv}_\tau$. Moreover,
\begin{align}
\nonumber\int_0^T\!\!\int_\Omega |\barbfv_\tau-\bfv_\tau|^2\dx\dt&=\int_0^T\!\!\int_\Omega(\tau-t_\tau)( \bfv_\tau-\barbfv_\tau)\cdot \DT{\bfv}_\tau\dx\dt
\\
&\le 
\tau
\big\| \bfv_\tau-\barbfv_\tau\big\|_{L^2(I;H^1(\Omega;\R^3))}
\big\| \DT{\bfv}_\tau\big\|_{L^2(I;H^1(\Omega;\R^3)^*)}\to 0.\label{eq:72}
\end{align}
By \eqref{eq:47}, $\barbfv_\tau\to\DT{\bfu}$ weakly
in 
{$L^2(I;H^1(\Omega;\R^3)$}. Thus, \eqref{eq:72} implies that also ${\bfv}_\tau\to\DT{\bfu}$ weakly${}^*$ in 
$L^2(I;H^1(\Omega;\R^3))$. On taking into account that $\DDT{\bfu}_\tau^{\rm i}=\DT{\bfv}_\tau$ and {using a standard argument to identify time derivatives (see for instance \cite[Theorem 8.9]{Roub13NPDE})}, we arrive at
\begin{equation}
  \label{eq:55}
  \DDT{\bfu}^{\rm i}_\tau\to\DDT{\bfu}\textrm{ weakly in }L^2(I;H^1(\O;R^3)^*).
\end{equation}

By Rellich's theorem we have the continuous and the compact embeddings 
$H^1(I;L^2(\Omega))\cap L^\infty(I;H^1(\Omega))
\subset H^1(Q)\Subset L^2(Q)$ so that (\ref{eq:46}c,f) imply
\begin{subequations}
\begin{align}
  \label{eq:56}
&\text{  $\chi_\tau\to\chi$ strongly in $L^2(Q;\R^N)$}.
\end{align}
Again, arguing as \eqref{eq:72}, we have that  $\|\barchi_\tau{-}\chi_\tau\|_{L^2(Q;\R^N)}\to0$ {and similarly also for 
$\underline\chi_\tau$}, cf.\ \cite[Rem.\,8.10]{Roub13NPDE}, hence
\begin{align}\label{eq:57}
&\text{ $\barchi_\tau\to\chi\ $ {and $\ \underline\chi_\tau\to\chi\ $} strongly in $\ L^2(Q;\R^N)$.}
\end{align}
\end{subequations}
Using the Aubin-Lions theorem with the estimates \eqref{apriori-Ia++} and \eqref{a-priori-IIe}, 
we obtain, for a subsequence,
\begin{align}
  c_\tau\to c\quad\textrm{strongly in}\quad L^2(I;L^q(\Omega))\quad \forall 1\le q<6.
\end{align}
Moreover, we have
$\int_0^T\!\!\int_\Omega  |c_\tau-\underline c_\tau|^2\dx{\rm d }t
\le \tau 
\|\DT c_\tau\|_{L^2(I;H^1(\Omega)^*)} \|c_\tau-\underline c_\tau\|_{L^2(I;H^1(\Omega))}\to 0$
and similarly also for  $\barc_\tau$. Hence,
\begin{align}\label{eq:84}
  \barc_\tau\to c 
\ \text{ and }\ \underline c_\tau\to c\ \text{ strongly in }\ L^2(Q).
\end{align}
Similarly, by using the generalized Aubin-Lions theorem
relying on the boundedness of $\{\underline{\DT d}_\tau\}_{\tau>0}$ in 
Meas($I;L^1(\Omega)$), see \cite[Corollary~7.9]{Roub13NPDE},
we also have at disposal 
\begin{align}\label{eq:84-for-d}
\underline d_\tau\to d\ \text{ strongly in }\ L^2(Q).
\end{align}

Now, let $1\le r<5/4$ and $1\le q<r^*=\frac{3r}{3-r}$. Then $L^q(\Omega)$ 
is compactly embedded in $W^{1,r}(\Omega)$. By \eqref{eq:73} and \eqref{eq:b58}, we have that $\|w_\tau\|_{L^r(I;W^{1,r}(\Omega))}\le C$. Thus, by \eqref{a-priori-IId}, thanks to the Aubin-Lions Lemma, there exists a subsequence such that $\|w_\tau-w\|_{L^r(I;L^q(\Omega))}\to 0$. Now, for any such $q$ and $r$, let $\lambda(r,q)=(1+\frac 1 r-\frac 1 q)^{-1}$. Then, we have $\frac{\lambda(r,q)}{r}=1-\lambda(r,q)+\frac {\lambda(r,q)}q$. Therefore, by interpolating between $L^\infty(I;L^1(\Omega))$ and $L^r(I;L^q(\Omega))$ (see \cite[Proposition 1.41]{Roub13NPDE}), we have the inequality $\|w_\tau-w\|_{L^{r/\lambda(r,q)}(Q)}^{}\le C  \|w_\tau-w\|^{1-\lambda(r,q)}_{L^\infty(I;L^1(\Omega))} \|w_\tau-w\|^{\lambda(r,q)}_{L^r(I;L^q(\Omega))}$. Now, for $1\le r<5/4$ fixed, 
we have $ \inf_{1\le q<r^*}\lambda(r,q)=(1+\frac 1 r-\frac 1 {r^*})^{-1}
=(1+\frac 1 r-\frac {3-r}{3r})^{-1}=\frac34$. Thus, $\sup_{1\le r<5/4\atop 1\le q<r^*}\frac{r}{\lambda(r,q)}=5/3$. This gives 
\begin{align}\label{eq:86}
  w_\tau\to w\text{ strongly in }L^s(Q)\quad\text{with any }1\le s<5/3.
\end{align}
{Next, we observe that $\DT{\barw}_\tau$ is bounded in the space of $W^{1,r/(r-1)}(\Omega)^*$-valued Radon measures on $I$, since:
\begin{align}
   \big\|\DT{\barw}_\tau\big\|_{\mathrm{Meas}(
I;W^{1,r/(r-1)}(\Omega)^*)}&=\sup_{\|\varphi\|_{{\rm C}(
I;W^{1,r/(r-1)}(\Omega)^*)}=1}\int_Q \barw_\tau\DT\varphi \dx\dt\nonumber
\\
&=\sup_{\|\varphi\|_{{\rm C}(
I;W^{1,r/(r-1)}(\Omega)^*)}=1}  \sum_{k=1}^{T/\tau}\int_\Omega w_\tau^k (\varphi(k\tau)-\varphi((k-1)\tau))\dx\nonumber
\\
&=\sup_{\|\varphi\|_{{\rm C}(
I;W^{1,r/(r-1)}(\Omega)^*)}=1}\int_\Omega\!w_{\tau}^{T/\tau}\varphi(T)
-\!\sum_{k=1}^{T/\tau-1}\!
\int_\Omega\big(w_\tau^{k+1}{-}w_\tau^k\big)\varphi(k\tau)\dx
-w_\tau^1\varphi(0)\dx\nonumber
\\&
\le\big\|\DT w_\tau\big\|_{L^1(I;W^{1,r/(r-1)}(\Omega)^*)}\le C,
\end{align}
where the last inequality follows from \eqref{a-priori-IId}.} We can now use the generalized version of the Aubin-Lions Lemma in 
\cite[Corollary~7.9]{Roub13NPDE} and interpolate with the estimate to 
conclude that 
{there exists $\barw\in L^s(Q)$ such that}
\begin{align}\label{eq:85}
  \barw_\tau\to \barw\ \ \text{ strongly in }\ L^s(Q)
\quad\textrm{ with any }\quad 1\le s<5/3.
\end{align}
{Thanks to \eqref{eq:86} and \eqref{eq:85}, in order to prove the first convergence statement in  \eqref{15.-theta-strong} it suffices for us to show that $\barw=w$. To this aim, we argue as in \cite[Remark 8.10]{Roub13NPDE}:} 
\begin{align}
\nonumber
&\big\|w_\tau-\barw_\tau\big\|_{L^1(I;W^{1,r/(r-1)}(\Omega)^*)}
=\sum_{k=1}^{T/\tau}\int_{(k-1)\tau}^k\Big\|\frac{t{-}k\tau}\tau(w_\tau^k-w_\tau^{k-1})\Big\|_{W^{1,r/(r-1)}(\Omega)^*}\dt
\\[-.5em]\nonumber
&\quad=\frac \tau 2 \sum_{k=1}^{T/\tau}
\big\|w_\tau^k-w_\tau^{k-1}\big\|_{W^{1,r/(r-1)}(\Omega)^*}
=\frac {\tau} 2 \sum_{k=1}^{T/\tau}
\int_{(k-1)\tau}^k\big\|\DT w_\tau\big\|_{W^{1,r/(r-1)}(\Omega)^*}\dt
=\frac {\tau}2 \int_0^T \big\|\DT w_\tau\big\|_{W^{1,r/(r-1)}(\Omega)^*}\dt\to 0,
\end{align}
{where we have used the bound \eqref{a-priori-IId}. The second convergence statement in \eqref{15.-theta-strong} is arrived at using a similar argument.}

In order to obtain the convergences in \eqref{eq:29ff}, we invert \eqref{heat-eq} with respect to $\bartheta_\tau$ to obtain (\emph{cf.} \eqref{eq:20}):
\begin{align}\label{theta=vartheta(w)}
  \bartheta_\tau=\vartheta(\barchi_\tau,
\barw_\tau),\qquad{\textrm{and}\qquad
\underline\theta_\tau=\vartheta(\underline\chi_\tau,\underline w_\tau).}
\end{align}
{Then, \eqref{eq:29ff} follows from the already-established convergences \eqref{15.-theta-strong} and \eqref{eq:57} by the contiuity of the Nemytski\u\i\ 
mapping associated to $\vartheta$.}
 
% \COMMENT{DO WE NOW NEED THE FOLLOWING PARAGRAPH? AND EVEN 
% $C_\text{weak}(I;H^2(\Omega))$ DOES NOT HOLD!}
% Eventually, by using the Arzel\`a-Ascoli modification of the Aubin-Lions
% theorem, cf.\ \cite[Lemma~7.10]{Roub13NPDE}, we have the 
% compact embedding $C_\text{weak}(I;H^2(\Omega))\cap H^1(I;L^2(\Omega))
% \Subset C(I;H^{2-\epsilon}(\Omega))\subset C({\,\overline{\!Q\!}\,})$ for 
% any $0<\epsilon<1/2$. 
% From \eqref{est-of-Delta-m} and \eqref{eq:28} together with the embedding 
% $\{v\in H^1(\Omega);\ \Delta v\in L^2(\Omega),\ \nabla v{\cdot}{\mathbf n}=0
% \text{ on }\Gamma\}\subset H^2(\Omega)$, we then obtain $\chi_\tau\to\chi$ 
% in $C({\,\overline{\!Q\!}\,};\R^N)$, as claimed in \eqref{15.-m}. 
% Note that here we used the assumption $\Omega$ smooth to have the 
% $H^2$ regularity at disposal.

\addcontentsline{toc}{subsection}{Weak convergence of the energetic stresses}

\medskip

\addcontentsline{toc}{subsection}{strong convergence of strain}
\phantomsection
\addtocounter{step}1
\normalsize

\noindent\emph{Step \thestep: Strong convergence of $\barEe_\tau$.}
In this step we prove:
\begin{equation}\label{eq:61}
  \barEe_\tau\to\Ee\quad{\rm strongly\  in\  }L^2(Q;\R^{3\times 3}).
\end{equation}
Note that we already have the weak convergence. 
Thus, by \eqref{eq:88+},
\begin{align}
\nonumber
&\epsilon
\|\barEe_\tau-\Ee\|_{L^2(Q;\R^3)}^2
\le{}
\int_Q \big(\partial_\Ee^{}\phimech(\barEe_\tau,\barchi_\tau,\underline\bbm_\tau)
-\partial_\Ee^{}\phimech(\Ee,\barchi_\tau,\underline\bbm_\tau)\big):(\barEe_\tau-\Ee)\dx\dt
\\
\nonumber
&\qquad=\int_Q \partial_\Ee^{}\phimech(\barEe_\tau,\barchi_\tau,\underline\bbm_\tau):(\bfeps(\bar\bfu_\tau)
-\bfeps(\bfu))-\partial_\Ee^{}\phimech(\barEe_\tau,\barchi_\tau,\underline\bbm_\tau):\bbE(\barchi_\tau{-}\chi)
\,\dx\dt
\\
\label{eq:59}
&\qquad\quad-\int_Q\partial_\Ee^{}\phimech(\Ee,\barchi_\tau,\underline\bbm_\tau)
:(\barEe_\tau-\Ee)\,\dx\dt.
\end{align}
We are going to show that the right-hand 
side of \eqref{eq:59} converges to 0 as $\tau\to 0$.

By \eqref{eq:57} and \eqref{eq:84-for-d} and the continuity of the 
Nemytski\u\i\ mapping induced by $\partial_\Ee^{}\phimech(\Ee,\cdot,\cdot)$
and by \eqref{eq:48}, we have 
$\partial_\Ee^{}\phimech(\Ee,\barchi_\tau,\underline\bbm_\tau){:}(\Ee_\tau{-}\Ee)\to 0$ weakly in $L^1(Q)$. Also, by the boundedness of 
$\{\partial_\Ee^{}\phimech(\barEe_\tau,\barchi_\tau,\underline\bbm_\tau)\}_{\tau>0}$
in $L^2(Q;\R^{3\times3})$ and again by \eqref{eq:57}, we have
$\partial_\Ee^{}\phimech(\barEe_\tau,\barchi_\tau,\underline\bbm_\tau){:}\bbE(\barchi_\tau{-}\chi)\to0$ strongly in $L^1(Q)$. Hence, 
relying on the discrete equation \eqref{mombal-disc},
we can continue in estimation \eqref{eq:59} as follows:
\begin{align}
\nonumber
&\epsilon\limsup_{\tau\to 0}\|\barEe_\tau-\Ee\|_{L^2(Q;\R^3)}^2
\le \limsup_{\tau\to 0} 
\int_Q\!\partial_\Ee^{}\phimech(\barEe_\tau,\barchi_\tau,\underline\bbm_\tau){:}(\bfeps(\bar\bfu_\tau)-\bfeps(\bfu))
\, \dx\dt.
\\\nonumber
&\qquad
=\limsup_{\tau\to 0}
\int_Q (\barbff_\tau-\varrho\DDT\bfu_\tau^{\rm i})\cdot(\bar\bfu_\tau-\bfu)
\dx\dt
-\int_Q\bbD\DTEEtau:(\barEe_\tau-\Ee)\dx\dt
\\\nonumber
&\qquad
\le\lim_{\tau\to 0}
\bigg(\int_\O\varrho \DT\bfu_\tau(0)\cdot\bar\bfu_\tau(\tau)
-\varrho\DT\bfu_\tau(T)\cdot\bar\bfu_\tau(T)\,\d x
+\int_\tau^T\!\!\!\int_\Omega\varrho\DT\bfu_\tau(\cdot-\tau){\cdot}
\DT\bfu_\tau\,\d x\d t
+\int_Q\varrho\DDT\bfu_\tau^{\rm i}{\cdot}\bfu\,\d x\d t\bigg)
\\[-.3em]\nonumber
&\qquad\qquad
+\lim_{\tau\to 0}\int_\O\barbff_\tau\cdot(\bar\bfu_\tau-\bfu)\,\d x\d t
-\liminf_{\tau\to 0}\int_Q\bbD\DT\Ee_\tau:(\Ee_\tau-\Ee)\dx\dt\nonumber
\\&\nonumber
\qquad=\int_\O\varrho \DT\bfu(0){\cdot}\bfu(0)
-\varrho\DT\bfu(T){\cdot}\bfu(T)\,\d x
+\int_Q\varrho|\DT\bfu|^2+\varrho\DDT\bfu{\cdot}\bfu\,\d x\d t
\\[-.3em]&\nonumber
\qquad\qquad-\liminf_{\tau\to 0}\int_Q\bbD\DT\Ee_\tau:\Ee_\tau\dx\dt
+\lim_{\tau\to 0}\int_Q\bbD\DT\Ee_\tau:\Ee\dx\dt\nonumber\\
\label{eq:81}
&\qquad=\liminf_{\tau\to 0}\frac12\int_\Omega\big(\bbD\Ee_\tau(T){:}\Ee_\tau(T)
-\bbD\Ee_0{:}\Ee_0\big)\dx
-\frac12\int_\Omega\big(\bbD\Ee(T){:}\Ee(T)-\bbD\Ee_0{:}\Ee_0\big)\dx\le0.
\end{align}
Here we also used the discrete by-part summation, cf.\ e.g.\ 
\cite[Remark~11.38]{Roub13NPDE} and that, since 
$\barEe_\tau-\Ee_\tau=(\tau k-t)\DT\Ee_\tau$ for $t\in((k-1)\tau,k\tau)$ and 
since  $\bbD$ is positive, we have 
$\int_Q\bbD\DT\Ee_\tau{:}(\barEe_\tau-\Ee_\tau)\dx\dt\ge 0$. Also we used 
the weak convergence $\Ee_\tau(T)\rightharpoonup\Ee(T)$ in 
$L^2(\Omega;\R^{3\times 3})$, which is readily verified through 
$\int_\Omega \Ee_\tau(T){:}\widetilde\Ee\dx
=\int_0^T\!\!\int_\Omega \DT\Ee_\tau{:}\widetilde\Ee\dx\dt
+\int_\Omega \Ee_\tau(0){:}\widetilde\Ee\dx
\to \int_0^T\!\!\int_\Omega \DT\Ee{:}\widetilde\Ee\dx\dt
+\int_\Omega \Ee(0){:}\widetilde\Ee\dx=\int_\Omega \Ee(T){:}\widetilde\Ee\dx$
holding for  $\widetilde\Ee\in L^2(\Omega;\R^{3\times 3})$ arbitrary.
Moreover, we also used that $\{\varrho\DDT\bfu_\tau^{\rm i}\}_{\tau>0}$
converges weakly* in a space which is in duality to the space where 
$\DT\bfu$ lives, and in particular also the 
$\varrho\DDT\bfu$ is in duality with 
$\DT\bfu$:
\begin{align}\label{quality-of-u}
\DT\bfu\in L^2(I;H^1(\O;\R^3))\quad \text{and}\quad 
\r\DDT\bfu\in L^2(I;H^1(\O;\R^3)^*).
\end{align}

\medskip

\addtocounter{step}1
\noindent
\emph{Step \thestep: Convergence in the semilinear mechanical part.}
Because of the smoothness of $\phimech$ and of the strong convergence of 
$\barEe_\tau$, $\barchi_\tau$, and $\underline\bbm_\tau$ already established 
and stated, respectively, in \eqref{eq:61}, \eqref{eq:57}, and 
\eqref{eq:84-for-d}, we have
\begin{align}
\partial_\Ee^{}
\phimech(\barEe_\tau,\barchi_\tau,\underline\bbm_\tau)
\to\partial_\Ee^{}
\phimech(\Ee,\chi,\bbm)\ \ \textrm{ 
in }\ L^2(Q;\R^{3\times 3}).
\end{align}
by continuity of the Nemytski\u\i\ mapping induced 
by $\phimech$. The limit passage in \eqref{mombal-disc} is then
done.

\medskip

\addtocounter{step}1
\noindent
\emph{Step \thestep: Limit passage in the phase-field 
equation.} 
We rewrite \eqref{damage-eq-disc} as two variational inequalities:
\begin{subequations}
\begin{align}\nonumber
&
\int_Q\!
\big(\barsigma_{\rm r,\tau}+
\partial_{\chi}
\phimechchem(\barEe_\tau,\barchi_\tau,\underline\bbm_\tau,\underline c_\tau)
+\partial_\chi^{}\phiterm(\underline\chi_\tau,
\underline\theta_\tau)
-
\bbE^\top{:}(\partial_\Ee^{}\phimech(\barEe_\tau,\barchi_\tau,\underline\bbm_\tau)
+\bbD\DTEEtau)\big){\cdot}(v{-}\DT \chi_\tau)\nonumber
\\\nonumber
&\hspace{4em}
+{\kappa_1}\nabla\barchi_\tau{:}(\nabla v{-}\nabla\DT\chi_\tau)
+\zeta(\underlineEe_\tau,\underline\chi_\tau,\underline
d_\tau,\underline c_\tau,\underline\theta_\tau,v)
\,\d x\d t
\\&\hspace{11em}
\ge\!\int_Q\!
-\zeta(\underlineEe_\tau,\underline\chi_\tau,\underline d
_\tau,\underline c_\tau,\underline\theta_\tau,
\DT\chi_\tau)\,\d x\d t
\qquad\ \forall v\!\in\!L^2(I;H^1(\Omega;\R^N)), 
\label{mm-ineq-1}
\\&
\label{mm-ineq-2}
\int_Q\!\barsigma_{\rm r,\tau}{\cdot}(v{-}\barchi_\tau)\,\d x\d t\ge0
\qquad\qquad\qquad\qquad\qquad\qquad\qquad
\forall v\!\in\!L^2(Q;\R^N),\ \ v\!\in\!K\text{ a.e. in } Q.
\end{align}
\end{subequations}\hypertarget{2013-11-16-a}{}
The limit passage in \eqref{mm-ineq-2} is easy because 
$\barsigma_{\rm r,\tau}\to \sigma_{\rm r}$ weakly in $L^2(Q;\R^N)$ and 
$\barchi_\tau\to\chi$ strongly in $L^2(Q;\R^N)$ has already been proved
in Step 2; thus $\sigma_{\rm r}\in N_{K}(\chi)$ is shown. Now we can make 
a limit passage in \eqref{mm-ineq-1}. Here on the left-hand side we have 
collected all terms that need to be handled through a continuity or a weak 
upper semicontinuity, while the right-hand side is to be treated by weak 
lower semicontinuity. We benefit from the already proven strong convergence 
of $\underline\bbm_\tau$, $\underline c_\tau$, and $\barEe_\tau$.
The limit passage in $\partial_{\chi}\phimechchem(\barEe_\tau,\barchi_\tau,\underline\bbm_\tau,\underline c_\tau){\cdot}\DT\chi_\tau$
and $\partial_\chi^{}\phiterm(\underline\chi_\tau,
\underline\theta_\tau){\cdot}\DT\chi_\tau$ is simply by continuity.
Furthermore, we have
\begin{align}\nonumber
\limsup_{\tau\to0}\int_Q\!-{\kappa_1}\nabla\barchi_\tau{\cdot}\nabla\DT\chi_\tau
\,\d x\d t&\le\int_\Omega\!\frac{\kappa_1}2\big|\nabla\chi_0\big|^2\,\d x
-\liminf_{\tau\to0}\int_\Omega\!\frac{\kappa_1}2\big|\nabla\chi(T)\big|^2\,\d x
\\&\label{by-part-chi}
\le\int_\Omega\frac{\kappa_1}2\big|\nabla\chi_0\big|^2
-\frac{\kappa_1}2\big|\nabla\chi(T)\big|^2\,\d x.
\end{align}
The only difficult term is $\bbE^\top{:}\bbD\DT\Ee_\tau\DT\chi_\tau$
because so far we know only a weak convergence of both $\DT\Ee_\tau$ and
$\DT\chi_\tau$. This requires quite tricky chain of arguments:
\begin{align}\nonumber
&\limsup_{\tau\to 0}\int_Q 
\bbE^\top:\big(\partial_\Ee^{}
\phimech(\barEe_\tau,\barchi_\tau,\underline\bbm_\tau)
+\bbD\DT\Ee_\tau\big)\DT\chi_\tau\dx\dt
\\\nonumber&\quad
=\limsup_{\tau\to 0}
\int_Q\big(\partial_\Ee^{}\phimech(\barEe_\tau,\barchi_\tau,\underline\bbm_\tau)
+\bbD\DT\Ee_\tau\big):\big(\bfeps(\DT{\bfu}_\tau)-\DT\Ee_\tau\big)
\dx\dt\\
\nonumber &\quad=-\liminf_{\tau\to 0}\int_Q 
\big(\varrho\DDT{\bfu}^{\rm i}_\tau-\barbff_\tau\big){\cdot}\DT{\bfu}_\tau
+(\partial_\Ee^{}\phimech(\barEe_\tau,\barchi_\tau,\underline\bbm_\tau)+\bbD\DT\Ee_\tau):\DT\Ee_\tau\dx\dt
+\lim_{\tau\to 0}\int_\Sigma\barbff_{\rm s,\tau}{\cdot}\DT{\bfu}_\tau\dS\dt
\\
\nonumber&\quad\le
-\liminf_{\tau\to 0}\Big(\frac12\int_\Omega\varrho|\DT{\bfu}_\tau(T)|^2\dx
+\int_Q\bbD\DT\Ee_\tau:\DT\Ee_\tau\dx\dt\Big) 
\\
\nonumber&\qquad+\frac 12 \int_\Omega \varrho|\DT{\bfu}(0)|^2\dx
+\lim_{\tau\to 0}\Big(\int_Q \barbff_\tau{\cdot}\DT{\bfu}_\tau
-\partial_\Ee^{}\phimech(\barEe_\tau,\barchi_\tau,\underline\bbm_\tau)
:\DT\Ee_\tau\dx\dt
+\int_\Sigma\barbff_{\rm s,\tau}{\cdot}\DT{\bfu}_\tau\dS\dt\Big)
\\
\nonumber&\quad\le
-\frac12\int_\Omega\varrho|\DT{\bfu}(T)|^2\dx-\int_Q\bbD\DT\Ee:\DT\Ee\dx\dt
\\
\nonumber&\qquad
+\frac 12 \int_\Omega \varrho|\DT{\bfu}(0)|^2\dx
+\int_Q \bff{\cdot}\DT\bfu
-\partial_\Ee^{}\phimech(\Ee,\chi,\bbm):\DT\Ee\dx\dt
+\int_\Sigma\bff_{\rm s}{\cdot}\DT\bfu\dS\dt
\\
\nonumber&\quad
=-\int_0^T  \big\langle\varrho\DDT{\bfu},\DT{\bfu}\big\rangle\dt
+\int_Q\bff{\cdot}\DT\bfu-\big(\partial_\Ee^{}\phimech(\Ee,\chi,\bbm)
+\bbD\DT\Ee\big):\DT\Ee\dx\dt
+\int_\Sigma\bff_{\rm s}{\cdot}\DT\bfu\dS\dt
\\
&\quad=\int_Q 
\bbE^\top:\big(\partial_\Ee^{}\phimech(\Ee,\chi,\bbm)+\bbD\DT\Ee\big)\DT\chi\dx\dt.
\end{align}
Here, the first equality used just $\DT\chi_\tau=\bfeps(\DT{\bfu}_\tau)
-\DT\Ee_\tau$, the second one used the force balance \eqref{mombal-disc} 
with the boundary conditions \eqref{BC-t-1} tested by 
$\DT u_\tau$, then we used the discrete by-part integration 
(or, in fact, summation) 
$\frac 12|\DT{\bfu}_\tau(T)|^2-\frac 12|\DT{\bfu}(0)|^2\le
\int_0^T\DDT{\bfu}^{\rm i}_\tau{\cdot}\DT{\bfu}_\tau\d t$
on $\Omega$, then semicontinuity arguments, then the by-part integration
formula
\begin{align}\label{by-part-for-u}
\int_0^T\!\big\langle\varrho\DDT\bfu,\DT\bfu\big\rangle\,\d t=
\int_\Omega \frac\varrho2\big|\DT\bfu(T)\big|^2
-\frac\varrho2\big|\DT\bfu(0)\big|^2\dx.
\end{align}
relying on the fact that, by \eqref{quality-of-u}, $\varrho\DDT\bfu$ and 
$\DT\bfu$ are in duality, then the limit force equilibrium we proved already in 
Step~3, and at the end still $\bbE\DT\chi=\bfeps(\DT{\bfu})-\DT\Ee$. The 
limit in $\int_Q\barsigma_{\rm r,\tau}{\cdot}\DT\chi_\tau\,\d x\d t$
is simple because, for any $\barsigma_{\rm r,\tau}\in N_{K}(\DT\chi_\tau)$,
this integral equals to 
$\int_\Omega\delta^{}_K(\chi_\tau(T))-\delta^{}_K(\chi_0)\,\d x=0
$. 
Eventually, the limit passage in the right-hand side of 
\eqref{mm-ineq-1} is by convexity of $\zeta(\Ee,\chi,d,c,\theta,\cdot)$ and the
weak lower semi-continuity 
\begin{align}
\liminf_{\tau\to0}\int_Q\!
\zeta(\underlineEe_\tau,\underline\chi_\tau,\underline d
_\tau,\underline c_\tau,\underline\theta_\tau,
{\DT\chi_\tau})\,\d x\d t\ge
\int_Q\!\zeta(\Ee,\chi,d,c,\theta,\DT\chi)\,\d x\d t.
\end{align}
Here we also used that $\underlineEe_\tau\to\Ee$, which follows from 
\eqref{eq:61} and from 
\begin{equation}\label{eq:9}
\|\barEe_\tau-\underlineEe_\tau\|_{L^2(Q;\R^{3\times3})}
\le\tau\|\DT\Ee_\tau\|_{L^2(Q;\R^{3\times3})}\to0.    
\end{equation}

\medskip

\noindent
\hypertarget{limit-passage-diffusion-equation}{\emph{Step 5: 
Limit passage in the diffusion equation.}}
By the strong convergence of $\barEe_\tau$, $\barchi_\tau$, 
$\underline \bbm_\tau$, and 
$\underline c_\tau$ established in \eqref{eq:61}, \eqref{eq:57}, 
\eqref{eq:84}, and \eqref{eq:29ff},
and by assumption 
of boundedness \eqref{eq:14} of $\mathbf M$  and 
by 
the Nemytskii-mapping continuity argument, we have that
\begin{align}\label{eq:91}
\mathbf M(\barEe_\tau,\barchi_\tau,\underline\bbm_\tau,\underline c_\tau)
\to\mathbf M(\Ee,\chi,c,\bbm)\ \ \ \textrm{ strongly in }\ L^p(Q)\quad\forall 1\le p<+\infty.
\end{align}
Now, owing to \eqref{eq:46e}, \eqref{eq:51}, and \eqref{eq:91}, we can pass to 
the limit in the first of \eqref{NS-disc}. 
In order to pass to the limit in the second of \eqref{NS-disc}, we observe 
that by the aforementioned a.e.\ convergence of $\barchi_\tau$, 
$\underline\bbm_\tau$, and $\underline c_\tau$ in $Q$ and by the continuity of 
$\partial_c^{}\phichem$, we have
\begin{align}\label{eq:90}
  \barmu_\tau=\partial_c^{}\phichem(\barchi_\tau,
\barc_\tau)\to\partial_c^{}\phichem(\chi,c)
\text{ a.e. in }Q,
\end{align}
for some subsequence. By comparing \eqref{eq:90} with \eqref{eq:51} we 
conclude that
\begin{align}
  \mu=\partial_c^{}\phichem(\chi,c).
\end{align}

\medskip

\noindent
\emph{Step 6: Limit passage in the semi-stability \eqref{damage-eq-disc} towards \eqref{damage-eq}.}
The mutual recovery sequence in the sense of \cite{MiRoSt08GLRR} for \eqref{damage-eq-disc}
uses the sophisticated construction of 
{M.\,Thomas} 
\cite{Thom10PhD,ThoMie10DNEM}. For all $t\in I$, we have 
\begin{equation}
  \label{eq:107}
  \bard_\tau(t)\to\bbm(t)\ \ \ \textrm{  weakly in }\ H^1(\Omega).
\end{equation}
 Consider a competitor $\widetilde\bbm$  of $\bbm(t)$ in \eqref{damage-eq}. It suffices to consider the case
\begin{equation}
  \label{eq:101}
  \widetilde\bbm(x)\le\bbm(x,t)\quad\textrm{ for a.e. } x\in\Omega,
\end{equation}
since otherwise the right-hand side the inequality \eqref{damage-eq} is $+\infty$.

We define the sequence
\begin{equation}
  \label{eq:100}
  \widetilde\bbm_\tau(x,t)=\min\big\{(\widetilde\bbm(x)-\varepsilon_\tau)^+,\bard_\tau(x,t)\big\}\qquad \textrm{with}\qquad\varepsilon_\tau=\big\|\bard_\tau(t)-\bbm(t)\big\|^{1/2}_{L^q(\Omega)}.
\end{equation}
with $1\le q<6$. It is immediate that 
$\widetilde d_\tau\in H^1(\Omega)$ and $\widetilde d_\tau(t)<\bard_\tau(t)$. Moreover, for $A_\tau(t)=\{x\in\Omega:\widetilde\bbm(x)-\varepsilon_\tau\le\bard_\tau(x,t)\}$, we have:
\begin{equation}
  \label{eq:103}
  \nabla \widetilde d_\tau(x,t)=
\begin{cases}
\nabla\widetilde d(x)\,\qquad\forall x\in A_\tau(t),\\[0.5em]
\nabla\bard_\tau(x,t)\quad\forall x\in \Omega\setminus A_\tau(t),
\end{cases}
\end{equation}
Consequently
\begin{equation}\label{eq:106}
\begin{aligned}
  \liminf_{\tau\to 0}\int_\Omega \big|\nabla\bard_\tau(x,t)\big|^2-\big|\nabla\widetilde\bbm_\tau(x,t)\big|^2\dx=\liminf_{\tau\to 0}\int_{A_\tau} \big|\nabla\bard_\tau(x,t)\big|^2-\big|\nabla\widetilde\bbm(x)\big|^2\dx.
\end{aligned}
\end{equation}
Because of \eqref{eq:101}, we have $|\bard_\tau(x,t)-\bbm(x)|\le\varepsilon_\tau\Rightarrow x\in A_\tau(t)$. Thus, $\Omega\setminus A_\tau(t)\subset\{x\in\Omega:|\bard_\tau(x,t)-\widetilde\bbm(x)|\ge \varepsilon_\tau\}$. Using Markov's inequality and the weak convergence of $\bard_\tau(t)$ to $\bbm(t)$ in $H^1(\Omega)$, we obtain
\begin{equation}
\big|\Omega{\setminus}A_\tau(t)\big|\le
\frac1{\varepsilon_\tau^q}\int_\Omega \big|\bard_\tau(x,t)-\bbm(x,t)\big|^q\d x
=\big\|\bard_\tau(t)-\bbm(t)\big\|^{q/2}_{L^q(\Omega)}\to 0
\end{equation}
where $|\Omega{\setminus}A_\tau(t)|$ denoted the Lebesgue measure of
the set $\Omega{\setminus}A_\tau(t)$.
Given a set $C$, let  {$\delta_{C}$}  be its indicator  function. It is shown in \cite{ThoMie10DNEM} that
\begin{equation}
  \label{eq:105}
  \delta_{A_\tau(t)}\nabla\bard_\tau(t)\to\nabla
%\bbm_\tau(t)
\bbm(t)\qquad\textrm{ weakly in }\ L^{2}(\Omega;\mathbb R^3).
\end{equation}
On the other hand \COL{we have, trivially, that} $\delta_{A_\tau(t)}\nabla\widetilde\bbm$ converges strongly 
to $\nabla\widetilde\bbm$.
%\COMMENT{CHECK WHAT IS IT? MAYBE RATHER $\nabla\bbm$???}
Thus, from \eqref{eq:106} we find, by lower semicontinuity,
\begin{align}
  \liminf_{\tau\to 0}\int_\Omega |\nabla\bard_\tau(x,t)|^2-|\nabla\widetilde\bbm_\tau(x,t)|^2\dx&=\liminf_{\tau\to 0}\int_{\Omega} \delta_{A_\tau(t)}(x)|\nabla\bard_\tau(x,t)|^2\dx-\mathop{\textrm{lim}}_{\tau\to 0}\int_{A_\tau(t)}|\nabla\widetilde\bbm(x)|^2\dx
\nonumber\\
&\ge \int_{\Omega} |\nabla{\bbm}_\tau(x,t)|^2\dx-\int_{A}|\nabla\widetilde\bbm(x)|^2\dx.
\end{align}
With this result, the limit passage in the semi-stability condition is easily achieved for a.e. $t\in I$, on taking into account that  $\barEe(t)\to\Ee(t)$ strongly in $L^2(\Omega;\mathbb R^{3\times 3})$; furthermore, $\barchi_\tau(t)\to\chi(t)$ strongly in $L^q(\Omega;\mathbb R^N)$, and  $\barc_\tau(t)\to c(t)$ strongly in $L^q(\Omega;\mathbb R)$.
\EEE

\medskip

\noindent
\emph{Step 7: Mechanical/chemical energy conservation \eqref{engr-equality}.} 
This is standardly achieved by testing the mechano-chemical equations 
(\ref{systemb}a--d)
respectively by $\DT\bfu$, $\DT\chi$, $\mu$, and $\DT\bbm$, and by using the 
chain rule to integrate with respect to $t$. Here, however, \eqref{damage-eqb} 
has to be specially treated because  $\DT\bbm$ can be a measure.

Making the first test 
legal, we again need $\varrho\DDT\bfu$ to be in duality with $\DT\bfu$, 
cf.\ \eqref{quality-of-u}, and 
make use of \eqref{by-part-for-u}.
For the second mentioned test, we need $\Delta\chi\in L^2(Q)$ 
to have the integration-by-part formula at our disposal. The regularity of 
$\Delta\chi$ follows from the estimate \eqref{est-of-Delta-m} and weak 
convergence. The proof of the by-part integration formula is rather technical 
because $\chi:I\to H^1(\O;\R^N)$ is actually only a weakly continuous 
function but not necessarily strongly continuous. The desired formula is
\begin{align}\label{by-part-for-chi}
\int_Q\Delta\chi\cdot\dott\chi\,\d x\d t=
\frac12\int_\O|\nabla \chi_0|^2-|\nabla \chi(T)|^2\,\d x.
\end{align}
Its proof is a bit tricky and can be done
either by a mollification in space \cite[Formula (3.69)]{PoRoTo10TCTF}
and or in time by a time-difference technique \cite[Formula (2.15)]{Gru95DPEF}.
Also we use that $\sigma_{\rm r}\in L^2(Q;\R^N)$ 
is in duality with $\DT\chi\in L^2(Q;\R^N)$, 
so that the integral 
$\int_Q \sigma_{\rm r}\cdot\DT\chi
\,\d x\d t$ has a sense and simply equals to 0 because 
$\sigma_{\rm r}\in\partial \delta^{}_K(\chi)$ 
has been proved in Step~4 and because 
$\delta^{}_K(\chi_0)=0$
has been assumed, cf.\ \eqref{IC-ass-1}.

Also, $\DT\CHI\in L^2(I;H^1(\Omega)^*)$ is in duality with 
$\mu\in L^2(I;H^1(\Omega))$ as well as $\DT\chi\in L^2(Q;\R^N)$ is in duality 
with $\partial_\chi^{}\phichem(\chi,\CHI)
\in L^2(Q;\R^N)$,
cf.\ \eqref{a-priori-IIe} with \eqref{15.-mu-strong}
and \eqref{growth-of-phi-c-d}
so that we obtain
\begin{align}\label{chain-rule-for-c}
\int_0^T\!\!\Big(\langle\DT\CHI,\mu\rangle+
\int_\Omega\!\partial_\chi^{}\phichem(\chi,c){\cdot}\DT\chi
\,\d x\Big)\d t
=\int_\Omega\phichem(
\chi(T),\CHI(T))-\phichem(\chi_0,\CHI_0)\,\d x.
\end{align}

Eventually, we use 
%\TTT 
the \EEE Riemann-sum approximation of Lebesgue integrals 
and semi-stability as devised in 
\cite{DaFrTo05QCGN,MaiMie05EREM}, cf.\ also 
\cite[Formulas (4.68)--(4.74)]{Roub10TRIP}
for a combination with rate-dependent mechanical part. In fact, 
the Riemann sums can range only a.a.\ points from $I$ and thus 
the semi-stability need not hold \COL{at} every time but only 
at almost each times. By this way we obtain
%\COMMENT{SIGNS CHANGED CONSISTENTLY TO E.G. \eqref{lim-inf-sup} -- BETTER TO CHECK!}
\begin{align}\nonumber
&\int_\Omega \phimech(\Ee(T),\chi(T),\bbm(T))+
\frac{\kappa_2}2|\nabla d(T)|^2
%\TTT 
+ \EEE\alpha(\chi(T))d(T)
\,\d x+\int_{{\,{\!Q\!}\,}}\alpha'(\chi)\DT\chi d \,\d x\d t
\\&\label{balance-of-d}\ \ \ 
=\int_\Omega\!\phimech(\Ee_0,\chi_0,\bbm_0)
+\frac{\kappa_2}2|\nabla d_0|^2
%\TTT 
+ \EEE\alpha(\chi_0)d_0
\,\d x
-\int_Q\!\!\partial_\Ee^{}\phimech(\Ee,\chi,\bbm)\DT\Ee
+\partial_\chi^{}\phimech(\Ee,\chi,\bbm)\DT\chi\,\d x\d t.
\end{align}
Again we used that $\DT\Ee\in L^2(Q;\R^{3\times 3})$ 
and $\DT\chi\in L^2(Q;\R^N)$ are in duality with 
$\partial_\Ee^{}\phimech(\Ee,\chi,\bbm)\in L^2(Q;\R^{3\times 3})$
and $\partial_\chi^{}\phimech(\Ee,\chi,\bbm)\DT\chi$, respectively.
Here we also used the semi-stability of the initial condition $d_0$ assumed
in \eqref{damage-IC}.

Eventually, by summing up all four obtained partial balances, we obtain 
\eqref{engr-equality}.

\medskip

\addcontentsline{toc}{subsection}{Convergence in the mechanical equations}
\phantomsection
\noindent
\emph{Step 8: Strong convergence of ${\DT\Ee_\tau}$, $\DT\chi_\tau$,
and $\nabla\barmu_\tau$.}
Using the discrete mechano-chemical energy imbalance 
\eqref{engr-equality-disc} and 
eventually the energy equality 
\eqref{engr-equality}, we can write
\begin{align}\nonumber
  &\!\!\int_Q
\bbD\DT\Ee{:}\DT\Ee
+\partial_{\DT\chi}\zeta(\Ee,\chi,c,\bbm,\theta,\DT\chi)\cdot\DT\chi
+\mathbf M(\Ee,\bbm,\CHI,\theta)\nabla\mu{\cdot}\nabla\mu\,\d x\d t
\\[-.4em]&\ \ \ \nonumber\le
\liminf_{\tau\to0}
\int_Q\!\bbD{\DT\Ee}_\tau{:}{\DT\Ee}_\tau
+\partial_{\DT\chi}\zeta(\underlineEe_\tau,\underline\chi_\tau,\underline\bbm_\tau,\underline c_\tau,\underline\theta_\tau,\DT\chi_\tau)\cdot\DT\chi_\tau
+
\mathbf M(\barEe_\tau,\barchi_\tau,\underline\bbm_\tau,\underline\CHI_\tau,\underline\theta_\tau)
\nabla\barmu_\tau
{\cdot}\nabla\barmu_\tau
\\&\ \ \  \nonumber\le\limsup_{\tau\to0}
\int_Q\!
\bbD{\DT\Ee}_\tau{:}{\DT\Ee}_\tau
+\partial_{\DT\chi}\zeta(\Ee_\tau,\underline\chi_\tau,\underline\bbm_\tau,\underline\CHI_\tau,\underline\theta_\tau,\DT\chi_\tau)\cdot\DT\chi
+\mathbf M(\barEe_\tau,\barchi_\tau,\underline\bbm_\tau,\underline\CHI_\tau,\underline\theta_\tau)
\nabla\barmu_\tau{\cdot}\nabla\barmu_\tau\,\d x\d t
\\\nonumber
&\ \ \ \le\limsup_{\tau\to0}\bigg(
\mathcal E_{_{\rm MC}}(0)
-\!\int_\O\!\frac\varrho2\big|\DT\bfu_\tau(T)\big|^2\!
+{\phimechchem(\Ee_\tau(T),\chi_\tau(T),\bbm_\tau(T),c_{\tau}(T))}
\\[-.4em]\nonumber
&\hspace{5em}
+\frac{\kappa_1}2\big|\nabla\chi_\tau(T)\big|^2+\frac{\kappa_2}2\big|\nabla\bbm_\tau(T)\big|^2\d x
-\int_\Sigma\!\barbff_{{\rm s},\tau}{\cdot}\DT\bfu_\tau\dS\d t
-\int_Q\barbff_{\tau}{\cdot}\DT\bfu_\tau
+\alpha(\barchi_\tau)\DT d_\tau
\dx\d t\bigg)
\\[-.3em]\nonumber
&\ \ \ \le\mathcal E_{_{\rm MC}}(0)-\mathcal E_{_{\rm MC}}(T)
-\int_\Sigma \bff_{\rm s}{\cdot}\DT\bfu\dS\d t
-\int_Q\bff{\cdot}\DT\bfu\dx\d t
-\int_{Q}\alpha'(\chi)\DT \chi d\dx\dt-\int_\Omega
\alpha(\chi(T))d(T)-\alpha(\chi_0)d_0\dx
\\[-.5em]&\ \ \ =\int_Q\bbD\DT\Ee{:}\DT\Ee
+\partial_{\DT\chi}\zeta(\Ee,\chi,c,\bbm,\theta,\DT\chi)\cdot\DT\chi
+\mathbf M(\Ee,\bbm,\CHI,\theta)\nabla\mu{\cdot}\nabla\mu\,\d x\d t.
\label{lim-inf-sup}
\end{align}
Note that in the last inequality we have made use of the discrete by-part
integration (=\,summation) formula for the following calculations, being
a discrete analogue of \eqref{by-part-d} with $z=1$\EEE: 
\begin{align}\nonumber
\lim_{\tau\to0}\int_Q\alpha(\barchi_\tau)\DT d_\tau\dx\dt
&=
\lim_{\tau\to0}\sum_{k=1}^{T/\tau}
\int_\Omega\alpha(\chi_\tau^k)(d_\tau^k{-}d_\tau^{k-1})\dx
\\[-.6em]\nonumber&=\lim_{\tau\to 0}\bigg(
\int_\Omega\alpha(\chi_\tau^{T/\tau})d_\tau^{T/\tau}\dx
-\sum_{k=1}^{T/\tau}\int_\Omega
\big(\alpha(\chi_\tau^k)-\alpha(\chi_\tau^{k-1})\big)d_\tau^{k-1}\dx\bigg)
-\int_\Omega\alpha(\chi^0_\tau)d^0_\tau\dx
\\[-.3em]\nonumber&=
\lim_{\tau\to0}\bigg(\int_\Omega\alpha(\chi_\tau(T))d_\tau(T)\dx
-\int_Q\!\!\big(\alpha'(\chi_\tau)\DT\chi_\tau\underline d_\tau
+r_\tau\underline d_\tau\big)\dx\dt\bigg)-\int_\Omega\alpha(\chi_0)d_0\dx
\\[-.3em]\nonumber&=
\int_\Omega\alpha(\chi(T))d(T)\dx
\TTT-\EEE\int_Q\!\!\big(\alpha'(\chi)\DT\chi d+0\big)\dx\dt
\TTT-\EEE\int_\Omega\alpha(\chi_0)d_0\dx,
%\\\nonumber&
%\int_0^{T-\tau}\!\!\int_\Omega\DT{\overline{\alpha(\chi_\tau)}}\bard_\tau\dx\dt-\int_\Omega\alpha(\barchi_\tau(T))\bard_\tau(T)-\alpha(\chi_0)d_0\dx
%\\&\to \int_Q\alpha'(\chi_\tau)\DT\chi d\dx\dt-\int_\Omega\alpha(\chi(T))d(T)-\alpha(\chi_0)d_0\dx.
%\label{eq:109}
\end{align}
where we have set $d_\tau^{-1}=d_\tau^0=d_0$ and where $r_\tau$ denotes the 
difference between $\displaystyle{[\alpha(\chi_\tau)]\!\DT{\!^{}}}$ and the 
piece-wise constant-in-time function with values 
$(\alpha(\chi_\tau^k){-}\alpha(\chi_\tau^k))/\tau$ on the interval 
$((k{-}1)\tau,k\tau)$; here we used differentiability assumption on $\alpha$ 
stated in \eqref{assum-dissi-poten-b} and the estimate 
$|r_\tau|\le\tau^2(\sup_{\R^N}\alpha'')|\DT\chi_\tau|^2$
so that
\begin{align*}
\Big|\int_Qr_\tau\underline d_\tau\dx\dt\Big|
\le\tau^2\big(\sup_{\ \R^N}\alpha''\big)\int_Q\big|\DT\chi_\tau\big|^2d_\tau\dx\dt
\le\tau^2\big(\sup_{\ \R^N}\alpha''\big)\big\|\DT\chi_\tau\big\|_{L^2(Q;\R^N)}^2
\big\|d_\tau\big\|_{L^\infty(Q)}=\mathscr{O}(\tau^2)\to0
\end{align*}
for $\tau\to0$. To converge $\alpha'(\chi_\tau)\DT\chi_\tau\underline d_\tau$ 
to $\alpha'(\chi)\DT\chi d$ weakly in $L^1(Q)$, we used 
\eqref{eq:57} together with \eqref{eq:49}, and \eqref{eq:84-for-d}.
The last equality in \eqref{lim-inf-sup} has been proved
in Step 7. Altogether, we can write ``lim'' and ``='' everywhere in 
\eqref{lim-inf-sup} and, together with the already proved weak convergence, 
we obtain the desired strong convergence of ${\DT\Ee}_\tau$
and $\DT\chi_\tau$ and $\nabla\barmu_\tau$ in $L^2(Q)$-spaces. Here we rely
%\TTT
on \EEE the well-known concept of compactness via convexity 
(\cite{visintin1984strong,visintin1996models}) with some modifications. 
In particular, for technical details about the term 
$\mathbf M\nabla\mu{\cdot}\nabla\mu$ 
with the nonconstant coefficient 
$\mathbf M=\mathbf M(\Ee,\bbm,\CHI,\theta)$, we refer to 
\cite[Formula (4.25)]{Roubivcek2010}. 

\smallskip
\addcontentsline{toc}{subsection}{Limit passage in the heat equation}
\phantomsection
\noindent
\emph{Step 9: Limit passage in the heat equation \eqref{heat-eq}.}
Having proved the strong convergence in Steps 3 and 9, the right-hand 
side of \eqref{heat-eq} converges strongly in $L^1(Q)$ 
except the term $\alpha(\chi)\DT d$ but even this term converges 
weakly$*$ in $\mathrm{Meas}({\,\overline{\!Q\!}\,})$, 
cf.\ Step~7, which is sufficient to  
the limit passage towards \eqref{heat-eq-cont}, which is then easy.

\smallskip

\noindent
\emph{Step 10: Total-energy conservation \eqref{eq:99} at almost each time.} 
Considering $t_*\in[0,T)$ fixed, we use the test function 
\begin{align}
z_\epsilon(t)=
\begin{cases}1&\text{if }t\le t_*,\\[-.2em]
1
+(t_*{-}t)/\epsilon
\!\!&
\text{if }t_*\le t\le t_*+\epsilon,\\[-.2em]
0&\text{if }t\ge t_*+\epsilon\end{cases}
\end{align}
for \eqref{heat-eq-weak}. Using that $\DT z=0$ for 
$t\in[0,t_*]\cup[t_*+\epsilon,T]$
and that $\nabla z=0$, this test leads to 
\begin{align}
&%-\int_Q w \DT z\dx\dt+
\frac1\epsilon\int_{t_*}^{t_*+\epsilon}\!\!\!\int_\Omega w\, \d x\dt
=
\int_0^{t_*}\!\int_\Omega
\Big(\big(\partial_{\DT\chi}\zeta(\Ee,\chi,c,\bbm,\theta,\DT\chi)
{+}\partial_\chi^{}\phiterm(\chi,\theta)
{+}\alpha'(\chi)d\big)
{\cdot}\DT\chi
+\mathbf M(\Ee,\chi,c,\bbm,\theta)\nabla\mu{\cdot}\nabla\mu
\nonumber\\[-.3em]\label{heat-eq-weak+}
&\qquad\quad
+\bbD\DTEE\!:\!\DTEE\Big)\,\d x\dt
+\frac1\epsilon\int_{t*}^{t_*+\epsilon}\!\!\!\int_\Omega\alpha(\chi)d\,\d x\dt
+\int_\Omega\big(w_0{+}\alpha(\chi_0)d_0\big)\dx
+\int_\Sigma q_{\rm s}\, \d S\dt
+o_t(\epsilon)
\end{align}
with $o_{t_*}(\epsilon)$ abbreviating the corresponding integrals 
over $[t_*,t_*+\epsilon]$. We have $\lim_{\epsilon\to0}o_{t_*}(\epsilon)=0$ for
any $t_*$ due to absolute continuity of the Lebesgue integral. Considering 
however $t_*$ as a right Lebesgue point of the functions
$t\mapsto\int_\Omega w(x,t)\,\d x$ and  
$t\mapsto\int_\Omega\alpha(\chi(x,t))d(x,t)\,\d x$, 
in the limit for $\epsilon\to0+$ we obtain 
\begin{align}&%-\int_Q w \DT z\dx\dt+
\int_\Omega w(t_*)\, \d x\dt
=
\int_0^{t_*}\!\!\!\int_\Omega
\Big(\big(\partial_{\DT\chi}\zeta(\Ee,\chi,c,\bbm,\theta,\DT\chi)
{+}\partial_\chi^{}\phiterm(\chi,\theta)
{+}\alpha'(\chi)d\big)
{\cdot}\DT\chi
+\mathbf M(\Ee,\chi,c,\bbm,\theta)\nabla\mu{\cdot}\nabla\mu
\nonumber\\[-.3em]
&\qquad\qquad\qquad\qquad
+\bbD\DTEE\!:\!\DTEE\Big)\,\d x\dt
+\!\int_\Omega\!\alpha(\chi(t_*))d(t_*)\,\d x
+\int_\Omega\big(w_0{+}\alpha(\chi_0)d_0\big)\dx
+\int_0^{t_*}\!\!\!\int_\Gamma q_{\rm s}\, \d S\dt.
\label{heat-balance}
\end{align}
Such points $t_*$'s has a full Lebesgue measure on $I$. 
Eventually, we get \eqref{eq:99} by summing 
\eqref{heat-balance} with mechanical/chemical energy balance 
\eqref{engr-equality} 
written for $t_*$ instead of $T$ obtained already in 
Step 8 by obvious modification of the arguments there. 
\QED

\begin{remark}[
%\TTT
{\sl Damageable \EEE
%Variable 
viscosity}]
\upshape
One may want to include damageable viscosity in the model through a constitutive 
equation of the form $\bfS_{\rm d}=\bbD(d)\DT\Ee$ for the viscous part of the 
stress. With this modification, when performing integration by parts in 
\eqref{eq:81}, the additional term $-\int_Q\frac {\partial}{\partial t}\bbD(
d_\tau)\Ee_\tau:\Ee_\tau\dx\d t$ would appear. Passage to the limit through lower semicontinuity would still be possible 
if the tensor $\frac {\partial}{\partial t}\bbD(
d_\tau)$ is non-positive. This could rely on the unidirectional evolution of 
$d$ adopted in this paper and monotone dependence of $\bbD$ on damage, in the sense of 
%\TTT 
the so-called \EEE L\"owner ordering, namely 
$(\mathbb D(d_1){-}\mathbb D(d_2))\Ee:\Ee\le 0$ if $d_1<d_2$.
\end{remark}\color{black}

\begin{remark}[{\sl Decoupling between concentration and strain}]\label{rem-dec1}
\color{gt} \upshape  
Our assumption \eqref{def-of-psi} rules out any direct coupling between strain and concentration. We need a decoupling between concentration and strain to avoid an explicit dependence of chemical potential on strain, which would lead to the appearance of the gradient of strain in the formulation. Indeed, in order to pass to the limit in the nonlinear equation: 
\begin{align}\label{eq:75}
\mu=\partial_c^{}\phichem(\chi,c)
\end{align}
which defines chemical potential, we exploit the strong convergence of $c$ in a suitable $L^p$ space, cf.\ \eqref{15.-chi} above. In order to obtain such convergence, we rely on the standard Aubin-Lions compactness theorem, whose application requires estimates on $\DT c$ and $\nabla c$. 
The natural energetic estimate provides us with a control on $\nabla\mu$ (\emph{cf.} \eqref{apriori-Ia+++}), but not on $\nabla c$. 
{To} control of
the latter (cf.\ the estimate \eqref{apriori-Ia++} above) 
we take the gradient of \eqref{eq:75} at the approximation level, see \eqref{est-of-nabla-concentration} in the proof of Lemma \ref{lem-2}. If we were allowed
$\partial_{c\Ee}^2\varphi\ne0$, then $\nabla\Ee$  would have appeared in that estimate.
To give a concrete example, let us  consider a toy model with no phase field $\chi$, no damage variable $\bbm$, and no temperature $\theta$. Let us assume that the following strain decomposition:
\begin{equation}
  \label{eq:6}
  \bfeps(\bfu)=\Ee+c\bbE
\end{equation}
holds, where $\bbE$ is a constant second-order symmetric tensor. In this case, the dissipation inequality takes the form
\begin{equation}
  \label{eq:7}
  \DT\psi-\mu\DT c\le \bfS:\DT\Ee+\bfS:
\bbE\DT c-\bfh\cdot\nabla\mu.
\end{equation}
Then, the application of the Coleman-Noll argument yields, for chemical potential, a constitutive equation of the form
\begin{equation}
  \label{eq:8}
  \mu=\partial_c^{}\psi(\Ee,c)+\bfS:\bbE.
\end{equation}
Suppose that we take $\psi=\frac 1 2 \mathbb C\Ee:\Ee$. Then, if we rule out viscoelastic dissipation the constitutive equation for the stress is
  $\bfS=\mathbb C\Ee$.
From \eqref{eq:8} we have 
\begin{equation}
  \label{eq:10}
  \partial^2_{cc}\psi(\Ee,c)\nabla c=\nabla\mu-\bbE^\top:\mathbb C\nabla\Ee
\end{equation}
Clearly, there is no hope to control the second term on the right-hand side of \eqref{eq:10}, unless we have some control on $\nabla\Ee$, which however cannot be expected. 
This issue may be circumvented by allowing a dependence of free energy on $\nabla c$, for instance by adding a term proportional to $|\nabla c|^2$. As discussed, for instance, in \cite{FriedG1999JoSP}, this term describes capillarity effects, and would lead to a system of Cahn--Hilliard type \cite{CahnH1958JCP}, similar to those studied in \cite{Garck2005AIPC}. Alternatively, one might think of reformulating the model in the framework of non-simple materials, by allowing for instance the free energy to depend on $\nabla\Ee$ and by introducing hyperstress. In this case, however, issues would arise from the choice of traction boundary conditions in non-smooth domain \cite{PodioV2010CMaT}.
\color{black}
\end{remark}

\begin{remark}[{\sl Quasistatic inviscid model}]\label{D=0}
\upshape
Some models in literature neglect inertia and viscosity by putting 
$\varrho=0$ and $\bbD=0$, see e.g.\ \cite{BHKS??MAPF,HaiKra14DCHS}.
%\COMMENT{TO COMPLETE...}
The semi-implicit scheme can then be spit naturally to five fractional step, 
each of them for each variable 
%\COL{$\Ee$},
%\TTT 
$\mathbf u$\EEE, $\chi$, $\bbm$, $c$, and $\theta$ 
separately by replacing $\chi\kt$ in \eqref{TGMd-1} by $\chi\kkt$. Instead of 
\eqref{assum-conve-e-chi}, it would suffice to require 
$\phimech(\cdot,\chi,\bbm)$ and $\phimech(\Ee,\cdot,d)+\frac12M|\cdot|^2$ convex. %\COL{Notice however that 
In this case, however, we 
%should 
%\TTT 
would have to 
%\EEE 
remove the dependence of $\zeta$ on $\Ee$ because we would lose the estimate \eqref{eq:9}, which relies on viscosity. 
%\TTT 
Also e.g.\ thermal expansion leading usually to adiabatic terms containing 
$\DT\Ee$ would have to be excluded, although some particular studies 
resulting to spatially constant temperature does exist, cf.\ \cite{Marita-new-paper}. \EEE
%\COMMENT{HERE A NEW PAPER OF MARITA \& CO.}
\end{remark}

\begin{remark}[{\sl Viscous damage and higher regularity of displacements}]
We already pointed out that in the case of 
%viscous
%\TTT 
rate-dependent \EEE evolution of damage (i.e.\ if the dissipation 
pseudopotential $\xi$ would be quadratic), the variable $d$ may be incorporated into the vectorial variable $\chi$, which would still obey an evolution equation having the structure of \eqref{chi-eqb}. This variant has recently been investigated in \cite{rocca2014degenerating} in the case of thermal coupling. It is worth pointing out that its mathematical analysis would however require $L^2(Q)$ regularity of the term $\partial^{}_d\phimech(\Ee,\chi,d)$. Such regularity would in general not be compatible with the quadratic growth 
%\TTT 
of $\phimech$ \EEE with respect to $\Ee$. 
In order to better elucidate this point, let us take $\phimech=\frac12\mathbb C(d)\Ee{:}\Ee$. In order to guarantee the above-mentioned regularity, we would need $L^4(Q)$ regularity of $\Ee$, which in general cannot be expected without additional assumptions. Yet, higher regularity of displacements may be proved through a sophisticated technique based on testing the standard-force balance by $-{\rm div}(\bfeps(\DT{\mathbf u}))$, as in \cite{BoScSe05DNME}, and also in \cite{rocca2014degenerating}. This estimate would however require the assumption that acceleration vanishes on the whole boundary, which in turn would restrict the applicability of the model to affine-in-time Dirichlet boundary conditions for $\mathbf u$.
\end{remark}

% \COMMENT{TO CITE AND COMPARE!!! - PERHAPS: a very special regularity technique relying
% on zero Dirichlet condition for $\bfu$ on the whole boundary $\Gamma$ -- in fact, only an
% affine-in-time Dirichlet condition would suffice so that $\DDT\bfu=0$ on $\Gamma$
% -- ALSO \cite{BoScSe05DNME} TO CITE!}

% \begin{remark}[Healing: reversible damage]\label{rem-healing}
% \upshape
% \COMMENT{TO BE MODIFIED -- PROBLEM WITH $\alpha(\chi)\DT d$???:}
% {\tiny For analytical reasons, in particular to have the regularity 
% \eqref{est-of-Delta-m2} at disposal, we allowed (and due to 
% \eqref{assum-dissi-poten-b} even required) 
% healing. However, in some applications, damage is rather irreversible, which 
% is reflected by that the driving force in \eqref{damage-eqb} in nonpositive, 
% i.e.\ $\partial_{\bbm}\phichem(\chi,c)
% {+}\partial_{\bbm}\phimech(\Ee,\chi,\bbm)\ge0$.
% Then the only source of a possible healing is the term 
% $\kappa_2\Delta \bbm$ in \eqref{damage-eqb} in the spots which are 
% surrounded by less damaged material. Yet, $\kappa_2>0$ is often 
% small and the slope of $\xi(\chi,\bbm,\cdot)$ can be large at $0$ for 
% positive arguments so that such healing effects can be negligible in 
% concrete simulation.}
% \end{remark}

% \COMMENT{MAYBE TO REFER:
% large strains:\\
% F.P. Duda, J.M. Barbosa, L.J. Guimar\~aes and A.C. Souza:
% Modeling of Coupled
% Deformation-Diffusion-Damage in Elastic Solids
% INTERNATIONAL JOURNAL OF MODELING AND SIMULATION FOR THE PETROLEUM INDUSTRY, 
% VOL. 1, NO. 1, AUGUST 2007
% }

\end{sloppypar}

\begin{thebibliography}{10}

\bibitem{AuReSt09MMSM}
F.~Auricchio, A.~Reali, and U.~Stefanelli.
\newblock A macroscopic 1{D} model for shape memory alloys including asymmetric
  behaviors and transformation-dependent elastic properties.
\newblock {\em Comput. Methods Appl. Mech. Engrg.}, 198:1631--1637, 2009.

\bibitem{bazant2004temperature}
Z.~P. Ba\v{z}ant, G.~Cusatis, and L.~Cedolin.
\newblock Temperature effect on concrete creep modeled by
  microprestress-solidification theory.
\newblock {\em J. Engrg. Mech.}, 130:691--699, 2004.

\bibitem{Bonetti2012}
E.~Bonetti, P.~Colli, and P.~Lauren\c{c}ot.
\newblock Global existence for a hydrogen storage model with full energy
  balance.
\newblock {\em Nonlinear Analysis: Theory, Meth. \& Appl.}, 75:3558--3573,
  2012.

\bibitem{Bonetti2007}
E.~Bonetti, M.~Fremond, and C.~Lexcellent.
\newblock Hydrogen storage: Modeling and analytical results.
\newblock {\em Appl. Math. Optim.}, 55:31--59, 2007.

\bibitem{BHKS??MAPF}
E.~Bonetti, C.~Heinemann, C.~Kraus, and A.~Segatti.
\newblock Modeling and analysis of a phase field system for damage and phase
  separation processes in solids.
\newblock {\em WIAS Preprint No. 1841, Berlin, 2013}, to appear.

\bibitem{BoScSe05DNME}
E.~Bonetti, G.~Schimperna, and A.~Segatti.
\newblock On a doubly nonlinear model for the evolution of damaging in
  viscoelastic materials.
\newblock {\em J. Diff. Eqns.}, 218:91--116, 2005.

\bibitem{CahnH1958JCP}
J.~W. Cahn and J.~E. Hilliard.
\newblock Free energy of a nonuniform system. {I}. {I}nterfacial free energy.
\newblock {\em J. Chem. Phys.}, 28:258--267, 1958.

\bibitem{Chiod2011MMAS}
E.~Chiodaroli.
\newblock A dissipative model for hydrogen storage: existence and regularity
  results.
\newblock {\em Math. Meth. Appl. Sci.}, 34:642--669, 2011.

\bibitem{DaFrTo05QCGN}
G.~{Dal Maso}, G.~Francfort, and R.~Toader.
\newblock Quasistatic crack growth in nonlinear elasticity.
\newblock {\em Arch. Rat. Mech. Anal.}, 176:165--225, 2005.

\bibitem{Anand2014cahn}
C.~V. Di~Leo, E.~Rejovitzky, and L.~Anand.
\newblock A {C}ahn-{H}illiard-type phase-field theory for species diffusion
  coupled with large elastic deformations: application to phase-separating
  {L}i-ion electrode materials.
\newblock {\em J. Mech. Phys. Solids}, 70:1--29, 2014.

\bibitem{duda2007modeling}
F.~P. Duda, J.~M. Barbosa, L.~J. Guimar{\~a}es, and A.~C. Souza.
\newblock Modeling of coupled deformation-diffusion-damage in elastic solids.
\newblock {\em Int. J. Model. Simul. Pet. Ind.}, 1, 2007.

\bibitem{FraMar98RBFE}
G.~Francfort and J.-J. Marigo.
\newblock Revisiting brittle fracture as an energy minimization problem.
\newblock {\em J. Mech. Phys. Solids}, 46:1319--1342, 1998.

\bibitem{Fr2002a}
M.~Fr{\'e}mond.
\newblock {\em Non-smooth Thermomechanics}.
\newblock Springer, 2002.

\bibitem{FriedG1999JoSP}
E.~Fried and M.~Gurtin.
\newblock Coherent solid-state phase transitions with atomic diffusion: A
  thermomechanical treatment.
\newblock {\em J. Stat. Phys.}, 95:1361--1427, 1999.

\bibitem{fujita2003itinerant}
A.~Fujita, S.~Fujieda, Y.~Hasegawa, and K.~Fukamichi.
\newblock Itinerant-electron metamagnetic transition and large magnetocaloric
  effects in la ({F}e${}_x$ {S}i${}_{1-x}$)${}_{13}$ compounds and their
  hydrides.
\newblock {\em Phys. Rev. B}, 67:104416, 2003.

\bibitem{Garck2005AIPC}
H.~Garcke.
\newblock On a {C}ahn-{H}illiard model for phase separation with elastic
  misfit.
\newblock {\em Ann. Inst. H. Poincar\'{e}}, 22:165--185, 2005.

\bibitem{gawin2007modelling}
D.~Gawin, F.~Pesavento, and B.~Schrefler.
\newblock Modelling creep and shrinkage of concrete by means of effective
  stresses.
\newblock {\em Materials and Structures}, 40:579--591, 2007.

\bibitem{Gru95DPEF}
G.~Gr\"{u}n.
\newblock Degenerate parabolic equations of fourth order and a plasticity model
  with nonlocal hardening.
\newblock {\em Zeits. Anal. u. Ihre Anwend.}, 14:541--573, 1995.

\bibitem{Gurtin1996a}
M.~E. Gurtin.
\newblock Generalized {G}inzburg-{L}andau and {C}ahn-{H}illiard equations based
  on a microforce balance.
\newblock {\em Physica D}, 92:178--192, 1996.

\bibitem{HaLyAg04EDPP}
Y.~Hamiel, V.~Lyakhovsky, and A.~Agnon.
\newblock Coupled evolution of damage and porosity in poroelastic media: Theory
  and applications to deformation of porous rocks.
\newblock {\em Geophys. J. Int.}, 156:701--713, 2004.

\bibitem{HaLyAg05DRDC}
Y.~Hamiel, V.~Lyakhovsky, and A.~Agnon.
\newblock Poroelastic damage rheology: Dilation, compaction, and failure of
  rocks.
\newblock {\em Geochem. Geophys. Geosyst.}, 6:Q01008, 2005.

\bibitem{havela2009f}
L.~Havela, K.~Miliyanchuk, and A.~Kolomiets.
\newblock f-{E}lement hydrides: structure and magnetism.
\newblock {\em Int. J. Mat. Res.}, 100:1182--1186, 2009.

\bibitem{HaiKra14DCHS}
C.~Heinemann and C.~Kraus.
\newblock A degenerating {C}ahn-{H}illiard system coupled with complete damage
  processes.
\newblock {\em Math. Bohem.}, 139:315--331, 2014.

\bibitem{jones2002soft}
R.~A. Jones.
\newblock Soft {C}ondensed {M}atter, 2002.

\bibitem{havela1999}
A.~V. Kolomiets, L.~Havela, V.~A. Yartys, and A.~V. Andreev.
\newblock Hydrogenation and its effect on crystal structure and magnetism in
  ${R}{E}${N}i{A}l intermetallic compounds.
\newblock {\em J. Phys. Studies}, 3:55--59, 1999.

\bibitem{kolwicz2005specific}
L.~Kolwicz-Chodak, Z.~Tarnawski, H.~Figiel, A.~Budziak, T.~Dawid, L.~Havela,
  A.~Kolomiets, and N.-T. Kim-Ngan.
\newblock Specific heat anomalies in {RMn}${}_2${(H, D)}${}_x$ hydrides.
\newblock {\em J. Alloy. Compd.}, 404:51--54, 2005.

\bibitem{KreSte16ENFT}
P.~Krej\v{c}\'{\i} and U.~Stefanelli.
\newblock Existence and nonexistence for the full thermomechanical
  {S}ouza-{A}uricchio model of shape memory wires.
\newblock {\em Math. Mech. Solids}, 16:349--365, 2011.

\bibitem{KrKoKr12MPAS}
J.~Kruis, T.~Koudelka, and T.~Krej\v{c}\'{\i}.
\newblock Multi-physics analyses of selected civil engineering concrete
  structures.
\newblock {\em Comm. Comput. Phys.}, 12:885--918, 2012.

\bibitem{Latro2004JPCS}
M.~Latroche.
\newblock Structural and thermodynamic properties of metallic hydrides used for
  energy storage.
\newblock {\em J. Phys. Chem. Solids.}, 65:517 -- 522, 2004.

\bibitem{Marita-new-paper}
G.~Lazzaroni, R.~Rossi, M.~Thomas, and R.~Toader.
\newblock Rate-independent damage in thermo-viscoelastic materials with
  inertia.
\newblock {\em WIAS Preprint No. 2025}, 2014.

\bibitem{LyaHam07DEFF}
V.~Lyakhovsky and Y.~Hamiel.
\newblock Damage evolution and fluid flow in poroelastic rock.
\newblock {\em Izvestiya, Physics of the Solid Earth}, 43:13--23, 2007.

\bibitem{MaiMie05EREM}
A.~Mainik and A.~Mielke.
\newblock Existence results for energetic models for rate-independent systems.
\newblock {\em Calc. Var. Part. Diff. Eq.}, 22:73--99, 2005.

\bibitem{MiRoSt08GLRR}
A.~Mielke, T.~Roub\'{\i}\v{c}ek, and U.~Stefanelli.
\newblock {${\Gamma}$}-limits and relaxations for rate-independent evolutionary
  problems.
\newblock {\em Calc. Var. Part. Diff. Eq.}, 31:387--416, 2008.

\bibitem{MiRoZe10CDEV}
A.~Mielke, T.~Roub\'\i\v{c}ek, and J.~Zeman.
\newblock Complete damage in elastic and viscoelastic media and its energetics.
\newblock {\em Comput. Meth. Appl. Mech. Engrg.}, 199:1242--1253, 2010.

\bibitem{MieThe04RIHM}
A.~Mielke and F.~Theil.
\newblock On rate-independent hysteresis models.
\newblock {\em Nonlin. Diff. Eqns. Appl.}, 11:151--189, 2004.

\bibitem{PoRoTo10TCTF}
P.~{Podio Guidugli}, T.~Roub\'\i\v{c}ek, and G.~Tomassetti.
\newblock A thermodynamically-consistent theory of the ferro/para\-magnetic
  transition.
\newblock {\em Archive Rat. Mech. Anal.}, 198:1057--1094, 2010.

\bibitem{tomassetti2004evolution}
P.~Podio-Guidugli and G.~Tomassetti.
\newblock On the evolution of domain walls in hard ferromagnets.
\newblock {\em SIAM J. Appl. Math.}, 64:1887--1906, 2004.

\bibitem{PodioT2006ITM}
P.~Podio-Guidugli and G.~Tomassetti.
\newblock Magnetization switching with nonstandard dissipation.
\newblock {\em IEEE Trans. Magn.}, 42:3652--3656, 2006.

\bibitem{PodioV2010CMaT}
P.~Podio-Guidugli and M.~Vianello.
\newblock Hypertractions and hyperstresses convey the same mechanical
  information.
\newblock {\em Cont. Mech. Thermodyn.}, 22:163--176, 2010.

\bibitem{RocRos??ESTC}
E.~Rocca and R.~Rossi.
\newblock ``{E}ntropic'' solutions to a thermodynamically consitent {PDE}
  system for phase transitions and damage.
\newblock Preprint arXiv:1403.2577.

\bibitem{rocca2014degenerating}
E.~Rocca and R.~Rossi.
\newblock A degenerating {PDE} system for phase transitions and damage.
\newblock {\em Math. Mod. Meth. Appl. Sci.}, 24:1265--1341, 2014.

\bibitem{roubivcek1989stefan}
T.~Roub{\'\i}{\v{c}}ek.
\newblock The {S}tefan problem in heterogeneous media.
\newblock {\em Ann. Inst. Henri Poincar{\'e}}, 6:481--501, 1989.

\bibitem{Roub10TRIP}
T.~Roub{\'{\i}}{\v{c}}ek.
\newblock Thermodynamics of rate independent processes in viscous solids at
  small strains.
\newblock {\em SIAM J. Math. Anal.}, 42:256--297, 2010.

\bibitem{Roub13NPDE}
T.~Roub{\'{\i}}{\v{c}}ek.
\newblock {\em Nonlinear Partial Differential Equations with Applications}.
\newblock Birkh\"auser, Basel, 2nd edition, 2013.

\bibitem{Roub13NCTV}
T.~Roub{\'{\i}}{\v{c}}ek.
\newblock Nonlinearly coupled thermo-visco-elasticity.
\newblock {\em Nonlin. Diff. Eq. Appl.}, 20:1243--1275, 2013.

\bibitem{RouSte??MSMA}
T.~Roub{\'{\i}}{\v{c}}ek and U.~Stefanelli.
\newblock Magnetic shape-memory alloys: thermomechanical modeling and analysis.
\newblock {\em Cont. Mech. Thermodyn.}, published online 28 Feb 2014,
  \texttt{doi:10.1007/s00161-014-0339-8}.

\bibitem{RoubT2011MMMAS}
T.~Roub{\'i}{\v{c}}ek and G.~Tomassetti.
\newblock Ferromagnets with eddy currents and pinning effects: their
  thermodynamics and analysis.
\newblock {\em Math. Mod. Meth. Appl. Sci.}, 21:29--55, 2011.

\bibitem{roubivcek2013phase}
T.~Roub{\'\i}{\v{c}}ek and G.~Tomassetti.
\newblock Phase transformations in electrically conductive ferromagnetic
  shape-memory alloys, their thermodynamics and analysis.
\newblock {\em Arch. Rat. Mech. An.}, 210:1--43, 2013.

\bibitem{Roubivcek2010}
T.~Roub\'i\v{c}ek and G.~Tomassetti.
\newblock Thermodynamics of shape-memory alloys under electric current.
\newblock {\em Z. Angew. Math. Phys.}, 61:1--20, 2010.

\bibitem{RoubT2014DCDS}
T.~Roub\'i\v{c}ek and G.~Tomassetti.
\newblock Thermomechanics of hydrogen storage in metallic hydrides: modeling
  and analysis.
\newblock {\em Discr. Cont. Dyn. Syst. B}, 14:2313--2333, 2014.

\bibitem{RoubTZ2009JMAA}
T.~Roub\'{\i}\v{c}ek, G.~Tomassetti, and C.~Zanini.
\newblock The {G}ilbert equation with dry-friction-type damping.
\newblock {\em J. Math. Anal. Appl.}, 355:453--468, 2009.

\bibitem{Thom10PhD}
M.~Thomas.
\newblock {\em Rate-independent damage processes in nonlinearly elastic
  materials}.
\newblock PhD thesis, Institut f\"ur Mathematik, Humboldt-Universit\"at zu
  Berlin, February 2010.

\bibitem{ThoMie10DNEM}
M.~Thomas and A.~Mielke.
\newblock Damage of nonlinearly elastic materials at small strain --
  {E}xistence and regularity results --.
\newblock {\em Zeit. Angew. Math. Mech.}, 90:88--112, 2010.

\bibitem{ubachs2004nonlocal}
R.~Ubachs, P.~Schreurs, and M.~Geers.
\newblock A nonlocal diffuse interface model for microstructure evolution of
  tin--lead solder.
\newblock {\em J. Mech. Phys. Solids}, 52:1763--1792, 2004.

\bibitem{visintin1984strong}
A.~Visintin.
\newblock Strong convergence results related to strict convexity.
\newblock {\em Comm. Part. Diff. Eq.}, 9:439--466, 1984.

\bibitem{visintin1996models}
A.~Visintin.
\newblock {\em Models of Phase Transitions}.
\newblock Birkh\"auser, Boston, 1996.

\bibitem{wang2004poroelasticity}
H.~Wang.
\newblock {\em Theory of {L}inear {P}oroelasticity with {A}pplications to
  {G}eomechanics and {H}ydrogeology}.
\newblock Princeton University Press, 2000.

\bibitem{YTLL06ESTP}
Q.~Yan, H.~Toghiani, Y.-W. Lee, K.~Liang, and H.~Causey.
\newblock Effect of sub-freezing temperatures on a {PEM} fuel cell performance,
  startup and fuel cell components.
\newblock {\em J. Power Sources}, 160:1242--1250, 2006.

\end{thebibliography}
\end{document}